\documentclass{aa}  
\usepackage{enumerate}
\usepackage{graphicx}
\usepackage{diagbox}
\usepackage{txfonts}
\def\iso#1{$^{#1}$}
\def\msun{M$_\odot$}

\begin{document}

   \title{Barium stars as tracers of $s$-process nucleosynthesis in AGB stars I. 28 stars with independently derived AGB mass \\
}

   \author{B. Cseh\inst{1}\fnmsep\thanks{cseh.borbala@csfk.org},
B. Világos\inst{1,2},
M. P. Roriz\inst{3},
C. B. Pereira\inst{3},
V. D'Orazi\inst{4,5},
A. I. Karakas\inst{5,6},
B. Soós\inst{1,2},
N. A. Drake\inst{7,8},
S. Junqueira\inst{3},
\and
M. Lugaro\inst{1,2,5}}

   \institute{Konkoly Observatory, Research Centre for Astronomy and Earth Sciences, Eötvös Loránd Research Network (ELKH), H-1121 Budapest, Konkoly Thege M. \'ut 15--17, Hungary
    \and
    ELTE E\"{o}tv\"{o}s Lor\'and University, Institute of Physics, Budapest 1117, P\'azm\'any P\'eter s\'et\'any 1/A, Hungary
    \and
    Observat\'orio Nacional/MCTI, Rua General Jos\'e Cristino, 77 Sao Cristovao, Rio de Janeiro, Brazil
    \and
    INAF Osservatorio Astronomico di Padova, Vicolo dell’Osservatorio 5, 35122 Padova, Italy
    \and
    School of Physics and Astronomy, Monash University, VIC 3800, Australia
    \and
    ARC Centre of Excellence for All Sky Astrophysics in 3 Dimensions (ASTRO 3D)
    \and
    Laboratory of Observational Astrophysics, Saint Petersburg State University, Universitetski pr. 28, 198504, Saint Petersburg, Russia
    \and
    Laboratório Nacional de Astrofísica/MCTI, Rua dos Estados Unidos 154, Bairro das Nações, 37504-364, Itajubá, Brazil
    \\
             }

   \date{Received ; accepted }

 
  \abstract
    {Barium (Ba) stars are polluted by material enriched in the $slow$ neutron capture ($s$-process) elements synthesised in the interior of their former asymptotic giant branch (AGB) companion star, which is now a white dwarf.}
   {We compare individual Ba star abundance patterns to AGB nucleosynthesis model predictions to verify if the AGB model mass is compatible with independently derived AGB mass previously estimated using binary parameters and Gaia parallax data.}
   {We selected a sample of 28 Ba stars for which both self-consistent spectroscopic observation and analysis are available and stellar mass determinations, via positioning the star on the HR diagram and comparing with evolutionary tracks.
   For this sample stars we considered both previously (Y, Zr, Ce, and Nd) and recently derived (Rb, Sr, Nb, Mo, Ru, La, Sm, and Eu) elemental abundances. Then, we performed a detailed comparison of these $s$-process elemental abundances to different AGB nucleosynthesis models from the Monash and the FRUITY theoretical data sets. We simplified the binary mass transfer by calculating dilution factors to match the [Ce/Fe] value of each star when using different AGB nucleosynthesis models, and we then compared the diluted model abundances to the complete Ba-star abundance pattern.}
   {Our comparison confirms that low mass (with initial masses roughly in the range 2--3 \msun), non-rotating AGB stellar models with \iso{13}C as the main neutron source are the polluters of the vast majority of the considered Ba stars. Out of the 28 stars, in 21 cases the models are in good agreement with both the determined abundances and the independently derived AGB mass, although in 16 cases higher observed abundances of Nb, Ru, Mo and/or Nd, Sm than predicted.  For 3 stars we obtain a match to the abundances only by considering models with masses lower than those independently determined. Finally, 4 stars show much higher first $s$-process peak abundance values than the model predictions, which may represent the signature of a physical (e.g. mixing) and/or nucleosynthetic process that is not represented in the set of models considered here.}
  {}
  
   
   \authorrunning{B. Cseh et al.}           
   \titlerunning{Ba stars as tracers of $s$-process nucleosynthesis in AGBs I.}
   \maketitle


\section{Introduction}
\label{sec:intro}

Barium (Ba) stars are of undisputable significance for the understanding of  nucleosynthetic processes in the interior of asymptotic giant branch (AGB) stars. A Ba star is a dwarf or a giant of F-K spectral class and typical metallicity from solar to roughly $-$0.6 dex \citep[see, e.g. Fig. 3a in][]{deC}, whose spectrum show signatures of enhancement relative to solar of $slow$ neutron-capture ($s$ process) elements heavier than Fe, as first reported by \citet{Bafirst}. Since the discovery of the binary nature of Ba stars \citep{RV1-McCFN, RV2-McC} it became widely accepted that the overabundance of these elements originates from mass transfer from a former AGB (now white dwarf) companion.

Two different neutron source reactions can provide neutrons for the $s$ process to produce the elements heavier than Fe in AGB stars \citep{gallino98,lugaro03,cristallo09,kaeppeler11}. The bulk of the neutrons in AGB stars of low-mass, in the range from $\simeq$ 1.5 \msun~(or maybe even lower, see discussion in Sect. \ref{sec:discussion} and \citealp{shetye19_1msun}) to 4 \msun, are produced within the He-rich intershell region (i.e. the region located between the He- and the H-burning shells) in a so-called \iso{13}C pocket through the \iso{13}C($\alpha$,n)\iso{16}O reaction. 
This reaction requires lower temperature ($\simeq$ 100 MK) to be activated than the \iso{22}Ne($\alpha$,n)\iso{25}Mg reaction, which needs T $\gtrsim$ 300 MK and is dominant instead in AGB stars of mass $\gtrsim$ 4 \msun. There temperatures can reach above this limit during the recurrent convective episodes of He burning (the thermal pulses). 
The \iso{22}Ne neutron source produces higher neutron densities than the \iso{13}C neutron source, therefore activating branching points along the $s$-process path, and leading for example to the overproduction of Rb with respect to neighbouring elements such as Sr and Zr. As such, Rb has been used as an indicator of the AGB stellar mass \citep{abia02,karakas12,vanraai12,roriz21}.

To characterise $s$-process abundance patterns, ratios of the elements at the first (neutron magic number of 50, i.e. Sr, Y, Zr) and second $s$-process peaks (neutron magic number of 82, i.e. Ba, La, Ce) have been widely used. This is because the relative abundances of these elements can provide a direct constraint to the total number of free neutrons available at the $s$-process site \citep[see, e.g.][for more details]{cseh18,lugaro20} 
Such ratios are particularly useful for Ba stars because the dilution due to the mass transfer affects all $s$-process elements in a similar way, and therefore their ratios closely reflect the intrinsically produced nucleosynthetic ratio.

In \citet{cseh18} we compared the large sample of 169 Ba stars observed by \citet{deC} to AGB models as a population and found that the distribution of the ratios of second-peak (Ce and Nd) to first-peak elements (Y and Zr) as function of [Fe/H]\footnote{throughout the paper we will use the usual spectroscopic notation [Y/X] = log$_{10}$(Y/X)$_{star}$$-$log$_{10}$(Y/X)$_{\odot}$} 
can be well described by non-rotating AGB models. We concluded that the metallicity is the best predictor of the $s$-process pattern, with lower metallicities favouring the production of the elements of the second peak relative to the first peak, as observed in the Ba stars. This results appear to be confirmed by recent observations of 10 Ba stars \citep{Shejeelammal20}. 

However, for each metallicity, a spread in the second to first-peak element ratios is present in the sample, which needs to be explained by variations in other stellar features, such as the initial mass, but also possibly rotation, and/or diffusive mixing. To disentangle these latter effects from the effect of the initial mass we need to compare the individual detailed stellar abundance pattern to single AGB models and verify if the mass of the stellar model corresponds to the mass of the AGB companion determined independently, when available. 

To remove the effects of binary mass transfer and mixing on the secondary Ba star
and select those effects directly related to nucleosynthesis, the observed [X/Fe] abundance ratios should be compared to the models relative to each other, rather than in absolutes terms. The mixing in our sample of giant Ba stars is mostly due to the fact that these stars undergo the first dredge up when ascending the first giant branch \citep{Becker79,Boothroyd99}, which mixes the material that was transferred onto the stellar surface with the whole envelope mass.
To mimic the dilution due to the mass transfer and mixing, a free parameter is therefore introduced, usually referred to as the {\it dilution factor (dil)}, representing the ratio between the total mass, i.e. the original envelope mass of the Ba star plus the mass transferred from the AGB star to its companion, and the transferred AGB material, and thus dil = M$_{\rm total}$ / M$_{\rm AGB,trans}$ = (M$_{\rm Ba,env}$ + M$_{\rm AGB,trans}$) / M$_{\rm AGB,trans}$ \citep[see e.g.][]{husti09, Bisterzo12}.
For Ba dwarf stars without extended convective envelopes the transferred $s$-process-rich material is deposited on top of the surface of the star and dilution is usually assumed to be inefficient\footnote{We note that mixing mechanism as thermohaline mixing could still play a role, see, e.g. \citet{Stancliffe07,Stancliffe08} for the case of carbon-enhanced metal-poor (CEMP) stars (the low-metallicity counterparts of Ba stars) and \citet{husti09} for Ba stars. Although, \citet{Denissenkov08} and \citet{Aoki08} suggest that thermohaline mixing is inefficient in dwarf CEMP stars.}, so that dil is practically 1. In the case of giant Ba stars, such as those analysed by \citet{deC} and considered here, the effect of dilution cannot be neglected due to the large convective envelope in which the $s$-process enriched material is mixed after being transferred, and thus dil $>$ 1.

Several studies have included comparison of individual AGB models and Ba star abundances including the estimate for calculations of dilution factors. For example, \cite{husti09} examined 34 Ba dwarfs and giants from two different studies and compared their abundance patterns to theoretical models calculated using evolutionary models from the FRANEC code \citep{FRANEC89} and the Torino $s$-process post-processing code. These authors found that AGB models in the mass range from 1.5 \msun~to 3.0 \msun~and dilution factors from 1 (for dwarf stars) to 30 were required to cover the observed abundances in the sample stars considered.
A recent study of \cite{Shejeelammal20} for Ba and CH stars provides abundances for many first and second $s$-process peak elements, and also including C, Rb and Eu. Comparing the derived abundances to FRUITY nucleosynthesis models these authors confirm that AGB stars with mass $\lesssim$ 3 \msun~are the typical former companions of  Ba stars. However, for these previous studies only the 
Ba star metallicity ([Fe/H]) is available from the observations, while the initial AGB stellar mass is unknown and therefore can be treated as a free parameter to match the observations. This hampers the possibility to determine if variations or mismatches are due to a different stellar mass or to other stellar physics or nucleosynthetic features. 

Thanks to the Gaia survey, it has been recently possible to provide better determinations for the masses of Ba stars \citep{jorissen19}, using the Gaia DR2 parallaxes \citep{Gaia18} and the spectral-energy distribution of the stars. It is therefore possible to estimate the mass of the white dwarf companion using assumptions on the inclination of the binary system, the mass ratio of the members (or Q value, where Q = M$_{\rm WD}^3$/(M$_{\rm Ba}$ + M$_{\rm WD}$)$^2$, see Sect. \ref{sec:stars} and \citet{jorissen19}, derived from the masses of the components, where M$_{\rm Ba}$ is the mass of the Ba star (Ba) and M$_{\rm WD}$ is that of the white dwarf (WD)). Then from the WD mass one can obtain an estimate for the initial mass of the AGB star that polluted the observed Ba star (hereafter, M$^{\rm{J}19}_{\rm{AGB}}$) using an initial-final mass relationship of \citet{El-Badry18}. By comparing the sample from \citet{jorissen19} to that of \citet{deC}, we have found 28 Ba stars in common, for which both an estimate of the initial mass of the AGB star and high-resolution spectroscopic abundances are available. Furthermore, the sample of \citet{deC} has been analysed for more elements \citep[Rb, Sr, Nb, Mo, Ru, La, Sm, and Eu,][]{roriz21, Roriz_heavy} including a re-analysis of La, which was previously found to be affected by the saturation of strong lines \citep{cseh18}.

Armed with such a set of unprecedented constraints, we can now provide the best comparison for Ba star data to AGB models, to verify to which extent AGB model predictions match the data and to attempt to disentangle the potential effects of different stellar features on the $s$ process. Furthermore, we provide estimates of the dilution factors, which, together with the known orbital parameters of these systems will also pave the way to a better understanding of mass transfer in these systems, which is still uncertain \citep[see, e.g. Sect. 5.4 of][]{demarco17}. 

The paper is organised as follows. In Sect. \ref{sec:stars} we present our sample data and summarise the new elemental abundances from \citet{roriz21, Roriz_heavy}. In Sect. \ref{sec:methods} we describe our methodology to find potential fits for each star using both the FRUITY and the Monash set of yields, and to calculate the dilution factor for each potential solution. In Sect. \ref{sec:results}, we present our comparison between models and determined abundances for each individual star.
Finally, we discuss our results in Sect. \ref{sec:discussion} and we summarise our results in Sect. \ref{sec:conclusions}.


\section{Sample stars and abundances}
\label{sec:stars}

\subsection{Sample stars}

\begin{figure*}[!ht]
 \caption{[Ce/Y] values as function of metallicity for the stars analysed in \cite{deC} and \cite{cseh18} (grey dots), with the 28 stars analysed in this work indicated by the names and highlighted as black, blue, and red dots, with colour depending on the Groups 1, 2, and 3 presented in Sect. \ref{sec:results}, respectively. For sake of clarity, the error bars are marked only for the stars analysed in this paper and the three Ba stars at [Fe/H] $\approx-$1 are not shown.}
 \label{fig:all_stars}
 \centering
 \includegraphics[width=0.8\hsize]{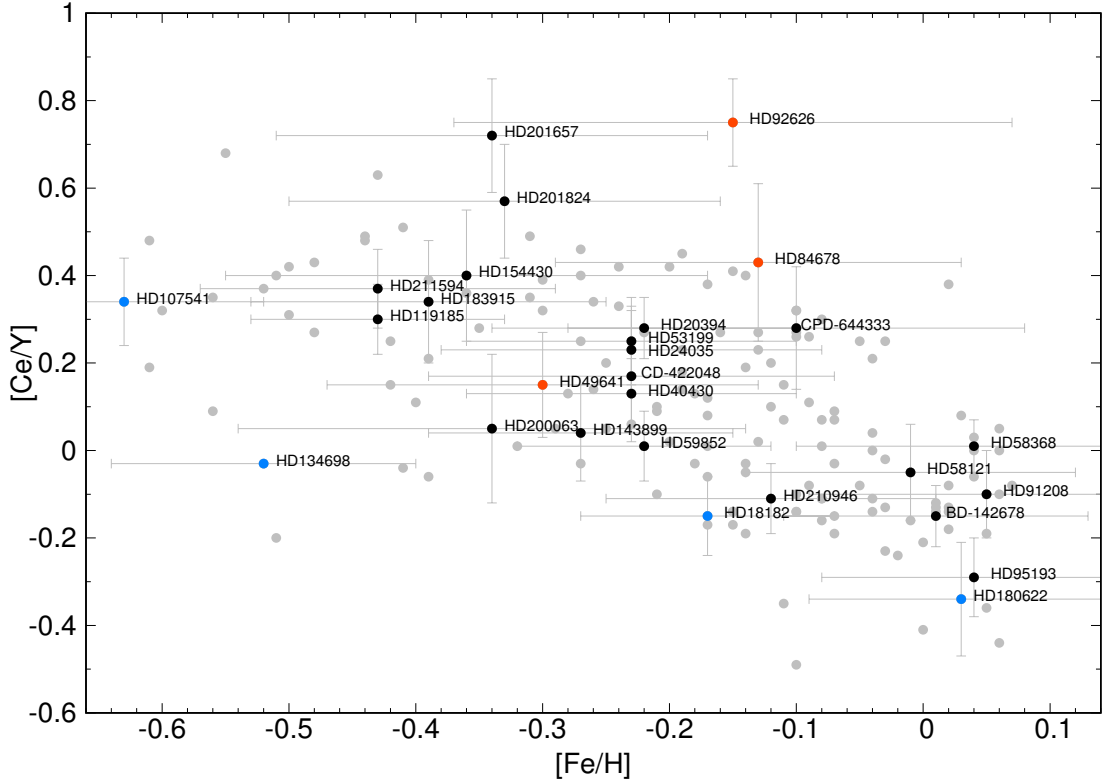}
 \end{figure*}

Out of the 169 Ba giants analysed in \citet{deC} and \citet{cseh18}, 28 stars were also included in the sample of \citet{jorissen19} aimed at determining the orbital parameters of Ba and related stars. The locations in the [Ce/Y] versus [Fe/H] plane of the stars analysed in this paper relative to the whole sample of \citet{cseh18} are shown in Fig. \ref{fig:all_stars}, which shows that the [Ce/Y] ratio of the 28 selected stars is typical of the entire sample, and it covers the whole range of the sample stars.
This means that our sample allows us to investigate the origin of the spread present at any given [Fe/H]. 
The masses of the Ba stars (M$_{\rm Ba}$) were determined by \citet{jorissen19} from the location of these stars on the HR diagram and using the STAREVOL evolutionary models \citep{starevol}. \citet{jorissen19} also calculated the companion white dwarf (WD) masses (M$_{\rm WD}$) from the mass function of the system assuming a random distribution of the orbital inclination and assuming that Q = M$_{\rm WD}^3$/(M$_{\rm Ba}$ + M$_{\rm WD}$)$^2$ is a constant (the 'constant Q assumption'). The initial AGB masses (M$_{\rm{AGB}}^{\rm{ J19}}$) were then determined from the estimated WD masses using the initial-final mass relation derived by \citet{El-Badry18}, as described in \citet{jorissen19}, see their Fig. 14, and assuming that the binary evolution did not affect this relation. For more details on this method see also \citet{escorza17} and \citet{vanderswaelmen17}.

Many uncertainties affect the derivation of this initial AGB mass. First, in relation to the mass of the Ba star, the Gaia DR2 parallaxes \citep{Gaia18} were derived using single-star solutions. Since our sample stars are in a binary system, this might introduce errors already in the derivation of M${_\mathrm{Ba}}$. As for the re-normalised unit-weight error (RUWE) -- 
a value of which above roughly 1--1.4 indicates potential problems with the single-star astrometric solution \citep[see, e.g.][]{El-Badry21,Stassun21}, only one star in our sample (BD $-$14$^{\circ}$2678) has RUWE < 1 (0.919), while 18 stars have RUWE > 1.4 \citep{gaia_ruwe}.
Second, there are several problems in deriving the AGB initial mass from the Ba stars mass. On top of the strong constant Q assumption, and the assumption that the orbital inclination is randomly distributed, neglecting the binary evolution in the initial-final mass relationship can introduce further uncertainties. In spite of these difficulties, a baseline evidence of the validity of the method is provided by the fact that, as shown by \citet{jorissen19}, most of the systems analysed have M$^{\rm J19}_{\mathrm{AGB}}$/M$_{\mathrm{Ba}}$ > 1, as expected. Overall, since the initial-final mass relationship of \citet{El-Badry18} gives an initial AGB mass of 2.75 $\pm^{0.36}_{0.31}$ for a 0.67 \msun~WD, and higher mass WDs have larger uncertainty, we assume that the uncertainty on the derived initial AGB mass is of at least $\pm$ 0.5 \msun.

Finally, we note that the classification criteria for Ba stars is still not homogeneous in the literature. In \cite{deC} a star is considered as a Ba star if the condition [s/Fe] $\geq$ 0.25 dex is fulfilled, where s is the average of Y, Zr, La, Ce, and Nd. In \citet{jorissen19} the [Ce/Fe] ratio was used to classify and distinguish between strong and mild Ba stars: if the value is over $\approx$ 1 dex, a Ba star is classified as strong, while mild Ba stars have [La/Fe] and [Ce/Fe] values between 0.2 and 1 dex.
In the below we will categorise the sample stars into 3 groups, depending on the match between the $s$-process elemental abundances and the AGB models. Stars with good agreement between the models and the $s$-process peak elements belong to Group 1, Group 2 includes 4 stars with low estimated Ba and AGB mass (< 2 \msun) and with higher first s-process peak than typically predicted, while Group 3 includes the 3 stars in our sample that have the highest initial AGB mass (> 3.8 \msun), as derived under the constant Q assumption \citep{jorissen19}.

\subsection{Elemental abundances}

On top of Y, Zr, Ce, and Nd from \citet{deC}, our present analysis has the advantage of the determination of new elemental abundances for Rb, Sr, Nb, Mo, Ru, La, Sm and Eu, which allow us a more detailed comparison between models and observational data.
Rb abundances have been published by \citet{roriz21} for the whole sample of \citet{deC}, while abundances for the other species are presented by \citet{Roriz_heavy}. As mentioned in Sect. \ref{sec:intro}, Rb is a key element to distinguish between reactions responsible for the neutron source in the polluting AGB star and also for the determination of the AGB mass, since its abundance depends on the activation of branching points on the $s$-process path. 

The final abundances were computed using the current version of the spectral analysis code MOOG \citep{SnedenPhDT}, which assumes a plane-parallel atmosphere model, under the conditions of local thermodynamic equilibrium (LTE), from either equivalent width measurements or spectral synthesis of the atomic lines considered. 
For the [X/Fe] ratios considered in this study, uncertainties were estimated from the variation in the abundances due to changes in the parameters of the atmospheric models, as well as from dispersion of the abundance, when three or more lines were available for a given element. See \citet{deC, roriz21, Roriz_heavy} for details of the analysis and the list of lines used to obtain the abundances considered in this work.

\begin{figure}[!ht]
 \caption{[Ce/Fe] (orange asterisks) and the new [La/Fe] (black dots) abundances for our sample stars.}
 \label{fig:LaCe_abund}
 \centering
 \includegraphics[width=\hsize]{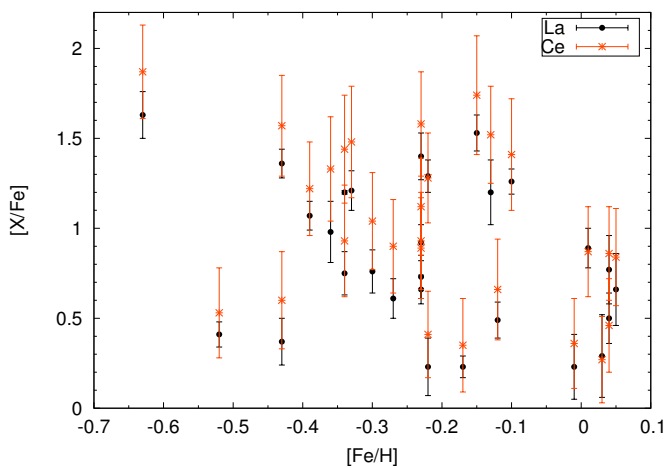}
 \end{figure}

\begin{figure*}[!ht]
 \caption{Heavy element abundances for all the sample stars. The colours are the same as in Fig. \ref{fig:all_stars}.}
 \label{fig:abund_multi}
 \centering
 \includegraphics[width=0.8\hsize]{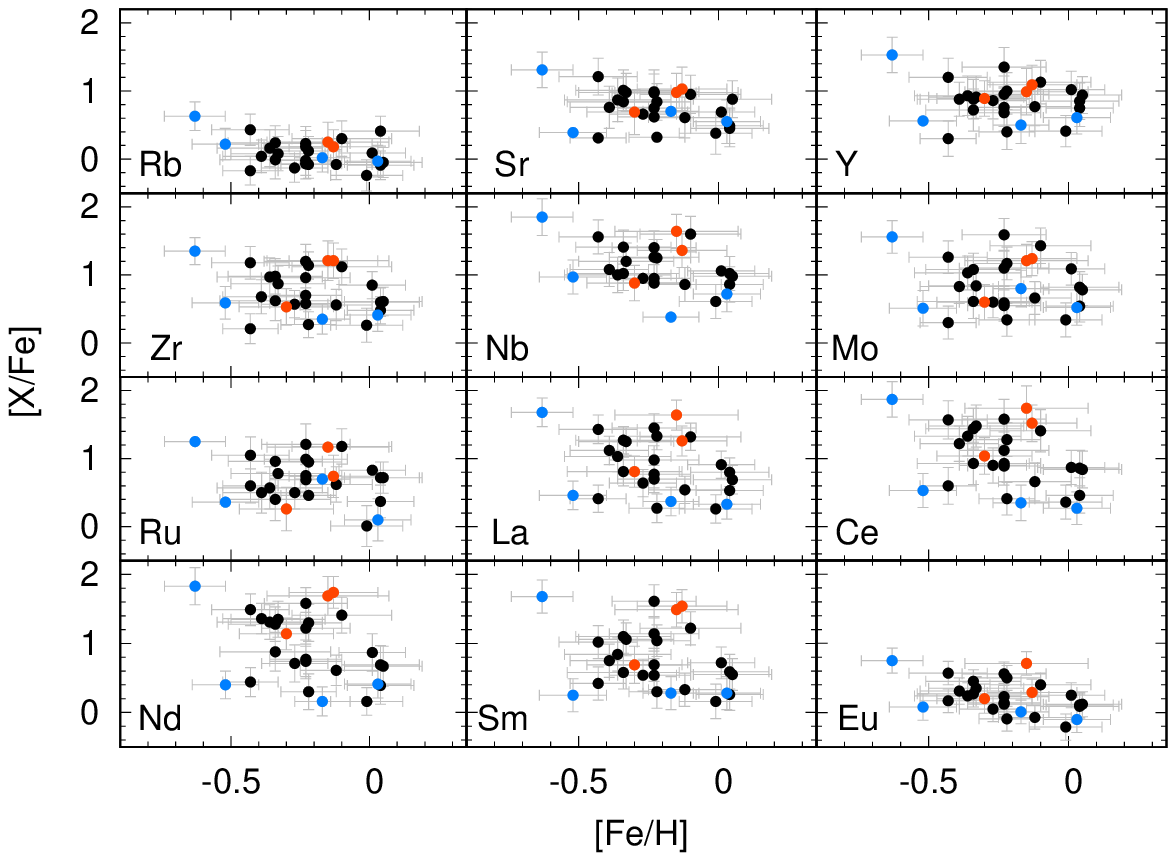}
 \end{figure*}

Although [La/Fe] ratios were published in \citet{deC}, some of the stars showed unexpectedly high values when compared to the ratios [Ce/Fe] and [Nd/Fe], as pointed out in \citet{cseh18}. Since La, Ce, and Nd belong to the same $s$-process peak, these elements are expected to exhibit similar overabundances relative to solar. This is, for example, the case of the star CPD $-64^{\circ}$4333, for which \citet{deC} reported [La/Fe] = 2.52 dex, approximately 1 dex higher than the values of [Ce/Fe] and [Nd/Fe] for the same object. Therefore, La abundances were re-determined \citep{Roriz_heavy}. 
We compare the new [La/Fe] ratios to the [Ce/Fe] ratios in Fig. \ref{fig:LaCe_abund} for the sample considered in the present study. The new [La/Fe] ratios are consistent with the [Ce/Fe] ratios.
We show in Fig. \ref{fig:abund_multi} the [X/Fe] ratios versus metallicity for elements between Rb and Eu, highlighting the stars belonging to our Groups 1, 2 and 3 classification (see Sect. \ref{sec:results}). 
 
\section{Methods}
\label{sec:methods}

\subsection{Stellar models}
\label{sec:models}

We compare the abundances of each individual Ba star with the two most extensive currently available set of AGB nucleosynthesis models:
Monash \cite[]{lugaro12,fishlock14,karakas16,karakas18} and FRUITY \cite[]{cristallo09,cristallo11,piersanti13,cristallo15,cristallo16}. Other sets of models 
were not taken into account because of the limited number of models in mass and metallicity.
Most of the Monash models with a given mass and metallicity are calculated also varying the masses of the partial mixing zone (M$_{\rm PMZ}$) leading to the formation of the \iso{13}C pocket different from the standard value reported in \citet{karakas16}. This parameter mostly affects the absolute amount of the produced $s$-process material (see also \citealt{buntain17}). If more models with different pocket sizes are applicable to a star, we show all of them in our comparison.
New unpublished Monash models with masses of 2 and 3 \msun~were calculated for Z = 0.01 ([Fe/H] $=-0.15$) and with standard M$_{\rm PMZ} = 2\times10^{-3}$ \msun, to have more Monash models closer in metallicity to several stars in our sample.

Within the FRUITY models the largest $s$-process production occurs in the models with an extended \iso{13}C pocket (labelled as 'TAIL'). Furthermore, FRUITY models were also considered here that include rotation, with different initial rotation velocities. As it was shown by \cite{cseh18}, rotating models have significantly lower values of the [Ce/Y] ratio than the large Ba star sample of \cite{deC}. Nevertheless, for the sake of completeness we include also the FRUITY rotating models when they are applicable as a match for a given Ba star. Note that we only selected models with  [$s$/Fe] $\geq$ 0.25 dex (where $s$ is the average of Y, Zr, La, Ce and Nd), since this criteria was applied as a limit for our sample of Ba stars.

The models we have used to compare to each individual star have metallicity [Fe/H] 
in the range of the uncertainties of the given star, and initial mass close to that estimated by \citet{jorissen19} for the same star. In some cases, we extended the range of mass of the considered models if the initial search for a good match with the derived [Rb/Fe] that, as described above, is the strongest diagnostic of stellar mass was not successful. For some stars (the 3 stars in Group 3 described in Sect. \ref{sec:high_mass}, specifically), there was no matching model found nearby the estimated values. In those cases we considered all the models in order to find the most appropriate match, even if they were far from the mass estimates by \citet{jorissen19}.

\begin{figure*}[!ht]
\caption{Example of selected models with the standard \iso{13}C pocket showing the final surface abundances without applying any dilution, ($\delta=1$ for all models, top panel) and with applying a dilution (bottom panel) such that [Ce/Fe] = 0.451 dex, which is the value produced by the FRUITY 4 \msun~model. Thus, this model has $\delta =$ 1. 
Specific properties of the FRUITY models are given in parentheses after the mass of each model: '-\mbox{-}-' indicates the standard models, while 'TAIL', 'rotating' and 'rotating TAIL' models with 60 km/s initial rotational velocity are indicated by (ext), (rNN) and (T60), respectively, where NN is the initial rotational velocity in km/s. The numbers given in parentheses after the mass of each Monash model is the M$_{\mathrm{PMZ}}$ in units of 10$^{-4}$ \msun.
We note that the high N and Na values shown by the undiluted 4 \msun~FRUITY model are due to production by the first dredge up, the other models do not show these high values because they are diluted. Also note that the FRUITY Nb and Sm abundances are lower than in the Monash models because $^{93}$Zr and $^{147}$Pm are not decayed.\newline}
  \label{fig:models}
 \centering
 \includegraphics[width=0.9\hsize]{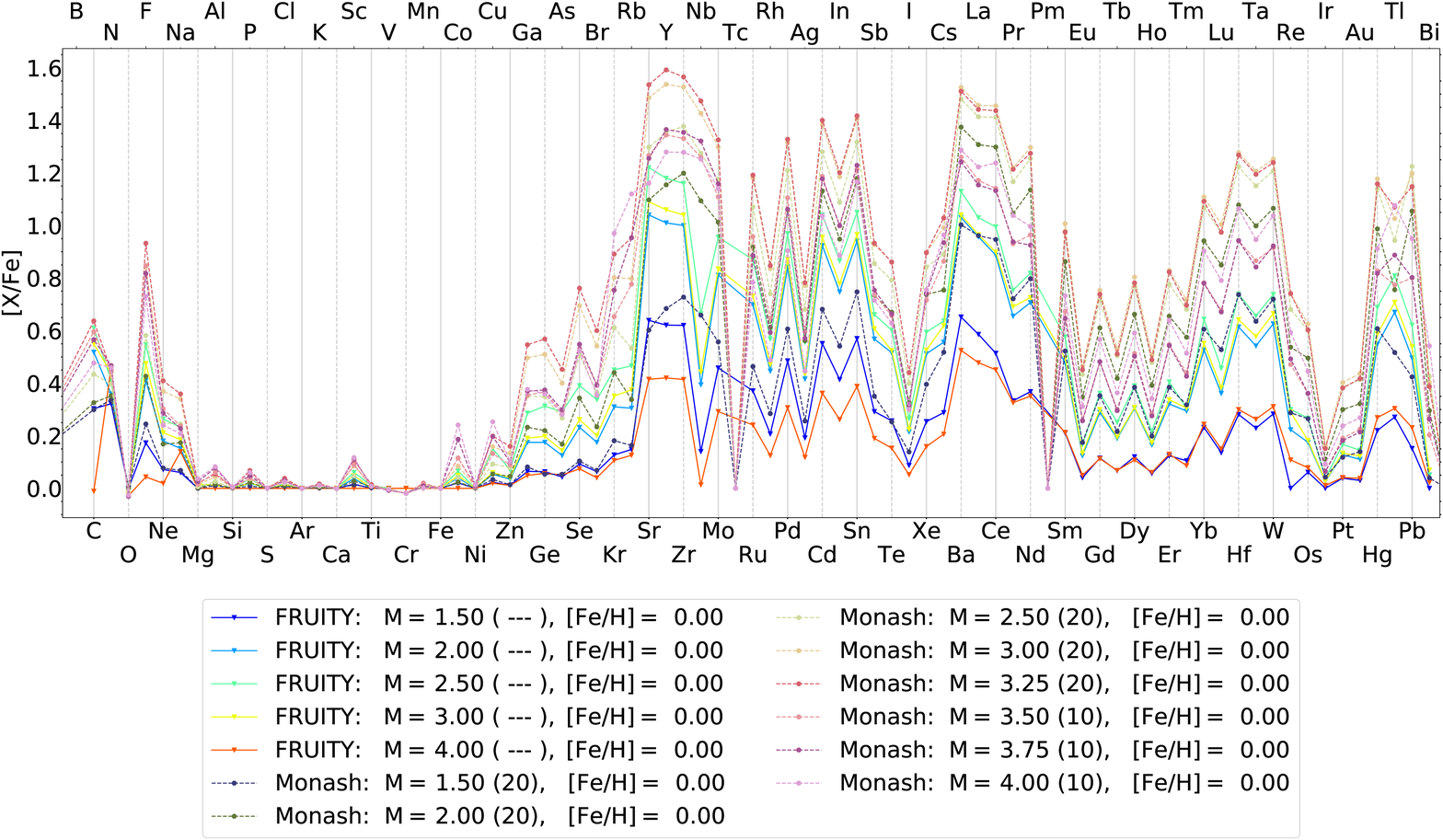}
 \includegraphics[width=0.9\hsize]{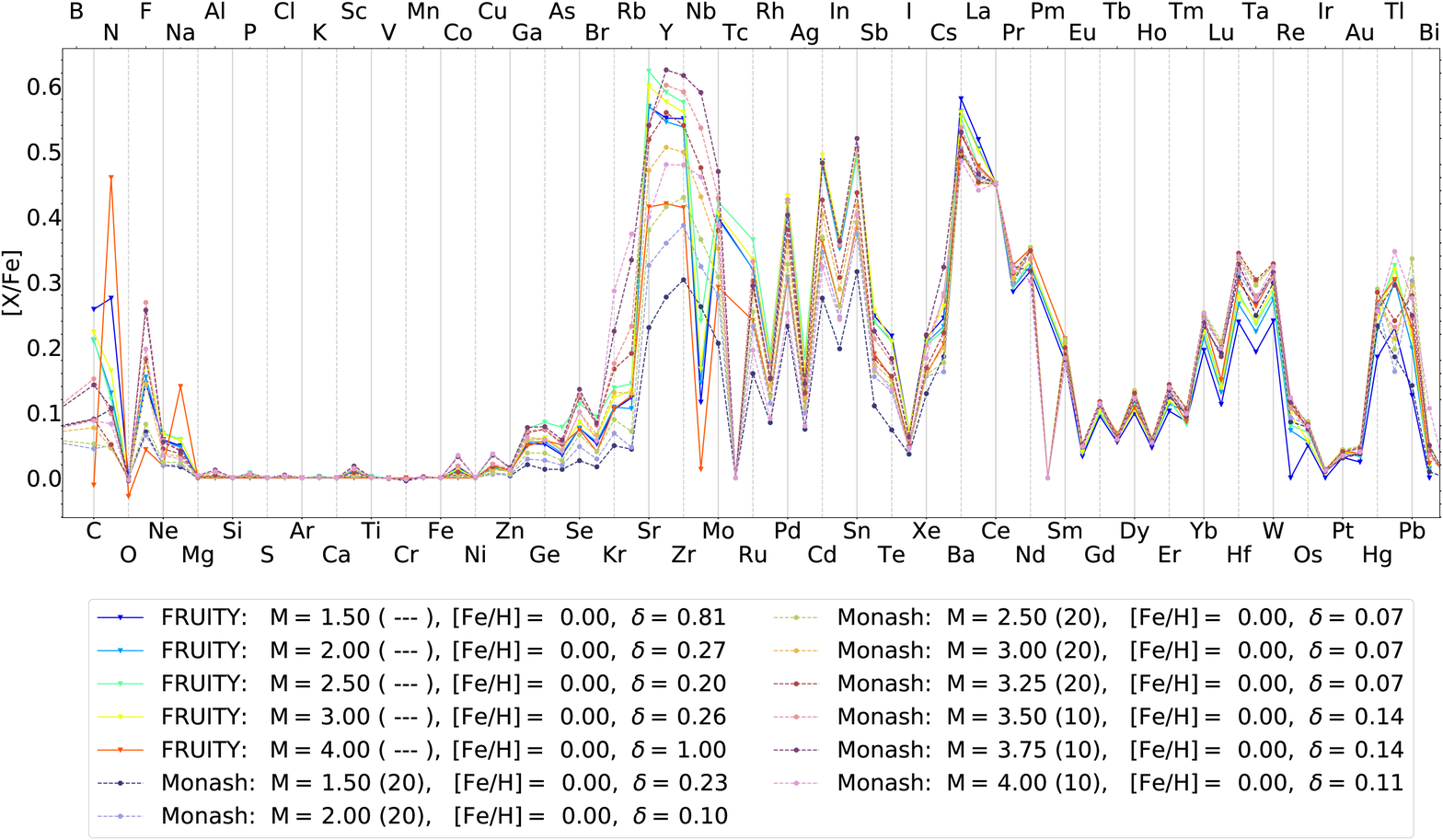}
 \end{figure*}

In Fig. \ref{fig:models} we show an example of solar metallicity ([Fe/H] = 0) Monash and FRUITY models between 1.5 and 4 \msun~with their standard \iso{13}C pockets. The top panel of the figure shows that the FRUITY models have typically lower abundances, which is likely due to smaller TDU efficiencies and smaller \iso{13}C pocket masses, on average. In fact, the Monash models typically experience more thermal pulses (TP) and show deeper TDU, along with hotter burning at the base of the TP and of the envelope. 
The largest difference is between the 4 \msun~models. This can be attributed to the fact that the \iso{22}Ne neutron source is more activated in the Monash models than in the FRUITY from mass $\approx$ 4 \msun. Also the decrease in the TDU efficiency has a large impact on the FRUITY 4 \msun~and higher mass models, strongly reducing the final surface abundances of the $s$-process elements compared to lower mass models \citep{cristallo15}. The 4 \msun~Monash model still has an active \iso{13}C neutron source, thus producing more $s$-process elements compared to the FRUITY models. For a detailed comparison of the two model sets in relation to the production of $s$-process elements we refer the reader to Sect. 5 of \citet{karakas16}.
Nevertheless, the relative abundance distributions are close to each other with similar height of the first two $s$-process peaks, and a lower third process peak. Rb is the element most sensitive to the stellar mass, with an abundance that increases with stellar mass, relative to the first peak elements.

Finally, we note that other stellar processes not explicitly considered in the models used here could affect our analysis. For example, the models used in this study do not include any core overshoot \citep{cristallo11,karakas14, cristallo15b}. Core overshoot would increase the mass of the H-exhausted core at the end of the main sequence and thus the star would behave as if it started with a slightly higher mass. The TDU may therefore occur at lower initial stellar masses than in the models used here leading to degeneracy between different masses. From studies of the core He-flash \citep[e.g.][]{Bertelli86} the maximum mass shift is from $\approx$ 2 \msun~to $\approx$ 1.8 \msun~(roughly 0.2 \msun~lower). This means that our models at around 1.5 \msun~would be lowered to $\approx$ 1.3 \msun, when including the core overshoot.

\subsection{The effect of dilution}
\label{sec:dil_fact_calc}

The effect of dilution after the mass transfer from the former AGB companion cannot be neglected, since all of the sample stars analysed in this paper are giant stars with an extended convective envelope. This means that the transferred material was mixed with the envelope of the current Ba star because of the first dredge up, lowering the abundances relative to Fe as compared to their value in the envelope of the polluting AGB star. 
The diluted model abundances ([X/Fe]$_\mathrm{dil}$) were calculated as following:
\begin{equation}
\label{eq:dil_eq}
\mathrm{[X/Fe]_{dil}} = \log \left( 10^{\mathrm{[X/Fe]_{ini}}} \times (1 - \delta) + 
    10^{\mathrm{[X/Fe]_{AGB}}} \times \delta \right), 
\end{equation}
where [X/Fe]$_\mathrm{ini}$ is the initial (solar) abundance ratio assumed in all the models and corresponds to the initial composition of the Ba star envelope, [X/Fe]$\mathrm{_{AGB}}$ is the final surface abundance ratio predicted by the AGB models, and $\delta$ = 1/dil. We derived the value of $\delta$ for each star and each model by equating [Ce/Fe]$_\mathrm{dil}$ to the [Ce/Fe] ratio observed in the Ba star. We subsequently applied this same value of $\delta$ to all the other elemental abundances calculated with Eq. \ref{eq:dil_eq} above. 

In the bottom panel of Fig. \ref{fig:models} we apply a dilution to normalise all the models to the same [Ce/Fe] value. This representation even more clearly shows the changes in [Rb/Fe], along with variations of the first peak elements with varying the mass. As we move from lower to higher mass models, [Rb/Fe] increases with increasing mass, for similar $\delta$ values. Generally, FRUITY models have higher $\delta$ values than Monash models, which is due to the lower predicted abundances at the AGB surface.

Since $\delta = 1$ means that the AGB material is deposited on top of the Ba star without any mixing, to allow some mixing of the transferred material with the envelope of the giant, we selected only models with $\delta \leq$ 0.9 as possible solutions. This value would imply that 90\% of the whole Ba star envelope is made of AGB material, which is still unrealistic, but allows us to investigate the behaviour of the $\delta$ values. We will discuss this point in more detail in Sect. \ref{sec:delta}. 

\subsection{Final output}
\label{sec:final_output}

Following the selection of the model and the dilution, our algorithm automatically generates a figure for each star, which shows, as function of the atomic number, [X/Fe] for all the available observational data (from the light to the heavy elements) and [X/Fe]$_\mathrm{dil}$ from all the models of the selected mass that match the first peak elements, especially Rb. The reason behind the choice of matching these elements is that we want to verify if the independent estimate of M$^{\rm{J}19}_{\rm{AGB}}$ is consistent with our mass estimate based on the chemical pattern. Since we are guided by the observation of the first peak elements only, we decided to not use the $\chi^2$ method usually employed to find the best fitting model to the derived abundance pattern. 
Furthermore, in our case because of the fine grid of models that we are using for the comparison, in terms of mass, metallicity, and \iso{13}C pocket size (see Tables \ref{tab:modelsFRUITY} and \ref{tab:modelsMonash} for FRUITY and Monash models, respectively) and the similarities between many of them (see Fig. \ref{fig:models}) there would be many degenerate solutions with similar $\chi^2$ values, and we would not obtain any strong indication.
We can see this for example in relation to the case of HD 95193 (Fig. \ref{fig:HD95193})
where 12 out of the 14 models plotted, of mass between 2.0 and 3.5 \msun, give practically the same results, especially for the Monash models.
For these reasons  we do not use the $\chi^2$ method in this study and prefer to first rely on the Rb abundance to independently verify the initial AGB mass of the sample stars. 

The residual difference [X/Fe]$-$[X/Fe]$_\mathrm{dil}$ is plotted in the bottom panel of each comparison figure. Below the figure, the legend lists the mass of the Ba star (M$_{\rm Ba}$) and the initial mass of the AGB star (M$_{\rm AGB}^{\rm J19}$) in \msun~units from \citet{jorissen19} and the [Fe/H] from \citet{deC}. All the plotted models with their corresponding line type in the figure and $\delta$ value used are also indicated by their set (Monash or FRUITY), mass M in \msun, M$_{\rm PMZ}$ in case of the Monash models (given in parentheses after the mass of each model in units of 10$^{-4}$ \msun) and metallicity [Fe/H]. For the FRUITY models the 'TAIL', 'rotating' and 'rotating TAIL' models with 60 km/s initial rotational velocity are indicated by (ext), (rNN) and (T60), respectively, where NN is the initial rotational velocity in km/s.

On the basis of the figures produced by our algorithm we classified the 28 stars in the 3 groups described in Sect. \ref{sec:results} below. Most of the stars (21) belong to the Group 1, we show three example figures of this class below and the remaining figures and detailed comments can be found in the Appendix \ref{appendix_figs}. The remaining 7 stars are split into Group 2 (4 stars) and Group 3 (3 stars) (see also Fig.~\ref{fig:all_stars}), for these more special cases we present and discuss all the corresponding figures in the main text.

\section{Results of the individual stars}
\label{sec:results}

We already showed in Fig. \ref{fig:all_stars} how our sample relates to the whole sample from \cite{deC} and the location of the stars also as separated into 3 groups. Group 1 (Sect. \ref{sec:OK_stars}, Figs. \ref{fig:HD154430}, \ref{fig:HD201657}, \ref{fig:CD-422048} and \ref{fig:HD58368}--\ref{fig:HD183915}) includes stars that are in agreement with the appropriate AGB models based on the mass of the initial AGB star estimated by \citet{jorissen19} and using a metallicity close to that of the Ba star, as described in Sect. \ref{sec:methods}.
Group 2 (Sect. \ref{sec:high_first_peak}, Figs. \ref{fig:HD18182}--\ref{fig:HD134698}) includes 4 stars which show higher abundances of their first $s$-process peak elements than typical predicted by models close to M$^{\rm{J}19}_{\rm{AGB}}$.
Group 3 (Sect. \ref{sec:high_mass}, Figs. \ref{fig:HD49641}--\ref{fig:HD92626}) includes 3 stars for which, based on our comparison, M$^{\rm{J}19}_{\rm{AGB}}$ may have been overestimated. 


\subsection{Group 1: abundance patterns in agreement with M$_{\rm AGB}^{\rm{J}19}$}
\label{sec:OK_stars}

The first group contains stars for which the $s$-process abundance pattern, and Rb in particular, is in agreement with the AGB mass derived from the binary parameters. We categorise these stars into 3 subgroups, depending on the deviations of heavy elements other than Rb. The first subgroup (a) includes stars that have the whole heavy elements pattern in agreement with the models. Stars showing higher Nb and sometimes Mo and Ru abundances than the model predictions belong to the second subgroup (b), while the third subgroup (c) shows abundances higher than the model predictions in the elements Nb, Mo, Ru, Nd, and Sm. Below we discuss one example of each subgroup, while the rest of the sample stars for this group are shown in Appendix \ref{appendix_figs}. We refer to Sect. \ref{sec:discussion} for the general discussion of the $s$-process abundance pattern seen in the sample stars compared to the models, along with the discussion of the light elements and the $\delta$ values.

\begin{enumerate}[(a)]

\begin{figure*}[!ht]
\caption{Top panel: Comparison of the abundance pattern of HD 154430 and the final surface abundances of different AGB models after applying dilution. The residuals for each model are shown in the bottom panel. The label box lists: the mass of the current Ba star (M$_{\rm Ba}$), the estimated initial mass of the AGB component (M$_{\rm AGB}^{\rm J19}$) from \citet{jorissen19} and the parameters of the plotted models. All the masses are given in \msun~units. 
Specific properties of the FRUITY models are given in parentheses after the mass of each model: “-\mbox{-}-” indicates the standard models, while “TAIL”, “rotating” and “rotating TAIL” models with 60 km/s initial rotational velocity are indicated by (ext), (rNN) and (T60), respectively, where NN is the initial rotational velocity in km/s. The numbers given in parentheses after the mass of each Monash model is the M$_{\mathrm{PMZ}}$ given in units of 10$^{-4}$ \msun.
\newline}
\label{fig:HD154430}
\centering
\includegraphics[width=\hsize]{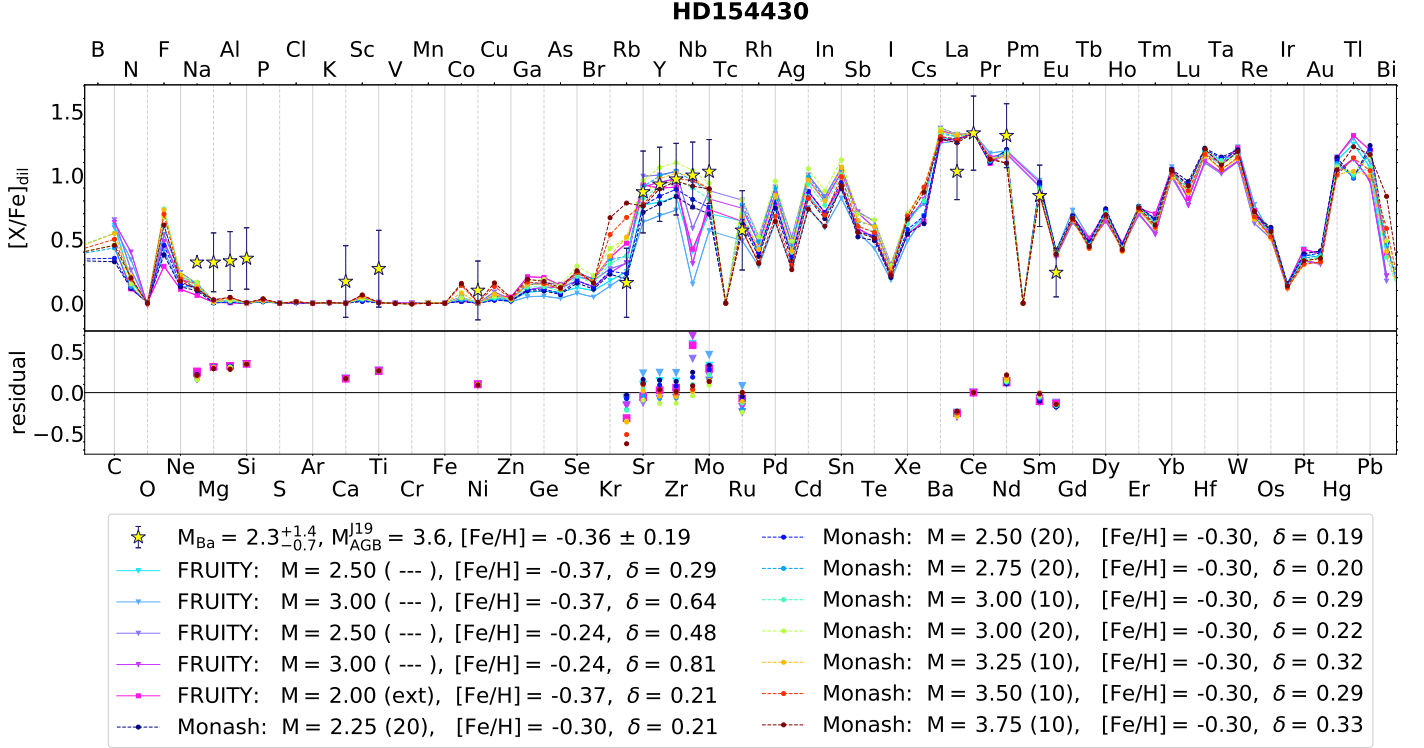}
\end{figure*}
 
\item The Ba stars HD 58368, HD 91208, HD 154430 and HD~201824 (Figs. \ref{fig:HD154430} and \ref{fig:HD58368}--\ref{fig:HD201824}) belong to this subgroup. As an example, HD 154430 (Fig. \ref{fig:HD154430}) is a representative of the subgroup for which most of the elements agree well with the diluted model pattern. Models between 2.0 and 3.75 \msun~are shown. [La/Fe] is somewhat lower than the model predictions, but all the other elements belonging to the $s$-process peaks are in agreement with the models. Although M$^{\rm{J}19}_{\rm{AGB}}$ is 3.6 \msun, the large uncertainty of the Ba star mass and the [Rb/Fe] value indicates a polluting AGB with initial mass of 2.5--3.0 \msun. We note that FRUITY models show much larger $\delta$ values than the Monash models for this star.


 \begin{figure*}[!ht]
  \caption{Same as Fig. \ref{fig:HD154430} but for HD 201657}
 \label{fig:HD201657}
 \centering
 \includegraphics[width=\hsize]{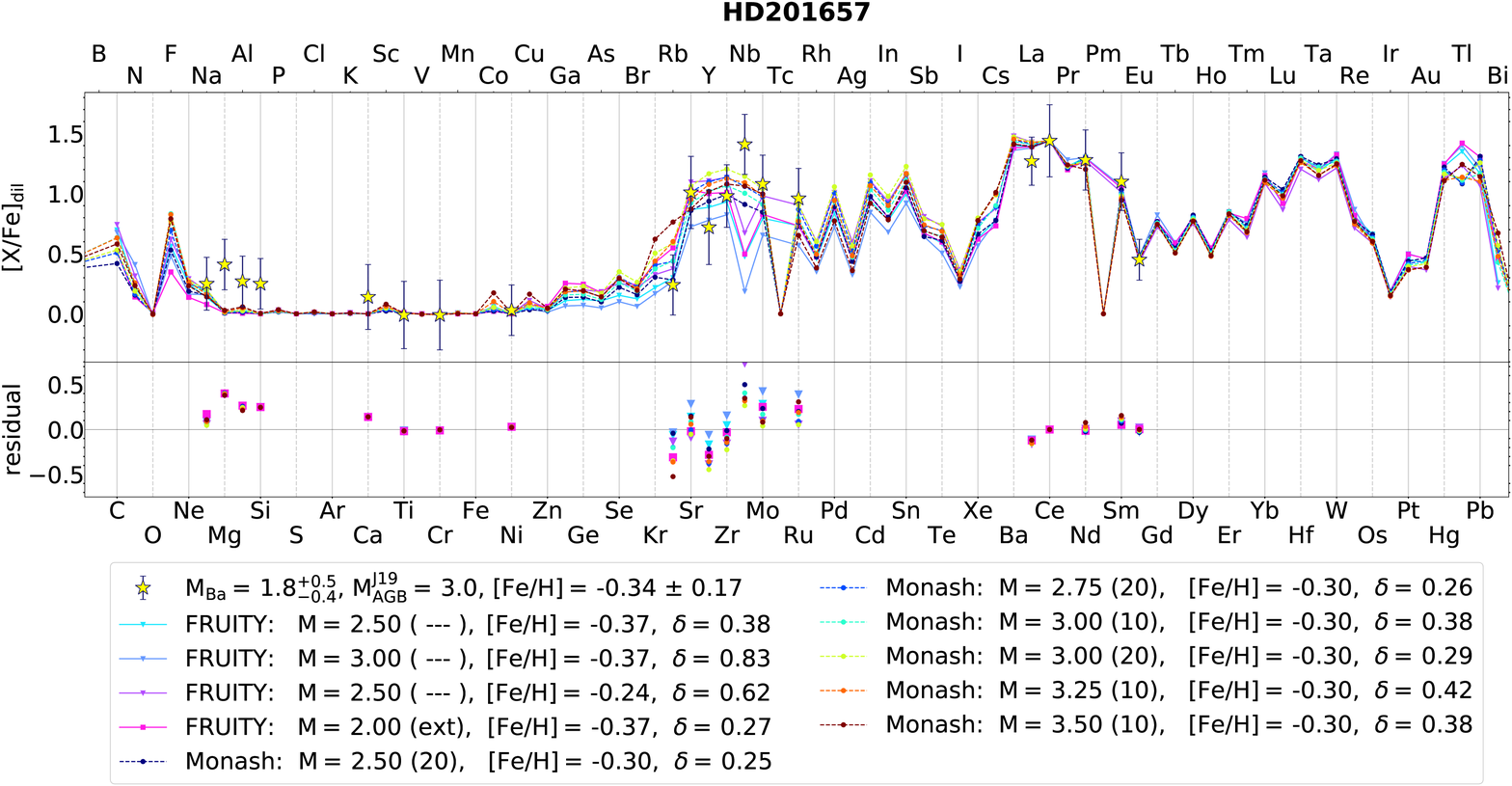}
 \end{figure*}

\item 12 stars (HD 20394, HD 40430, HD 53199, HD~58121, HD~59852, HD 95193, HD 119185, HD 143899, HD~200063, HD 201657, HD 210946, HD 211594, Figs. \ref{fig:HD201657} and  \ref{fig:HD20394}--\ref{fig:HD211594}) belong to the subgroup with higher Nb compared to the models and HD 201657 (Fig. \ref{fig:HD201657}) shows an example of this case.
We show different models between 2.0 (for the FRUITY 'TAIL' case) and 3.5 \msun. The abundance pattern (both $s$-process peaks and Rb) points to a 2.5--3.0 \msun~initial AGB, in agreement with M$^{\rm{J}19}_{\rm{AGB}}$. 


 \begin{figure*}[!ht]
 \caption{Same as Fig. \ref{fig:HD154430} but for CD $-$42$^{\circ}$2048}
 \label{fig:CD-422048}
 \centering
 \includegraphics[width=\hsize]{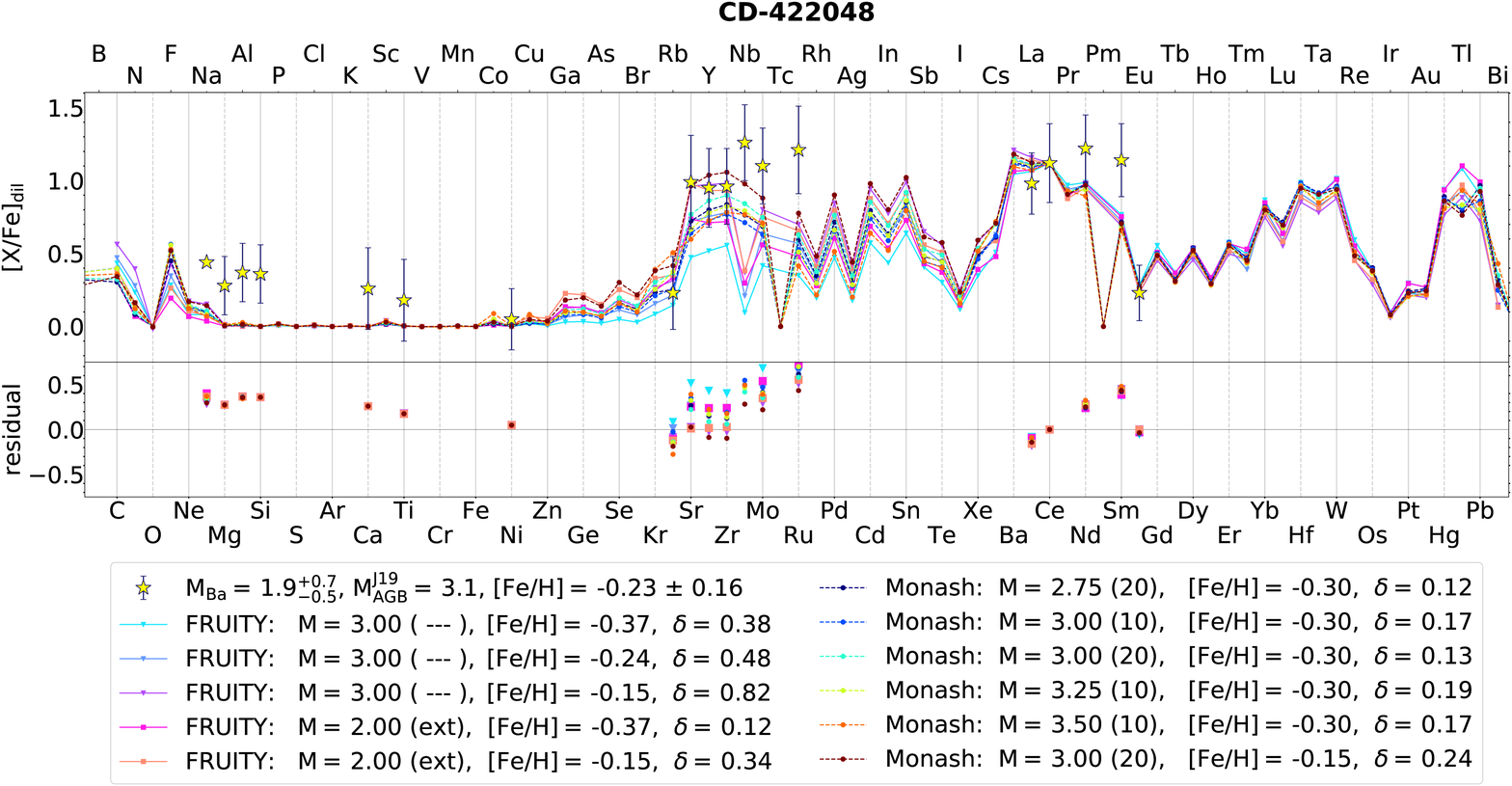}
 \end{figure*}
 
\item We found 5 stars belong to this subgroup (BD $-$14$^{\circ}$2678, CD~$-$42$^{\circ}$2048, CPD $-$64$^{\circ}$4333, HD 24035, HD 183915, Figs. \ref{fig:CD-422048} and \ref{fig:BD-142678}--\ref{fig:HD183915}). They show overabundance in the elements Nb, Mo, Ru, Nd, and Sm compared to the models and CD $-$42$^{\circ}$2048 (Fig. \ref{fig:CD-422048}) is selected as the example. 
Models in the range close to M$^{\rm{J}19}_{\rm{AGB}}$ agree well with the observations, although lower masses are allowed too when considering FRUITY models with extended pocket size. 
We also considered these models because there is a large error bar in the Ba star mass, which would translate into a higher uncertainty in the mass of the initial AGB as well. The 3 \msun~FRUITY models show much larger $\delta$ values than the Monash models for this star.
The [Rb/Fe] value is indicating that the mass of the polluting AGB was not above 3 \msun.
The 3 \msun~Monash models with larger \iso{13}C pocket have a higher first $s$-process peak and thus matching the abundances the best. Both of the $s$-process peaks are in good agreement with the models, however, the elements listed above and located just after the $s$-process peaks show somewhat higher abundances than the model predictions. 

\end{enumerate}


\subsection{Group 2: higher first $s$-process peak than predicted by the models}
\label{sec:high_first_peak}

The second group consists of stars for which the first $s$-process peak elements are typically higher than the model predictions, and specific solutions need to be applied. Interestingly, all of the stars belonging to this group have a low M$^{\rm{J}19}_{\rm{AGB}}$, $<$ 2 \msun. Two of them have the lowest AGB determined mass of the whole sample (1.25 and 1.35 \msun), and the third and fourth stars have AGB mass 1.8 \msun, but, as discussed below in Sect. \ref{sec:HD18182}, they behave differently from the other four stars with similar AGB masses, which belong instead to Group 1 (Figs. \ref{fig:HD119185}, \ref{fig:HD210946}, \ref{fig:HD24035}, and \ref{fig:HD183915}).  

\begin{figure*}[!ht]
 \caption{Same as Fig. \ref{fig:HD154430} but for HD 18182}
 \label{fig:HD18182}
 \centering
 \includegraphics[width=\hsize]{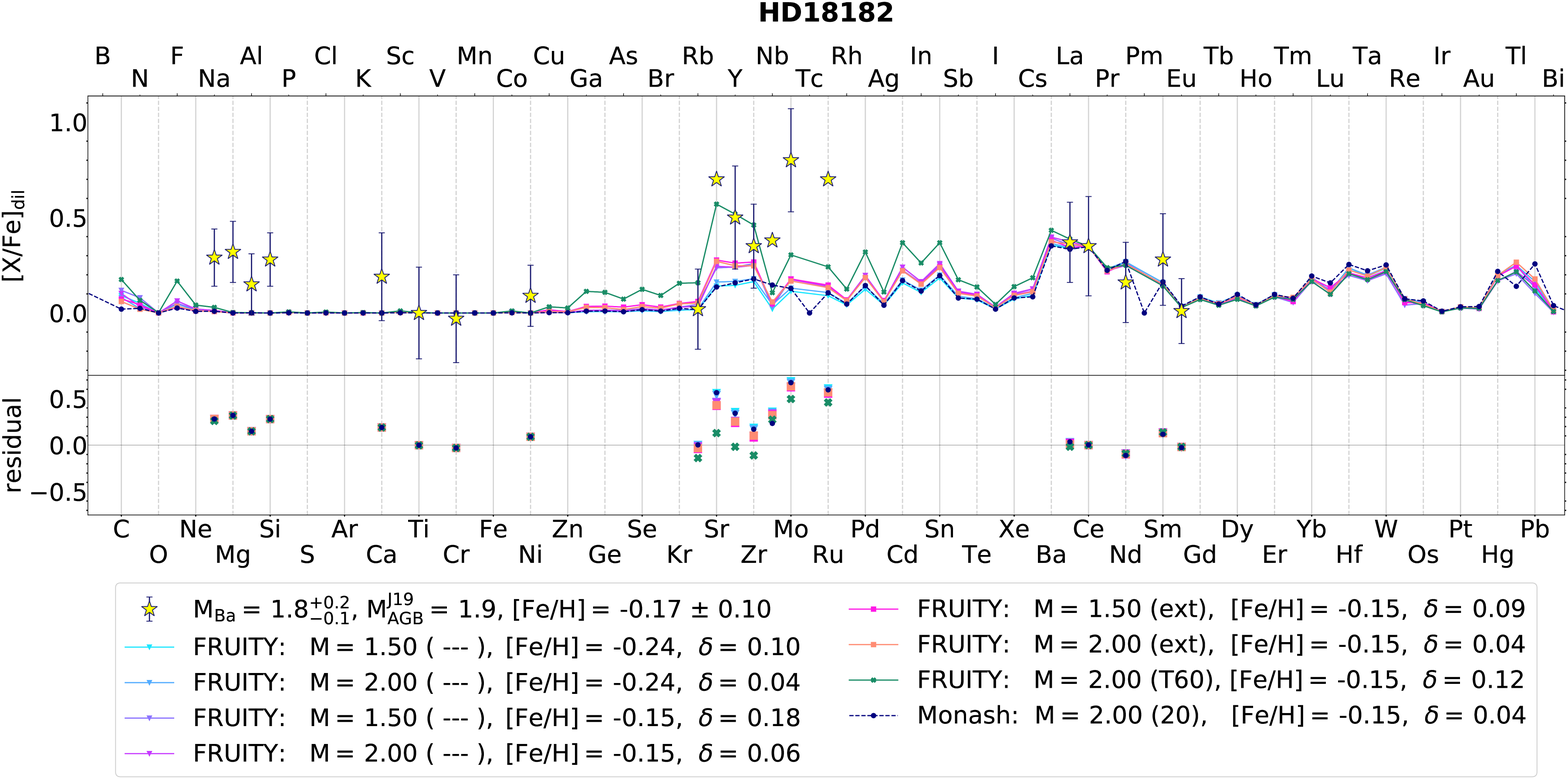}
 \end{figure*}

\begin{figure*}[!ht]
 \caption{Same as Fig. \ref{fig:HD154430} but for HD 180622}
 \label{fig:HD180622}
 \centering
 \includegraphics[width=\hsize]{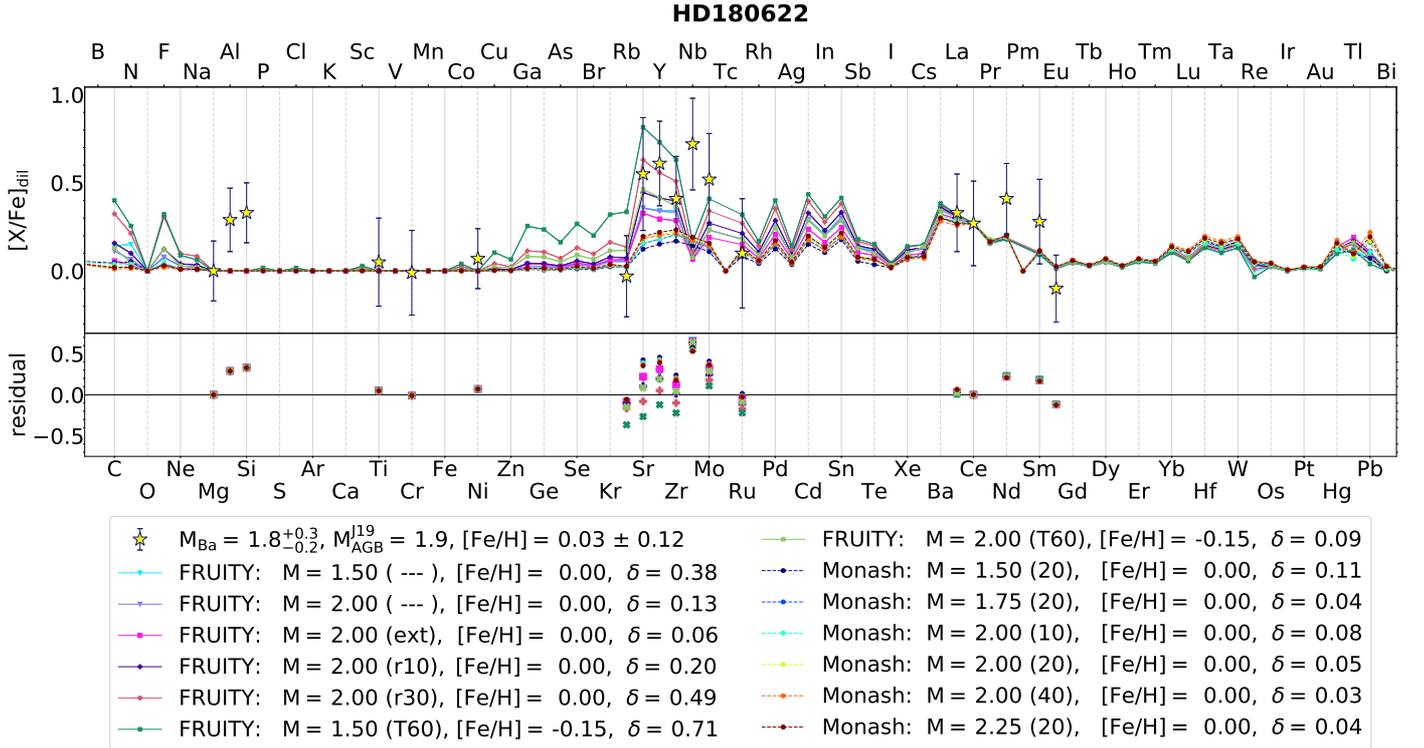}
 \end{figure*}

\subsubsection{HD 18182 (Fig.~\ref{fig:HD18182}) and HD 180622 (Fig.~\ref{fig:HD180622})}
\label{sec:HD18182}
 
For these stars we show models between 1.5 and 2.25 \msun. M$^{\rm{J}19}_{\rm{AGB}}$ is estimated to be around 1.8 \msun~and the nearly solar metallicity of HD 180622 makes it an ideal target to compare different models and model parameters. The low [Rb/Fe] is in agreement with a low mass polluter AGB. Both Ba stars have low [Ce/Y] ratio, with HD 180622 showing the lowest [Ce/Y] ratio in our sample. Monash models with different mass and \iso{13}C pocket sizes show very similar abundance pattern and $\delta$ values, but these models do not reach the high first $s$-process peak.
To match HD 18182 only a very specific solution is allowed: that of the 2.0 \msun~FRUITY T60. For HD 180622 also the r30, and the r10 models are a possible match and we show the [Fe/H] = $-$0.15 FRUITY T60 models for comparison, although this is outside of the metallicity range. 
The $\delta$ value is higher for the r30 model than for r10, with a more feasible value of $\delta =$ 0.2.

We conclude that an AGB around 2.0 \msun~with 'TAIL' or slow rotation could be the polluter of these Ba stars. These solutions are possible because these models have a lower neutron exposure, which favours the production of the first peak elements relative to the second peak elements. However, for HD 18182 these models are still factor of 3 to 5 lower from the observed [Mo/Fe] and [Ru/Fe] values, as for the stars of subgroup 1b, while HD 180622 has also higher Nd than the model predictions, similar to the stars in the subgroup 1c. 

\begin{figure*}[!ht]
 \caption{Same as Fig. \ref{fig:HD154430} but for HD 107541}
 \label{fig:HD107541}
 \centering
 \includegraphics[width=\hsize]{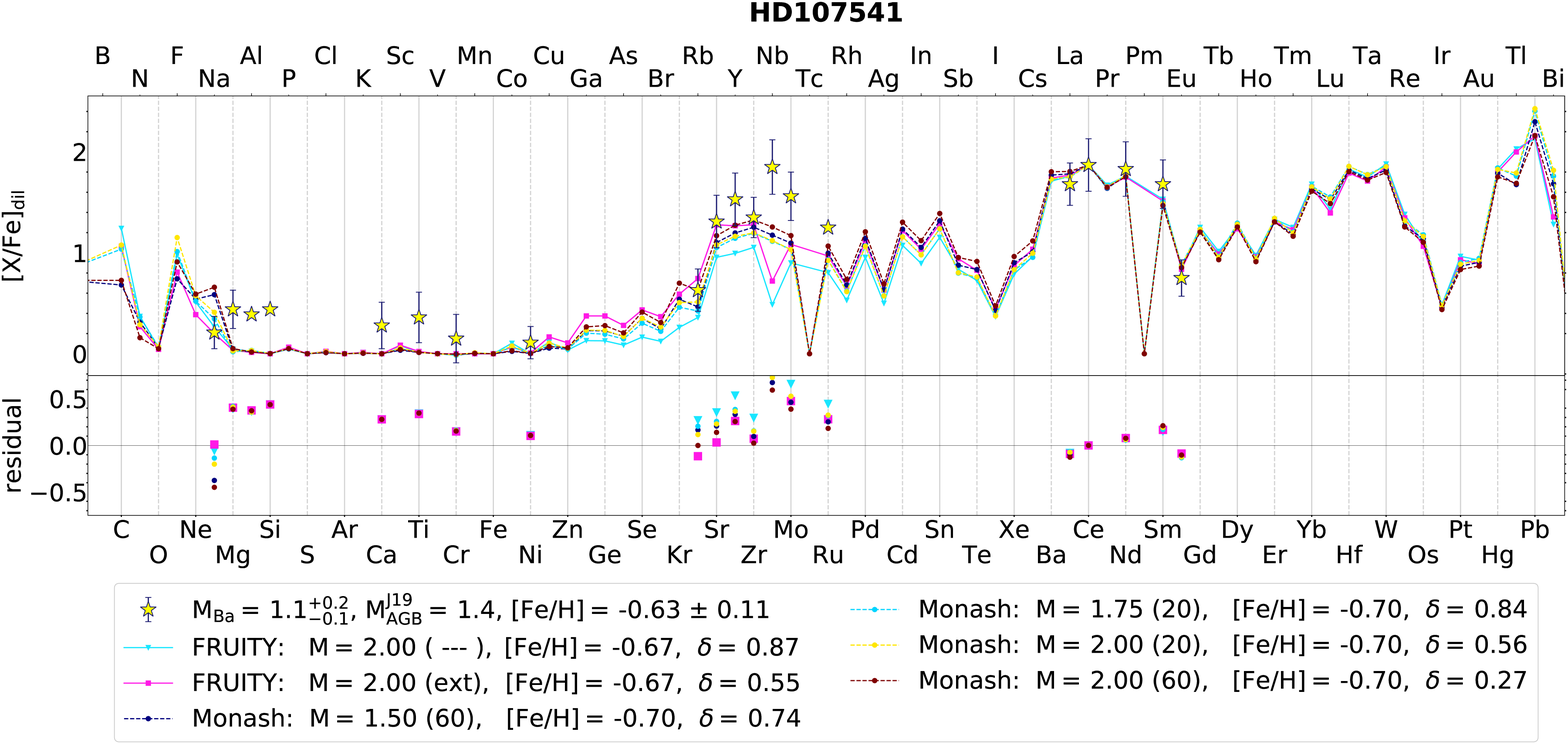}
 \end{figure*}

 
\subsubsection{HD 107541 (Fig.~\ref{fig:HD107541})}
 
The AGB companion of HD 107541 has a very low M$^{\rm{J}19}_{\rm{AGB}}$ of 1.4 \msun. Monash models are presented between 1.5 and 2.0 \msun, and we plot two FRUITY 2 \msun~model with different pocket sizes. Except from the 2 \msun~FRUITY model with standard pocket size, all the other models can match Rb, Sr and Zr within the error bars. Only the 2 \msun~Monash model with larger pocket size can match the abundances with a reasonable $\delta$ value, but this mass is far away from M$^{\rm{J}19}_{\rm{AGB}}$. Mo and Ru are overabundant compared to the models, which is similar to the stars belonging to the subgroup 1b. 

\begin{figure*}[!ht]
 \caption{Same as Fig. \ref{fig:HD154430} but for HD 134698}
 \label{fig:HD134698}
 \centering
 \includegraphics[width=\hsize]{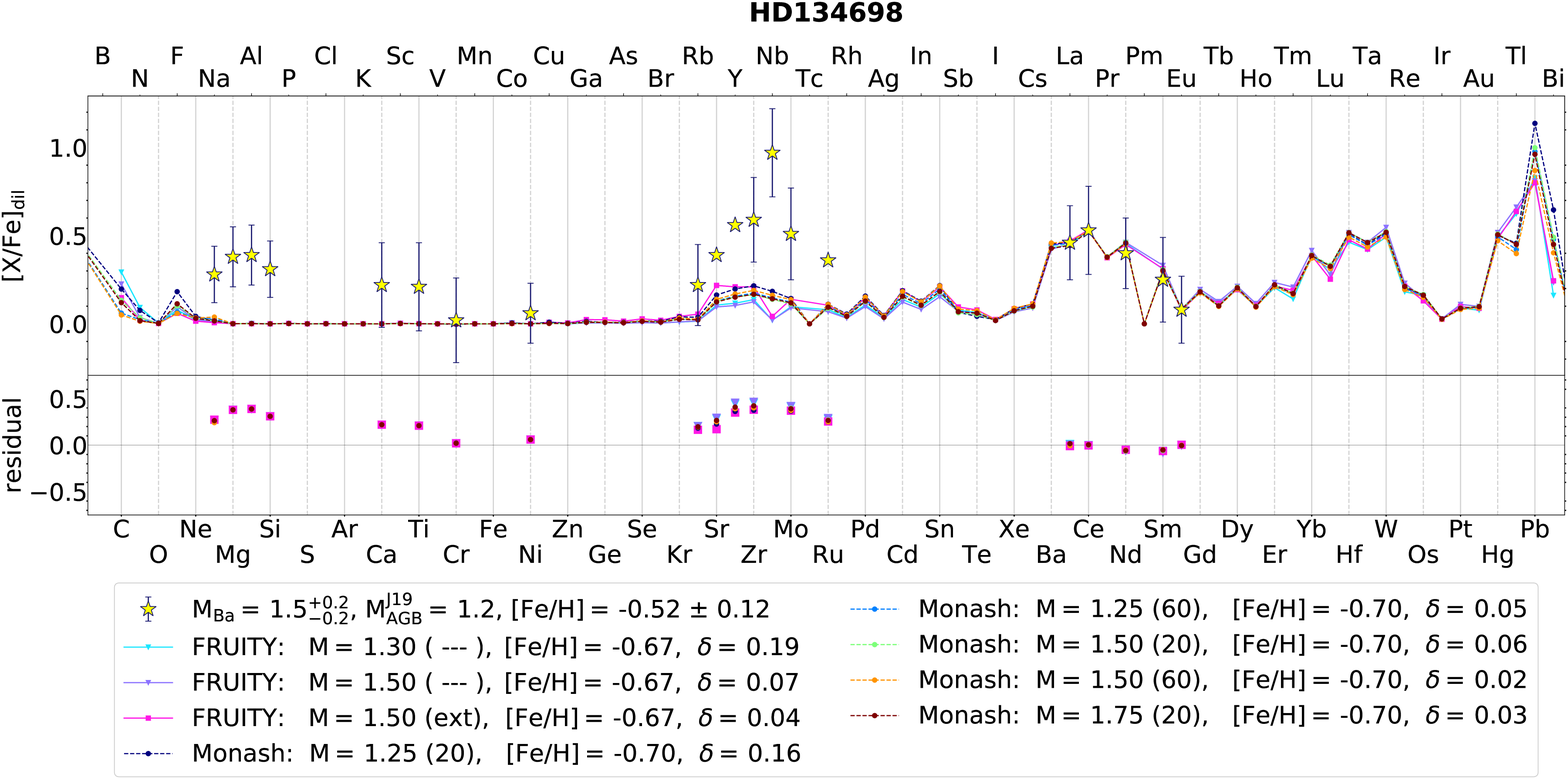}
 \end{figure*}

\subsubsection{HD 134698 (Fig.~\ref{fig:HD134698})}
 
HD 134698 has also a very low M$^{\rm{J}19}_{\rm{AGB}}$ of 1.2 \msun~and we show models up to 1.75 \msun. The [Rb/Fe] value and the first $s$-process peak shows a higher value than the presented model masses, close to M$^{\rm{J}19}_{\rm{AGB}}$.
The metallicity of the models is slightly lower than that of this Ba star, however, we checked the FRUITY models with [Fe/H]~$=-$0.37~and we found that they also predict a higher second-to-first peak ratio than observed in this star. A match to the [Rb/Fe] value can be found by higher mass models, although, none of these models can reproduce the high abundance values of the other light $s$-process elements. We also note that this star has the most eccentric orbit \citep[e = 0.95,][]{jorissen19} among our sample stars.


\subsection{Group 3: abundance patterns that require lower stellar masses than M$^{J19}_{AGB}$}
\label{sec:high_mass}

Three of the sample stars appear to have overestimated M$^{\rm{J}19}_{\rm{AGB}}$ (between 3.8 and 5.6 \msun) when comparing the abundance pattern with the models. 
This is because models of the corresponding masses have typically smaller \iso{13}C pockets than their lower masses counterparts \citep[see discussion in][]{cristallo09, karakas16} and therefore do not produce enough $s$-process element abundances to reach the very high observed values, for example, of [Ce/Fe] between 1 and 2 dex. Below we show models for each star that match the abundance pattern. Our suggested initial AGB masses are between 2 and 3 \msun~for each star in this group. Furthermore, Rb is available for 2 of these stars and it is lower than the higher mass model predictions.

\begin{figure*}[!ht]
 \caption{Same as Fig. \ref{fig:HD154430} but for HD 49641}
 \label{fig:HD49641}
 \centering
 \includegraphics[width=\hsize]{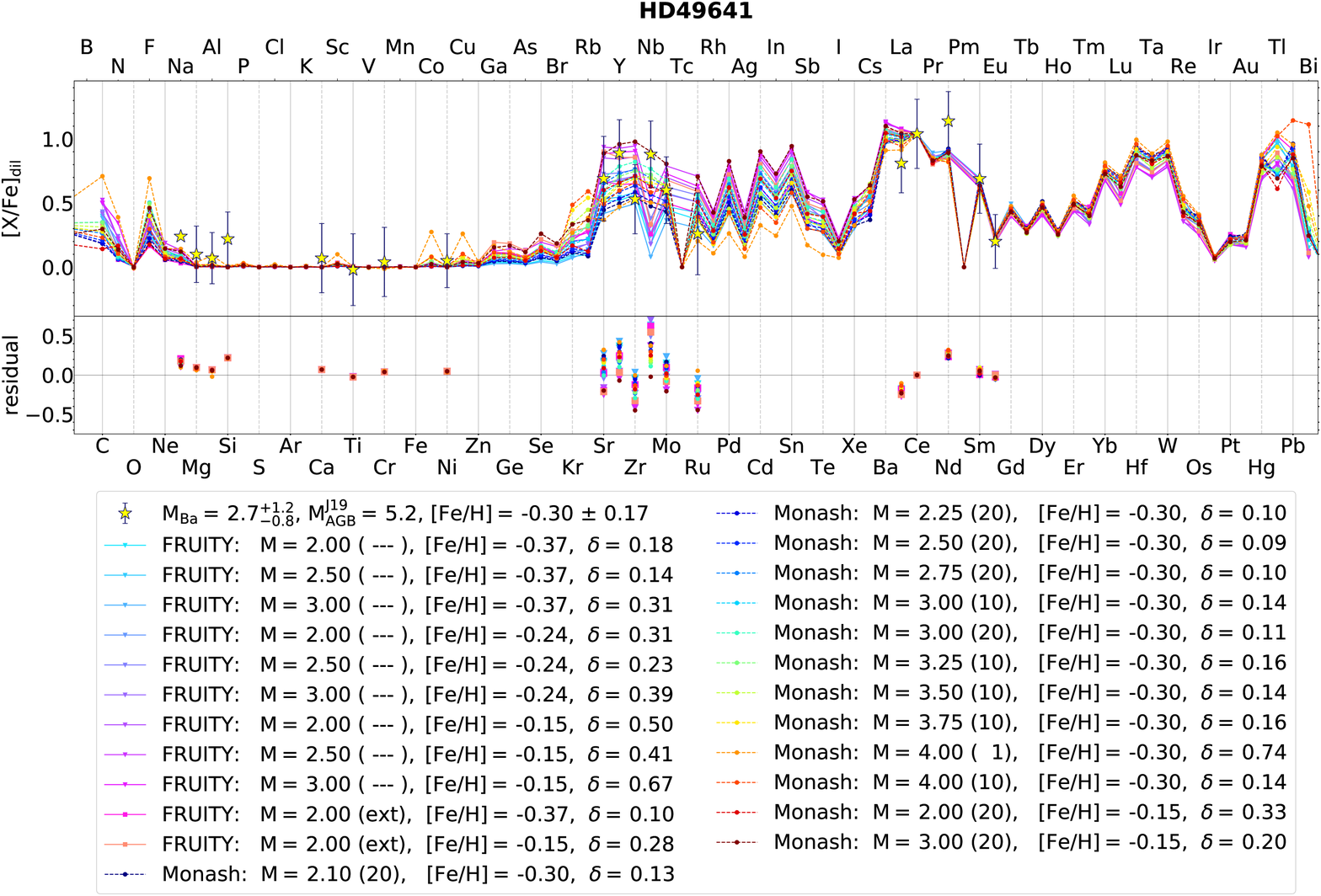}
 \end{figure*}

\subsubsection{HD 49641 (Fig.~\ref{fig:HD49641})}
 
HD 49641 is the first star in our sample which has high M$^{\rm{J}19}_{\rm{AGB}}$, 5.2 \msun. Considering the large error bar on the mass of the Ba star, we searched for matching AGB models with mass down to 2.0 \msun. 
Unfortunately, no [Rb/Fe] values are available for this star to better constrain the initial AGB mass. Additionally, the error bars of the first $s$-process peak elements are covering most of the models shown here, which does not allow us to constrain the mass of the polluting AGB.

\begin{figure*}[!ht]
 \caption{Same as Fig. \ref{fig:HD154430} but for HD 84678}
 \label{fig:HD84678}
 \centering
 \includegraphics[width=\hsize]{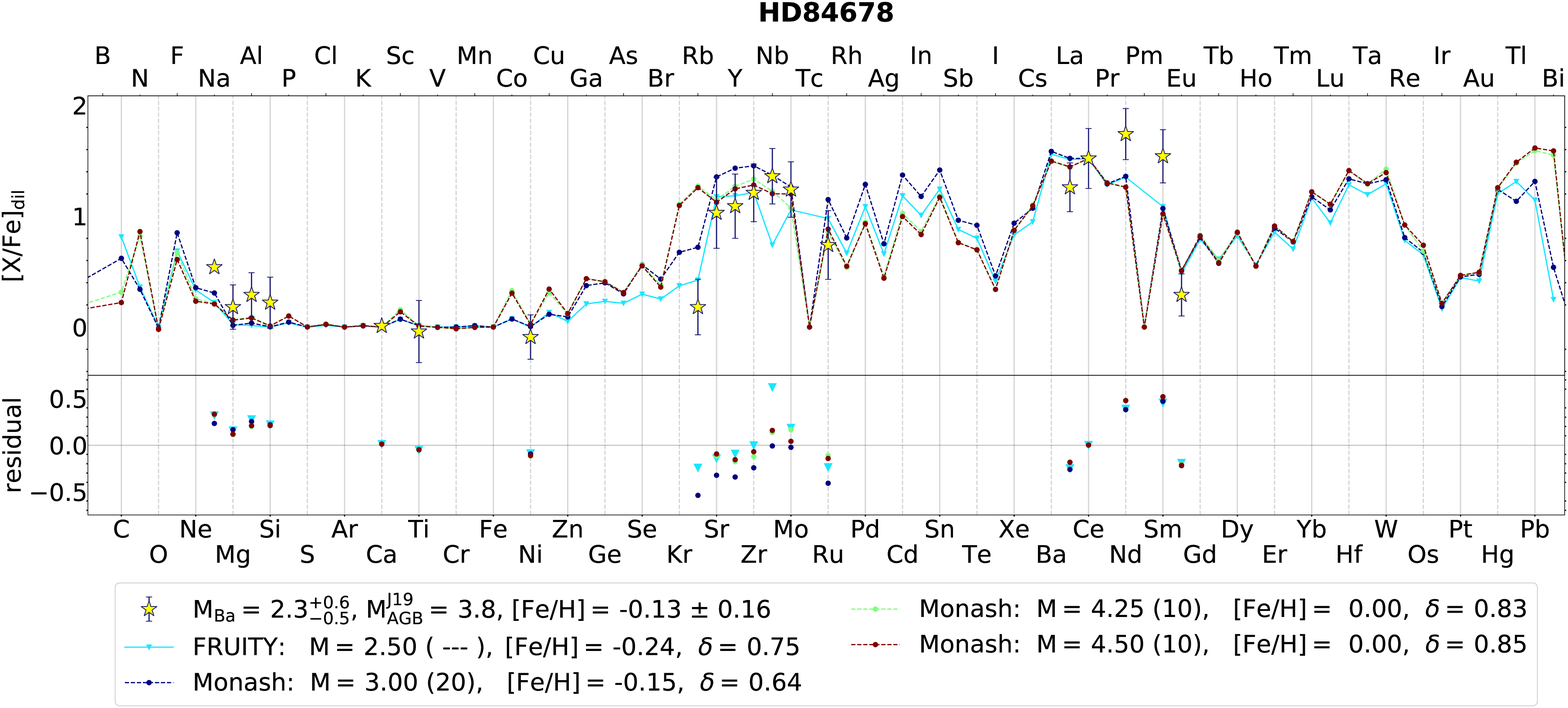}
 \end{figure*}

\subsubsection{HD 84678 (Fig.~\ref{fig:HD84678})}

We show 4 different models for HD 84678 between 2.5 and 4.5 \msun.  The larger mass models overestimate the [Rb/Fe] value, which points to a much lower initial AGB mass, below 2.5 \msun. Only the 2.5 \msun~FRUITY model is close to the abundance pattern, although it has a very high $\delta$. Furthermore, none of the models match the overabundance of Nd and Sm, as for stars in the subgroup 1b. 

\begin{figure*}[!ht]
 \caption{Same as Fig. \ref{fig:HD154430} but for HD 92626}
 \label{fig:HD92626}
 \centering
 \includegraphics[width=\hsize]{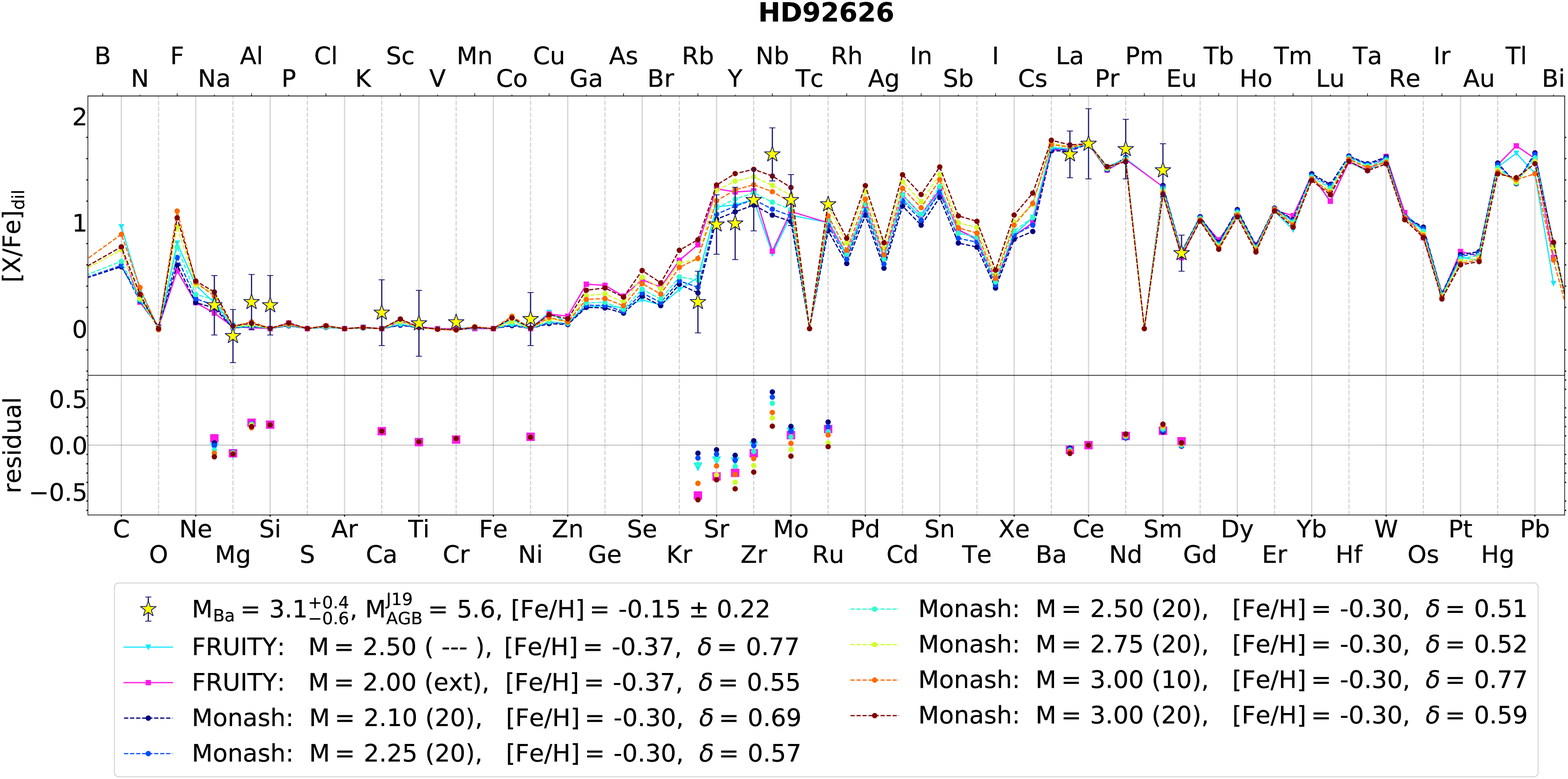}
 \end{figure*}

\subsubsection{HD 92626 (Fig.~\ref{fig:HD92626})}
 
HD 92626 has a M$^{\rm{J}19}_{\rm{AGB}}$ of 5.6 \msun. We show models between 2.0 and 3.0 \msun, since the [Rb/Fe] value is pointing to such low mass model, the 3 \msun~models already showing much higher Rb values than observed. 
We note that none of the models with metallicity higher than those shown in the figure can match the observed pattern within our dilution limit ($\delta \leq 0.9$). This indicates that the mass of the polluting AGB may be overestimated and was between 2--2.5 \msun, and with metallicity of $\approx -$0.30 dex, which is within the given error bar for this star. 
To match the observed abundance pattern, high $\delta$ values, thus large amount of accreted AGB mass are needed. 
Since the orbital period of this system is relatively short, 918 days, this could explain the higher amount of AGB material in the atmosphere of the Ba star, as compared to other, longer period ($\simeq$ 5--10 yr) systems.


\section{Discussion}
\label{sec:discussion}

As discussed above, the [Rb/Fe] value of the individual stars is a good proxy for the initial mass of the polluter AGB \citep{roriz21}. The [Rb/Fe] of the star generally does confirm M$^{\rm{J}19}_{\rm{AGB}}$.
Fifteen stars in our sample were also included in the work of \citet{Stancliffe21}, who considered
the Ba stars in the sample of \cite{deC} with [Fe/H] = $-$0.25 $\pm$ 0.125 dex and, using the Monash models, found that most of these stars can be reproduced by AGB companions of mass 2.5 and 3 \msun, similar to what we report here.

In the Group 3 stars specifically (see Sect. \ref{sec:high_mass}) instead there are discrepancies between the initial masses derived by \citet{jorissen19} using the constant Q method and the AGB mass indicated by the [Rb/Fe] value. When applying the constant Q assumption for the determination of the WD masses, some stars have WD mass close or above 1 \msun~(initial AGB mass $\approx 5.7$ \msun). However, we do not wait Ba stars enriched by such a high mass companion at metallicity close to solar. We found that the polluter AGB might have been between 2 and 3 \msun~also for these stars, instead of a more massive star.

The shape of the first $s$-process peak elements (Sr, Y, Zr) shows different pattern for each star, for example in some cases the Zr abundance is much lower than the neighbouring Y values. We should note that the Zr abundances used in this study were derived from Zr I lines, which are systematically lower than those derived from Zr II lines. Such fact is generally attributed to departures from the LTE approximation, as discussed by some authors \citep[e.g.][]{gratton_sneden94, AB2006, karinkuzhi18b}. Further investigation aiming to compare abundances from Y I and Zr II lines for the sample of \citet{deC} could be useful in order to quantify these differences.

The pattern of the second peak elements (La, Ce, Nd) is in good agreement with the models shown here. This is not surprising, since we are normalising the models to match the [Ce/Fe] values of the sample stars. However, for some stars the [La/Fe] value is somewhat lower than model predictions. Since we have not taken into account the uncertainties of [Ce/Fe], lowering the models within the uncertainties of [Ce/Fe] could solve this issue.

Eu is mostly produced by the $r$-process, the [Eu/Fe] abundances are in the most cases in agreement with the model predictions. The overall distribution of [Eu/Fe] abundances of the sample stars is similar to the disc population \citep{Roriz_heavy}. 

In relation to Group 2, recently, \citet{shetye19_1msun} found S-type stars with initial masses around 1 \msun~in the metallicity range between $-$0.27 and $-$0.54 dex and enhancements in $s$-process elements requiring the TDU to operate also at these masses. 
Therefore, AGB stars of around 1 \msun~may be able to produce heavy elements at the beginning of the AGB phase and the polluter AGB stars of the Ba stars belonging to our Group 2 could represent other examples of this group of low mass AGBs with active TDU. As it can be seen in Fig. \ref{fig:HD134698} for HD 134698, unlike the models of \citet{shetye19_1msun} the low mass models used in this study do not reach the large enhancement in the first $s$-process peak elements. In fact, the dredge-up at these lower masses is very model-dependent: higher metallicity Monash models below 1.5 \msun~do not have a PMZ leading to the formation of $s$-process elements at masses, while FRUITY 1.3 \msun~models have enrichment high enough in $s$-process elements ([s/Fe] $\geq$ 0.25) only for metallicities below [Fe/H] = $-$0.24.

\begin{figure}[!ht]
 \caption{Period-eccentricity distribution of the sample stars.}
 \label{fig:period_e_plot}
 \centering
 \includegraphics[width=\hsize]{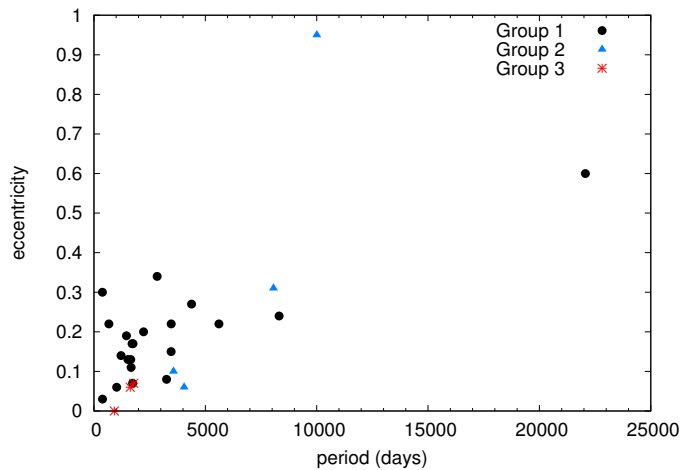}
 \end{figure}

When examining the orbital parameters of our sample stars (Fig. \ref{fig:period_e_plot}), we see that the 3 stars in Group 3 have eccentricities below 0.1 and periods below 2000 days. While there are 3 stars in Group 1 in the same region of Group 3, stars in Group 2 have instead all longer periods than 3000 days. HD 134698, the most eccentric system in our sample also belongs to Group 2. The other 3 stars in Group 2 share similar binary parameters as some stars in Group 1. We have to note that some of our sample stars have only preliminary orbital parameters \citep[see Table 4 in][]{jorissen19}.

\subsection{The problem of subgroups 1b and 1c}

The stars belonging to subgroups 1b and 1c, as well as the stars of Group 2 show excess in the heavy elements that immediately follow the first and second $s$-process peaks compared to the models. As already discussed in \cite{Roriz_heavy} (their Fig. 9) the explanation of the excess of Nb, Mo, and Ru relative to Sr, and of Nd and Sm relative to Ce, within the framework of the spectroscopic analysis is not trivial. 

The fact that these elements are produced not only via the $s$-process predicts that contributions from other stellar sources or nucleosynthetic processes might miss from the AGB models used in this analysis. For example, some stars show [Sm/Fe] value higher than the model predictions. Since Sm is produced by $\approx$ 70\% the $r$-process, the slight overabundance can be caused by a missing $r$-process component of this element in the AGB models. Nuclear uncertainties could also play a role in the deviations, although neutron capture cross sections are relatively well known for stable isotopes. However, as pointed out by \cite{Roriz_heavy}, the overabundance of Sm is so high, that probably there are no signs of the initial abundances left in the observed abundance pattern. 

The study of \citet{mishenina19} showed that the available galactic chemical evolution models do not produce enough Mo and Ru compared to the disc star observations in the metallicity range of our sample stars. 
In the galactic chemical evolution models of \citet{Kobayashi20} Mo and Ru are both overproduced compared to the observations when specific components are added to the model computations. These uncertainties still leave the question of the observed overabundance of certain elements open.

An explanation for these stars might be the activation of the intermediate neutron capture process ($i$-process), which was introduced to explain several peculiarities observed in different types of stellar objects \citep[e.g. CEMP stars, post-AGBs, and also the extreme Ba enhancement in young stars][]{baratella21}. The $i$-process has higher neutron densities compared to the $s$-process, which typically manifests in [Rb/Sr] $\geqslant$ 1.0. However, this is not the case in our sample Ba stars, all stars having negative [Rb/Sr]. 
Late neutron fluxes occurring in some AGBs might be able to keep Rb low compared to Sr and produce more proton-rich elements close to the first and second peaks, by shifting part of the elements to heavier ones, from Sr, Y, Zr to Nb, Mo, Ru and from Ba, La, Ce to Nd, Sm, for the first and second peak, respectively. The exact process needs further investigation.

\subsection{The light elements and C/O ratio}

Although C and O were not derived for the sample stars, we shortly discuss these elements to stress their importance in a future analysis.
As noticeable in the figures, some of the stars require high $\delta$ values to be matched, and this may create problems with respect to the C/O ratio. If [C/Fe] $>$ 0.27, this results in Ba stars that are C-rich (C$>$O), however, these stars are not observed to be C-rich. For example, if we wish to match HD 91208 (Fig. \ref{fig:HD91208}) using the FRUITY 2.5 \msun~model, we obtain $\delta=0.67$. However, such high value of $\delta$ results in [C/Fe] $\simeq$ 0.5, which correspond to C/O $\simeq$ 2 and would make this Ba stars observed as a C-star instead. Therefore, such large $\delta$ are not realistic also from the abundances point of view. 
This problem could be mitigated in some cases if the initial O abundance, which is not observed and is unaffected by AGB stars, is higher than solar. This would keep the C/O ratio lower and it applies mostly to stars of metallicity lower than solar, due to the effect of $\alpha$-enhancements. This effect can be derived from the observed abundances of the other $\alpha$-elements, especially Mg, Si, Ca, and Ti, which may be higher compared to AGB models. These elements reflect the initial composition of the stars and vary from star to star due to their different birthplaces.
This was already pointed out by \cite{deC} and they suggested that some stars are in transition between the thin and the thick disc. 
For example, HD 201657 (Fig. \ref{fig:HD201657}) has [Mg/Fe] $\simeq$ 0.4, and if we assume also [O/Fe] $\simeq$ 0.4, the 2.5 \msun\ FRUITY models with relatively high $\delta$ values and high [C/Fe] could still produce a solution with an O-rich star.
Instead, in relation to the example above of HD 91208, this star has solar metallicity and these elements are not enhanced relative to solar.

\begin{figure}
 \caption{$\alpha$-element abundances for the sample Ba stars (black dots) and literature data for solar neighbourhood giants (orange triangles) from \citet{jonsson17}}
 \label{fig:alpha-elements}
 \centering
 \includegraphics[width=\hsize]{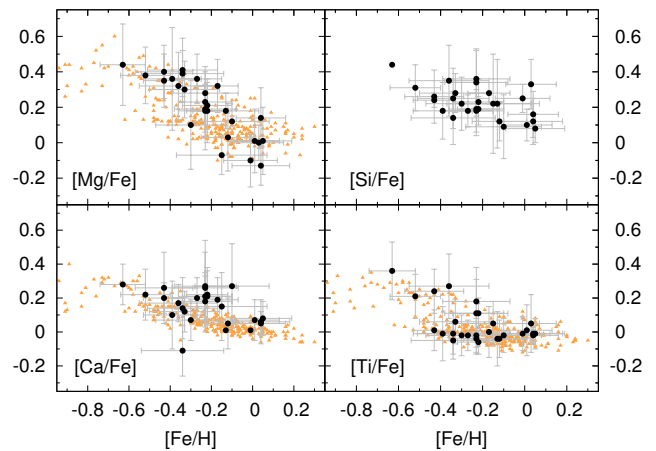}
 \end{figure}

Since the sample stars are all located in the solar neighbourhood, we can also compare the distribution of their light elements to that of normal giants. We found that the Ba stars examined in this study show similar $\alpha$-element abundances to the normal giants in the solar neighbourhood analysed by \citet{jonsson17}. Fig. \ref{fig:alpha-elements} shows that the abundances of these elements in the Ba stars increase when the metallicity decreases, in agreement with solar neighbourhood giants and galactic chemical evolution models. 


\subsection{The $\delta$ values}
\label{sec:delta}

\begin{figure}
 \caption{$\delta$ values as a function of the orbital period separately for FRUITY and Monash models. The error bar on the $\delta$ value is the standard deviation.}
 \label{fig:delta_period}
 \centering
 \includegraphics[width=\hsize]{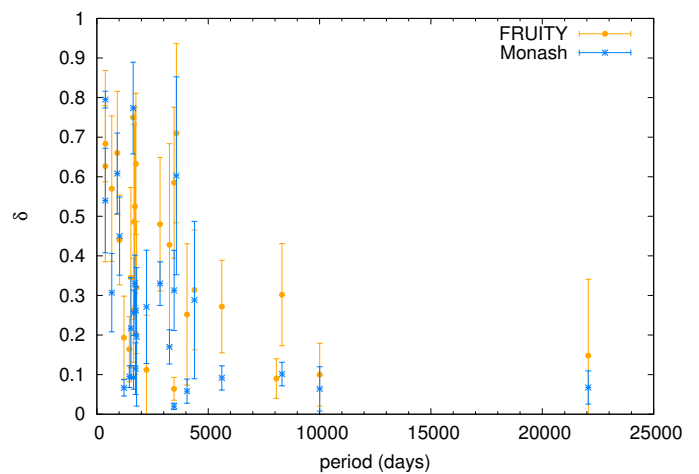}
 \end{figure}

Figure \ref{fig:delta_period} shows the average $\delta$ values over all the matching models versus period for the sample stars. The error bar on the $\delta$ value is the standard deviation, and some of the plotted FRUITY models were excluded from the calculation. The T60 models were not included for HD 40430, HD 53199, HD 59852, HD 143899, HD 180622 and HD 200063 since in these cases this model is only plotted for comparison and completeness, but it is not a good match to the abundance pattern. The removal of the FRUITY r30, and [Fe/H] = 0.15 dex models for HD 58121 and HD 95193, respectively, lowers the average $\delta$ value below the 0.3 limit for these two stars. Overall, the average $\delta$ value is lower than $\approx$ 0.3 for all the systems with period over $\approx$ 4000 days, while stars with short period (< 1000 days) have $\delta \gtrsim$ 0.3, although we should note that some longer period systems have uncertain periods.

Overall, though depending on the exact parameters, FRUITY models need higher $\delta$ values to match the observed abundances pattern. This is because, as shown in Fig. \ref{fig:models} and mentioned in Sect. \ref{sec:dil_fact_calc}, these models predict  typically lower abundances at the AGB surface. In the Monash models using larger \iso{13}C pocket of course leads to a reduction of the $\delta$ value, since there is more Ce produced.

Most of the stars with binary period in the typical range 1000 to 10000 days can be fitted by models with $\delta$ values below 0.2, except for HD 211594, HD 201824, and HD 201657 (all belonging to Group 1), which need $\delta$ values from 0.3 and above.
Instead, the three stars (CPD $-$64$^{\circ}$4333, HD 24035 and HD 92626) with the shortest orbital periods (of less than 3 yr) typically require models with $\delta$ values higher than roughly 0.4. For example, for CPD $-$64$^\circ$4333, all of the matching models have $\delta$ values above 0.6. Together with HD 24035, CPD $-$64$^\circ$4333 belongs to Group 1c, while the third star, HD 92626, belongs instead to Group 3. Also HD 84678, belonging to Group 3, needs $\delta$ values above 0.6 to be matched, even though its period is 1630 days. Within the Group 2 stars, which have the lowest masses, only HD 107541 requires relatively high $\delta$ values above 0.55. 
It would be interesting to further analyse individually the binary evolution of all these peculiar systems. 
The fact that these stars need higher $\delta$ values may be related to a different mass transfer and binary evolution, rather than the AGB model, as also discussed by \citet{Stancliffe21} for HD 24035.
For example, based on hydrodynamical simulations of binary AGB systems, \cite{chen20} found that with decreasing binary separation, the accretion rate increases. In their simulations the two least separated binaries have dusty circumbinary disks and these authors showed that if a circumbinary disc is present in the system, the mass transfer efficiency (defined as the ratio of the accretion rate of the secondary, the current Ba star, and the mass-loss rate of the AGB primary) through wind-Roche-lobe overflow can increase up to 31\%, 8 times larger as compared to the canonical Bondi-Hoyle-Lyttleton accretion mechanism. Thus, the presence of a circumbinary disc during the mass transfer may explain the higher $\delta$ values for the systems, where even the largest Monash \iso{13}C-pocket size model is not giving a satisfactory match. 

\section{Conclusions}
\label{sec:conclusions}

We have analysed the individual abundance patterns of 28 Ba stars against AGB models predictions of the $s$ process. For these stars, masses were independently derived by \cite{jorissen19} via investigation of binary interaction, and high-resolution spectroscopy abundances were reported by \citet{deC}. New elemental abundances for Rb \citep{roriz21}, Sr, Nb, Mo, Ru, La, Sm and Eu \citep{Roriz_heavy} allowed us to carry out a comparative study between the individual abundance patterns with different AGB models in the metallicity range of the stars using dilution to match the [Ce/Fe] values of each star. 

We grouped our sample stars into three main groups. The first group includes 21 stars that have matching AGB models, the second group includes 4 stars with higher first $s$-process peak abundances compared to most models, and the third group includes 3 stars with high M$^{\rm{J}19}_{\rm{AGB}}$ (> 3.8 \msun), while our investigation supports a lower mass. Our main conclusions are:

\begin{enumerate}

\item{Stars in Group 1 have good agreement with the models, but stars in subgroups b and c show overabundance in Nb and in Nb, Mo, Ru, Nd, and Sm, respectively. This overabundance was already pointed out in the large sample analysed by \citet{Roriz_heavy}, who suggested a different nucleosynthesis path to reproduce the excess of these heavy elements located just after the $s$-process peaks. Our analysis of individual stars confirms the need for such future investigation.}

\item{Group 2 includes 4 stars with low estimated Ba and AGB mass (< 2 \msun) and with higher first $s$-process peak than typically predicted, which requires somewhat lower neutron exposures, as found in the FRUITY rotating or 'TAIL' models. However, there are only 2 stars out of the 4 in Group 2, where these models can give a possible solution to match the observed abundance pattern.
The production of the light $s$-process elements can also be attributed to diffusive mixing \citep{battino19}, and these models will need to be included in a future study.}

\item{Group 3 includes 3 stars in our sample that have higher initial AGB mass (> 3.8 \msun) derived under the constant Q assumption \citep{jorissen19}. As these authors also indicated, these AGB masses might be somewhat overestimated due to the high WD mass. For these stars we found that lower mass AGB models, below 4 \msun, can explain the elemental patterns.}

\item{Most of the light elements in our sample stars show excess compared to the AGB models, but are in agreement with the overall trend with the solar neighbourhood giants and with the predicted trend of galactic chemical evolution.}

\item{The calculated $\delta$ values are mostly lower than 0.3. However, some systems need larger $\delta$ values, and thus probably higher mass transfer efficiency to reproduce the Ba star abundance pattern. These stars have also typically shorter orbital period, which might suggest an explanation to the larger amount of transferred AGB material.}

\end{enumerate}

In this paper we confirmed the findings of \cite{roriz21} and \cite{Roriz_heavy} using the abundance pattern of the individual stars rather than comparing the bulk sample of the Ba stars to AGB models. None of the stars show high Rb or [Rb/Sr] > 0 dex \citep{roriz21}, which means that AGBs below 4 \msun~are the contributors to the abundance pattern of that observed in this Ba star sample. This raises the question where are the Ba stars polluted by a higher mass AGB companion? To resolve this problem we should consider including in the Ba stars category stars to be with [$s$/Fe] < 0.25 dex, as expected from the theoretical models of mass over $\approx$4 \msun.

Future work involves further spectroscopic analysis of the stars belonging to Group 1b and 1c and Group 2, and the comparison with other sets of models.
Furthermore, elemental abundances of C and the third peak element Pb could help to distinguish between the processes in the AGB stars that lead to the formation of the observed Ba stars. Also the dilution due to mass transfer and its implication in relation to each individual binary system should be further investigated by connecting the $\delta$ values reported here for each star to a value of the accreted mass and to a possible mass transfer efficiency, which are the quantities used in the modelling of the binary interaction. For example, \citet{Stancliffe21} presented models of the accretion of AGB ejecta on to low-mass companions to form Ba stars of typical mass 2.5 \msun. Such models could be targeted at each individual systems considered here as the now available, more extended abundance pattern, together with the known binary properties and masses, could help to derive more precise accreted masses. The corresponding binary-derived dilution factors can be then compared to those derived here on the basis of AGB nucleosynthesis. The method used by \citet{abate15a} for CEMP-s stars could also be applied to the sample of Ba stars with known orbital periods. Finally, to be able to analyse each individual Ba star of the full sample of 169, we are implementing machine learning techniques, which will be presented in a second paper in this series.


\begin{acknowledgements}
The authors thank the referee for comments that helped to improve the clarity of the paper.
The authors would like to thank A. Jorissen for providing data.
This article is based upon work from the “ChETEC” COST Action (CA16117), supported by COST (European Cooperation in Science and Technology). AIK was supported by the Australian Research Council Centre of Excellence for All Sky Astrophysics in 3 Dimensions (ASTRO 3D), through project number CE170100013. B.Cs. and M.L. acknowledge the support of the Hungarian National Research, Development and Innovation Office (NKFI), grant KH\_18 130405. B.V. is supported by the ÚNKP-21-1 New National Excellence Program of the Ministry for Innovation and Technology from the source of the National Research, Development and Innovation Fund. M.P.R. acknowledges financial support by Coordenação de Aperfeiçoamento de Pessoal de Nível Superior (CAPES). N.A.D. acknowledges financial support by Russian Foundation for Basic Research (RFBR) according to the research projects 18\_02-00554 and 18-52-06004. V.D. is supported by the PRIN-INAF 2019 'Planetary systems at young ages (PLATEA)'.

\end{acknowledgements}

\bibliographystyle{aa} 
\bibliography{bibliography.bib} 

\begin{thebibliography}{62}
\expandafter\ifx\csname natexlab\endcsname\relax\def\natexlab#1{#1}\fi

\bibitem[{{Abate} {et~al.}(2015){Abate}, {Pols}, {Karakas}, \&
  {Izzard}}]{abate15a}
{Abate}, C., {Pols}, O.~R., {Karakas}, A.~I., \& {Izzard}, R.~G. 2015, \aap,
  576, A118

\bibitem[{{Abia} {et~al.}(2002){Abia}, {Dom{\'{\i}}nguez}, {Gallino}, {Busso},
  {Masera}, {Straniero}, {de Laverny}, {Plez}, \& {Isern}}]{abia02}
{Abia}, C., {Dom{\'{\i}}nguez}, I., {Gallino}, R., {et~al.} 2002, \apj, 579,
  817

\bibitem[{{Allen} \& {Barbuy}(2006)}]{AB2006}
{Allen}, D.~M. \& {Barbuy}, B. 2006, \aap, 454, 895

\bibitem[{{Aoki} {et~al.}(2008){Aoki}, {Beers}, {Sivarani}, {Marsteller},
  {Lee}, {Honda}, {Norris}, {Ryan}, \& {Carollo}}]{Aoki08}
{Aoki}, W., {Beers}, T.~C., {Sivarani}, T., {et~al.} 2008, \apj, 678, 1351

\bibitem[{{Baratella} {et~al.}(2021){Baratella}, {D'Orazi}, {Sheminova},
  {Spina}, {Carraro}, {Gratton}, {Magrini}, {Randich}, {Lugaro}, {Pignatari},
  {Romano}, {Biazzo}, {Bragaglia}, {Casali}, {Desidera}, {Frasca}, {de Silva},
  {Melo}, {Van der Swaelmen}, {Tautvai{\v{s}}ien{\.{e}}},
  {Jim{\'e}nez-Esteban}, {Gilmore}, {Bensby}, {Smiljanic}, {Bayo},
  {Franciosini}, {Gonneau}, {Hourihane}, {Jofr{\'e}}, {Monaco}, {Morbidelli},
  {Sacco}, {Sbordone}, {Worley}, \& {Zaggia}}]{baratella21}
{Baratella}, M., {D'Orazi}, V., {Sheminova}, V., {et~al.} 2021, \aap, 653, A67

\bibitem[{{Battino} {et~al.}(2019){Battino}, {Tattersall}, {Lederer-Woods},
  {Herwig}, {Denissenkov}, {Hirschi}, {Trappitsch}, {den Hartogh}, {Pignatari},
  \& {NuGrid Collaboration}}]{battino19}
{Battino}, U., {Tattersall}, A., {Lederer-Woods}, C., {et~al.} 2019, MNRAS,
  489, 1082

\bibitem[{{Becker} \& {Iben}(1979)}]{Becker79}
{Becker}, S.~A. \& {Iben}, I., J. 1979, \apj, 232, 831

\bibitem[{{Bertelli} {et~al.}(1986){Bertelli}, {Bressan}, {Chiosi}, \&
  {Angerer}}]{Bertelli86}
{Bertelli}, G., {Bressan}, A., {Chiosi}, C., \& {Angerer}, K. 1986, \aaps, 66,
  191

\bibitem[{{Bidelman} \& {Keenan}(1951)}]{Bafirst}
{Bidelman}, W.~P. \& {Keenan}, P.~C. 1951, \apj, 114, 473

\bibitem[{{Bisterzo} {et~al.}(2012){Bisterzo}, {Gallino}, {Straniero},
  {Cristallo}, \& {K{\"a}ppeler}}]{Bisterzo12}
{Bisterzo}, S., {Gallino}, R., {Straniero}, O., {Cristallo}, S., \&
  {K{\"a}ppeler}, F. 2012, \mnras, 422, 849

\bibitem[{{Boothroyd} \& {Sackmann}(1999)}]{Boothroyd99}
{Boothroyd}, A.~I. \& {Sackmann}, I.~J. 1999, \apj, 510, 232

\bibitem[{{Buntain} {et~al.}(2017){Buntain}, {Doherty}, {Lugaro}, {Lattanzio},
  {Stancliffe}, \& {Karakas}}]{buntain17}
{Buntain}, J.~F., {Doherty}, C.~L., {Lugaro}, M., {et~al.} 2017, \mnras, 471,
  824

\bibitem[{{Chen} {et~al.}(2020){Chen}, {Ivanova}, \&
  {Carroll-Nellenback}}]{chen20}
{Chen}, Z., {Ivanova}, N., \& {Carroll-Nellenback}, J. 2020, \apj, 892, 110

\bibitem[{{Chieffi} \& {Straniero}(1989)}]{FRANEC89}
{Chieffi}, A. \& {Straniero}, O. 1989, \apjs, 71, 47

\bibitem[{{Cristallo} {et~al.}(2015{\natexlab{a}}){Cristallo}, {Abia},
  {Straniero}, \& {Piersanti}}]{cristallo15b}
{Cristallo}, S., {Abia}, C., {Straniero}, O., \& {Piersanti}, L.
  2015{\natexlab{a}}, \apj, 801, 53

\bibitem[{{Cristallo} {et~al.}(2016){Cristallo}, {Karinkuzhi}, {Goswami},
  {Piersanti}, \& {Gobrecht}}]{cristallo16}
{Cristallo}, S., {Karinkuzhi}, D., {Goswami}, A., {Piersanti}, L., \&
  {Gobrecht}, D. 2016, \apj, 833, 181

\bibitem[{{Cristallo} {et~al.}(2011){Cristallo}, {Piersanti}, {Straniero},
  {Gallino}, {Dom{\'{\i}}nguez}, {Abia}, {Di Rico}, {Quintini}, \&
  {Bisterzo}}]{cristallo11}
{Cristallo}, S., {Piersanti}, L., {Straniero}, O., {et~al.} 2011, \apjs, 197,
  17

\bibitem[{{Cristallo} {et~al.}(2009){Cristallo}, {Straniero}, {Gallino},
  {Piersanti}, {Dom{\'{\i}}nguez}, \& {Lederer}}]{cristallo09}
{Cristallo}, S., {Straniero}, O., {Gallino}, R., {et~al.} 2009, \apj, 696, 797

\bibitem[{{Cristallo} {et~al.}(2015{\natexlab{b}}){Cristallo}, {Straniero},
  {Piersanti}, \& {Gobrecht}}]{cristallo15}
{Cristallo}, S., {Straniero}, O., {Piersanti}, L., \& {Gobrecht}, D.
  2015{\natexlab{b}}, \apjs, 219, 40

\bibitem[{{Cseh} {et~al.}(2018){Cseh}, {Lugaro}, {D'Orazi}, {de Castro},
  {Pereira}, {Karakas}, {Moln{\'a}r}, {Plachy}, {Szab{\'o}}, {Pignatari}, \&
  {Cristallo}}]{cseh18}
{Cseh}, B., {Lugaro}, M., {D'Orazi}, V., {et~al.} 2018, A\&A, 620, A146

\bibitem[{{de Castro} {et~al.}(2016){de Castro}, {Pereira}, {Roig}, {Jilinski},
  {Drake}, {Chavero}, \& {Sales Silva}}]{deC}
{de Castro}, D.~B., {Pereira}, C.~B., {Roig}, F., {et~al.} 2016, \mnras, 459,
  4299

\bibitem[{{De Marco} \& {Izzard}(2017)}]{demarco17}
{De Marco}, O. \& {Izzard}, R.~G. 2017, \pasa, 34, e001

\bibitem[{{Denissenkov} \& {Pinsonneault}(2008)}]{Denissenkov08}
{Denissenkov}, P.~A. \& {Pinsonneault}, M. 2008, \apj, 679, 1541

\bibitem[{{El-Badry} {et~al.}(2021){El-Badry}, {Rix}, \& {Heintz}}]{El-Badry21}
{El-Badry}, K., {Rix}, H.-W., \& {Heintz}, T.~M. 2021, \mnras, 506, 2269

\bibitem[{{El-Badry} {et~al.}(2018){El-Badry}, {Rix}, \& {Weisz}}]{El-Badry18}
{El-Badry}, K., {Rix}, H.-W., \& {Weisz}, D.~R. 2018, \apjl, 860, L17

\bibitem[{{Escorza} {et~al.}(2017){Escorza}, {Boffin}, {Jorissen}, {Van Eck},
  {Siess}, {Van Winckel}, {Karinkuzhi}, {Shetye}, \& {Pourbaix}}]{escorza17}
{Escorza}, A., {Boffin}, H.~M.~J., {Jorissen}, A., {et~al.} 2017, \aap, 608,
  A100

\bibitem[{{Fishlock} {et~al.}(2014){Fishlock}, {Karakas}, {Lugaro}, \&
  {Yong}}]{fishlock14}
{Fishlock}, C.~K., {Karakas}, A.~I., {Lugaro}, M., \& {Yong}, D. 2014, \apj,
  797, 44

\bibitem[{{Forsberg} {et~al.}(2019){Forsberg}, {J{\"o}nsson}, {Ryde}, \&
  {Matteucci}}]{Forsberg19}
{Forsberg}, R., {J{\"o}nsson}, H., {Ryde}, N., \& {Matteucci}, F. 2019, \aap,
  631, A113

\bibitem[{{Gaia Collaboration}(2020)}]{gaia_ruwe}
{Gaia Collaboration}. 2020, VizieR Online Data Catalog, I/350

\bibitem[{{Gaia Collaboration} {et~al.}(2018){Gaia Collaboration}, {Brown},
  {Vallenari}, {Prusti}, {de Bruijne}, {Babusiaux}, {Bailer-Jones}, {Biermann},
  {Evans}, {Eyer}, {Jansen}, {Jordi}, {Klioner}, {Lammers}, {Lindegren},
  {Luri}, {Mignard}, {Panem}, {Pourbaix}, {Randich}, {Sartoretti}, {Siddiqui},
  {Soubiran}, {van Leeuwen}, {Walton}, {Arenou}, {Bastian}, {Cropper},
  {Drimmel}, {Katz}, {Lattanzi}, {Bakker}, {Cacciari}, {Casta{\~n}eda},
  {Chaoul}, {Cheek}, {De Angeli}, {Fabricius}, {Guerra}, {Holl}, {Masana},
  {Messineo}, {Mowlavi}, {Nienartowicz}, {Panuzzo}, {Portell}, {Riello},
  {Seabroke}, {Tanga}, {Th{\'e}venin}, {Gracia-Abril}, {Comoretto},
  {Garcia-Reinaldos}, {Teyssier}, {Altmann}, {Andrae}, {Audard},
  {Bellas-Velidis}, {Benson}, {Berthier}, {Blomme}, {Burgess}, {Busso},
  {Carry}, {Cellino}, {Clementini}, {Clotet}, {Creevey}, {Davidson}, {De
  Ridder}, {Delchambre}, {Dell'Oro}, {Ducourant},
  {Fern{\'a}ndez-Hern{\'a}ndez}, {Fouesneau}, {Fr{\'e}mat}, {Galluccio},
  {Garc{\'\i}a-Torres}, {Gonz{\'a}lez-N{\'u}{\~n}ez}, {Gonz{\'a}lez-Vidal},
  {Gosset}, {Guy}, {Halbwachs}, {Hambly}, {Harrison}, {Hern{\'a}ndez},
  {Hestroffer}, {Hodgkin}, {Hutton}, {Jasniewicz}, {Jean-Antoine-Piccolo},
  {Jordan}, {Korn}, {Krone-Martins}, {Lanzafame}, {Lebzelter}, {L{\"o}ffler},
  {Manteiga}, {Marrese}, {Mart{\'\i}n-Fleitas}, {Moitinho}, {Mora}, {Muinonen},
  {Osinde}, {Pancino}, {Pauwels}, {Petit}, {Recio-Blanco}, {Richards},
  {Rimoldini}, {Robin}, {Sarro}, {Siopis}, {Smith}, {Sozzetti}, {S{\"u}veges},
  {Torra}, {van Reeven}, {Abbas}, {Abreu Aramburu}, {Accart}, {Aerts},
  {Altavilla}, {{\'A}lvarez}, {Alvarez}, {Alves}, {Anderson}, {Andrei},
  {Anglada Varela}, {Antiche}, {Antoja}, {Arcay}, {Astraatmadja}, {Bach},
  {Baker}, {Balaguer-N{\'u}{\~n}ez}, {Balm}, {Barache}, {Barata}, {Barbato},
  {Barblan}, {Barklem}, {Barrado}, {Barros}, {Barstow}, {Bartholom{\'e}
  Mu{\~n}oz}, {Bassilana}, {Becciani}, {Bellazzini}, {Berihuete}, {Bertone},
  {Bianchi}, {Bienaym{\'e}}, {Blanco-Cuaresma}, {Boch}, {Boeche}, {Bombrun},
  {Borrachero}, {Bossini}, {Bouquillon}, {Bourda}, {Bragaglia}, {Bramante},
  {Breddels}, {Bressan}, {Brouillet}, {Br{\"u}semeister}, {Brugaletta},
  {Bucciarelli}, {Burlacu}, {Busonero}, {Butkevich}, {Buzzi}, {Caffau},
  {Cancelliere}, {Cannizzaro}, {Cantat-Gaudin}, {Carballo}, {Carlucci},
  {Carrasco}, {Casamiquela}, {Castellani}, {Castro-Ginard}, {Charlot},
  {Chemin}, {Chiavassa}, {Cocozza}, {Costigan}, {Cowell}, {Crifo}, {Crosta},
  {Crowley}, {Cuypers}, {Dafonte}, {Damerdji}, {Dapergolas}, {David}, {David},
  {de Laverny}, {De Luise}, {De March}, {de Martino}, {de Souza}, {de Torres},
  {Debosscher}, {del Pozo}, {Delbo}, {Delgado}, {Delgado}, {Di Matteo},
  {Diakite}, {Diener}, {Distefano}, {Dolding}, {Drazinos}, {Dur{\'a}n},
  {Edvardsson}, {Enke}, {Eriksson}, {Esquej}, {Eynard Bontemps}, {Fabre},
  {Fabrizio}, {Faigler}, {Falc{\~a}o}, {Farr{\`a}s Casas}, {Federici},
  {Fedorets}, {Fernique}, {Figueras}, {Filippi}, {Findeisen}, {Fonti},
  {Fraile}, {Fraser}, {Fr{\'e}zouls}, {Gai}, {Galleti}, {Garabato},
  {Garc{\'\i}a-Sedano}, {Garofalo}, {Garralda}, {Gavel}, {Gavras}, {Gerssen},
  {Geyer}, {Giacobbe}, {Gilmore}, {Girona}, {Giuffrida}, {Glass}, {Gomes},
  {Granvik}, {Gueguen}, {Guerrier}, {Guiraud}, {Guti{\'e}rrez-S{\'a}nchez},
  {Haigron}, {Hatzidimitriou}, {Hauser}, {Haywood}, {Heiter}, {Helmi}, {Heu},
  {Hilger}, {Hobbs}, {Hofmann}, {Holland}, {Huckle}, {Hypki}, {Icardi},
  {Jan{\ss}en}, {Jevardat de Fombelle}, {Jonker}, {Juh{\'a}sz}, {Julbe},
  {Karampelas}, {Kewley}, {Klar}, {Kochoska}, {Kohley}, {Kolenberg},
  {Kontizas}, {Kontizas}, {Koposov}, {Kordopatis}, {Kostrzewa-Rutkowska},
  {Koubsky}, {Lambert}, {Lanza}, {Lasne}, {Lavigne}, {Le Fustec}, {Le
  Poncin-Lafitte}, {Lebreton}, {Leccia}, {Leclerc}, {Lecoeur-Taibi},
  {Lenhardt}, {Leroux}, {Liao}, {Licata}, {Lindstr{\o}m}, {Lister}, {Livanou},
  {Lobel}, {L{\'o}pez}, {Managau}, {Mann}, {Mantelet}, {Marchal}, {Marchant},
  {Marconi}, {Marinoni}, {Marschalk{\'o}}, {Marshall}, {Martino}, {Marton},
  {Mary}, {Massari}, {Matijevi{\v{c}}}, {Mazeh}, {McMillan}, {Messina},
  {Michalik}, {Millar}, {Molina}, {Molinaro}, {Moln{\'a}r}, {Montegriffo},
  {Mor}, {Morbidelli}, {Morel}, {Morris}, {Mulone}, {Muraveva}, {Musella},
  {Nelemans}, {Nicastro}, {Noval}, {O'Mullane}, {Ord{\'e}novic},
  {Ord{\'o}{\~n}ez-Blanco}, {Osborne}, {Pagani}, {Pagano}, {Pailler},
  {Palacin}, {Palaversa}, {Panahi}, {Pawlak}, {Piersimoni}, {Pineau}, {Plachy},
  {Plum}, {Poggio}, {Poujoulet}, {Pr{\v{s}}a}, {Pulone}, {Racero}, {Ragaini},
  {Rambaux}, {Ramos-Lerate}, {Regibo}, {Reyl{\'e}}, {Riclet}, {Ripepi}, {Riva},
  {Rivard}, {Rixon}, {Roegiers}, {Roelens}, {Romero-G{\'o}mez}, {Rowell},
  {Royer}, {Ruiz-Dern}, {Sadowski}, {Sagrist{\`a} Sell{\'e}s}, {Sahlmann},
  {Salgado}, {Salguero}, {Sanna}, {Santana-Ros}, {Sarasso}, {Savietto},
  {Schultheis}, {Sciacca}, {Segol}, {Segovia}, {S{\'e}gransan}, {Shih},
  {Siltala}, {Silva}, {Smart}, {Smith}, {Solano}, {Solitro}, {Sordo}, {Soria
  Nieto}, {Souchay}, {Spagna}, {Spoto}, {Stampa}, {Steele},
  {Steidelm{\"u}ller}, {Stephenson}, {Stoev}, {Suess}, {Surdej}, {Szabados},
  {Szegedi-Elek}, {Tapiador}, {Taris}, {Tauran}, {Taylor}, {Teixeira},
  {Terrett}, {Teyssandier}, {Thuillot}, {Titarenko}, {Torra Clotet}, {Turon},
  {Ulla}, {Utrilla}, {Uzzi}, {Vaillant}, {Valentini}, {Valette}, {van Elteren},
  {Van Hemelryck}, {van Leeuwen}, {Vaschetto}, {Vecchiato}, {Veljanoski},
  {Viala}, {Vicente}, {Vogt}, {von Essen}, {Voss}, {Votruba}, {Voutsinas},
  {Walmsley}, {Weiler}, {Wertz}, {Wevers}, {Wyrzykowski}, {Yoldas},
  {{\v{Z}}erjal}, {Ziaeepour}, {Zorec}, {Zschocke}, {Zucker}, {Zurbach}, \&
  {Zwitter}}]{Gaia18}
{Gaia Collaboration}, {Brown}, A.~G.~A., {Vallenari}, A., {et~al.} 2018, \aap,
  616, A1

\bibitem[{{Gallino} {et~al.}(1998){Gallino}, {Arlandini}, {Busso}, {Lugaro},
  {Travaglio}, {Straniero}, {Chieffi}, \& {Limongi}}]{gallino98}
{Gallino}, R., {Arlandini}, C., {Busso}, M., {et~al.} 1998, \apj, 497, 388

\bibitem[{{Gratton} \& {Sneden}(1994)}]{gratton_sneden94}
{Gratton}, R.~G. \& {Sneden}, C. 1994, \aap, 287, 927

\bibitem[{{Husti} {et~al.}(2009){Husti}, {Gallino}, {Bisterzo}, {Straniero}, \&
  {Cristallo}}]{husti09}
{Husti}, L., {Gallino}, R., {Bisterzo}, S., {Straniero}, O., \& {Cristallo}, S.
  2009, \pasa, 26, 176

\bibitem[{{J{\"o}nsson} {et~al.}(2017){J{\"o}nsson}, {Ryde}, {Nordlander},
  {Pehlivan Rhodin}, {Hartman}, {J{\"o}nsson}, \& {Eriksson}}]{jonsson17}
{J{\"o}nsson}, H., {Ryde}, N., {Nordlander}, T., {et~al.} 2017, \aap, 598, A100

\bibitem[{{Jorissen} {et~al.}(2019){Jorissen}, {Boffin}, {Karinkuzhi}, {Van
  Eck}, {Escorza}, {Shetye}, \& {Van Winckel}}]{jorissen19}
{Jorissen}, A., {Boffin}, H.~M.~J., {Karinkuzhi}, D., {et~al.} 2019, A\&A, 626,
  A127

\bibitem[{{K{\"a}ppeler} {et~al.}(2011){K{\"a}ppeler}, {Gallino}, {Bisterzo},
  \& {Aoki}}]{kaeppeler11}
{K{\"a}ppeler}, F., {Gallino}, R., {Bisterzo}, S., \& {Aoki}, W. 2011, Reviews
  of Modern Physics, 83, 157

\bibitem[{{Karakas} {et~al.}(2012){Karakas}, {Garc{\'{\i}}a-Hern{\'a}ndez}, \&
  {Lugaro}}]{karakas12}
{Karakas}, A.~I., {Garc{\'{\i}}a-Hern{\'a}ndez}, D.~A., \& {Lugaro}, M. 2012,
  \apj, 751, 8

\bibitem[{{Karakas} \& {Lattanzio}(2014)}]{karakas14}
{Karakas}, A.~I. \& {Lattanzio}, J.~C. 2014, \pasa, 31, e030

\bibitem[{{Karakas} \& {Lugaro}(2016)}]{karakas16}
{Karakas}, A.~I. \& {Lugaro}, M. 2016, \apj, 825, 26

\bibitem[{{Karakas} {et~al.}(2018){Karakas}, {Lugaro}, {Carlos}, {Cseh},
  {Kamath}, \& {Garc{\'\i}a-Hern{\'a}ndez}}]{karakas18}
{Karakas}, A.~I., {Lugaro}, M., {Carlos}, M., {et~al.} 2018, \mnras, 477, 421

\bibitem[{{Karinkuzhi} {et~al.}(2018){Karinkuzhi}, {Van Eck}, {Jorissen},
  {Goriely}, {Siess}, {Merle}, {Escorza}, {Van der Swaelmen}, {Boffin},
  {Masseron}, {Shetye}, \& {Plez}}]{karinkuzhi18b}
{Karinkuzhi}, D., {Van Eck}, S., {Jorissen}, A., {et~al.} 2018, ArXiv e-prints
  [\eprint[arXiv]{1807.06332}]

\bibitem[{{Kobayashi} {et~al.}(2020){Kobayashi}, {Karakas}, \&
  {Lugaro}}]{Kobayashi20}
{Kobayashi}, C., {Karakas}, A.~I., \& {Lugaro}, M. 2020, \apj, 900, 179

\bibitem[{{Lugaro} {et~al.}(2020){Lugaro}, {Cseh}, {Vil{\'a}gos}, {Karakas},
  {Ventura}, {Dell'Agli}, {Trappitsch}, {Hampel}, {D'Orazi}, {Pereira},
  {Tagliente}, {Szab{\'o}}, {Pignatari}, {Battino}, {Tattersall}, {Ek},
  {Sch{\"o}nb{\"a}chler}, {Hron}, \& {Nittler}}]{lugaro20}
{Lugaro}, M., {Cseh}, B., {Vil{\'a}gos}, B., {et~al.} 2020, \apj, 898, 96

\bibitem[{{Lugaro} {et~al.}(2003){Lugaro}, {Herwig}, {Lattanzio}, {Gallino}, \&
  {Straniero}}]{lugaro03}
{Lugaro}, M., {Herwig}, F., {Lattanzio}, J.~C., {Gallino}, R., \& {Straniero},
  O. 2003, \apj, 586, 1305

\bibitem[{{Lugaro} {et~al.}(2012){Lugaro}, {Karakas}, {Stancliffe}, \&
  {Rijs}}]{lugaro12}
{Lugaro}, M., {Karakas}, A.~I., {Stancliffe}, R.~J., \& {Rijs}, C. 2012, \apj,
  747, 2

\bibitem[{{McClure}(1983)}]{RV2-McC}
{McClure}, R.~D. 1983, \apj, 268, 264

\bibitem[{{McClure} {et~al.}(1980){McClure}, {Fletcher}, \&
  {Nemec}}]{RV1-McCFN}
{McClure}, R.~D., {Fletcher}, J.~M., \& {Nemec}, J.~M. 1980, \apjl, 238, L35

\bibitem[{{Mishenina} {et~al.}(2019){Mishenina}, {Pignatari}, {Gorbaneva},
  {Travaglio}, {C{\^o}t{\'e}}, {Thielemann}, \& {Soubiran}}]{mishenina19}
{Mishenina}, T., {Pignatari}, M., {Gorbaneva}, T., {et~al.} 2019, \mnras, 489,
  1697

\bibitem[{{Piersanti} {et~al.}(2013){Piersanti}, {Cristallo}, \&
  {Straniero}}]{piersanti13}
{Piersanti}, L., {Cristallo}, S., \& {Straniero}, O. 2013, \apj, 774, 98

\bibitem[{{Roriz} {et~al.}(2021{\natexlab{a}}){Roriz}, {Lugaro}, {Pereira},
  {Drake}, {Junqueira}, \& {Sneden}}]{roriz21}
{Roriz}, M.~P., {Lugaro}, M., {Pereira}, C.~B., {et~al.} 2021{\natexlab{a}},
  \mnras, 501, 5834

\bibitem[{{Roriz} {et~al.}(2021{\natexlab{b}}){Roriz}, {Lugaro}, {Pereira},
  {Sneden}, {Junqueira}, {Karakas}, \& {Drake}}]{Roriz_heavy}
{Roriz}, M.~P., {Lugaro}, M., {Pereira}, C.~B., {et~al.} 2021{\natexlab{b}},
  \mnras

\bibitem[{{Shejeelammal} {et~al.}(2020){Shejeelammal}, {Goswami}, {Goswami},
  {Rathour}, \& {Masseron}}]{Shejeelammal20}
{Shejeelammal}, J., {Goswami}, A., {Goswami}, P.~P., {Rathour}, R.~S., \&
  {Masseron}, T. 2020, \mnras, 492, 3708

\bibitem[{{Shetye} {et~al.}(2019){Shetye}, {Goriely}, {Siess}, {Van Eck},
  {Jorissen}, \& {Van Winckel}}]{shetye19_1msun}
{Shetye}, S., {Goriely}, S., {Siess}, L., {et~al.} 2019, \aap, 625, L1

\bibitem[{{Siess} {et~al.}(2000){Siess}, {Dufour}, \& {Forestini}}]{starevol}
{Siess}, L., {Dufour}, E., \& {Forestini}, M. 2000, \aap, 358, 593

\bibitem[{{Sneden}(1973)}]{SnedenPhDT}
{Sneden}, C.~A. 1973, PhD thesis, THE UNIVERSITY OF TEXAS AT AUSTIN.

\bibitem[{{Stancliffe}(2021)}]{Stancliffe21}
{Stancliffe}, R.~J. 2021, \mnras, 505, 5554

\bibitem[{{Stancliffe} \& {Glebbeek}(2008)}]{Stancliffe08}
{Stancliffe}, R.~J. \& {Glebbeek}, E. 2008, \mnras, 389, 1828

\bibitem[{{Stancliffe} {et~al.}(2007){Stancliffe}, {Glebbeek}, {Izzard}, \&
  {Pols}}]{Stancliffe07}
{Stancliffe}, R.~J., {Glebbeek}, E., {Izzard}, R.~G., \& {Pols}, O.~R. 2007,
  \aap, 464, L57

\bibitem[{{Stassun} \& {Torres}(2021)}]{Stassun21}
{Stassun}, K.~G. \& {Torres}, G. 2021, \apjl, 907, L33

\bibitem[{{Van der Swaelmen} {et~al.}(2017){Van der Swaelmen}, {Boffin},
  {Jorissen}, \& {Van Eck}}]{vanderswaelmen17}
{Van der Swaelmen}, M., {Boffin}, H.~M.~J., {Jorissen}, A., \& {Van Eck}, S.
  2017, \aap, 597, A68

\bibitem[{{van Raai} {et~al.}(2012){van Raai}, {Lugaro}, {Karakas},
  {Garc{\'{\i}}a-Hern{\'a}ndez}, \& {Yong}}]{vanraai12}
{van Raai}, M.~A., {Lugaro}, M., {Karakas}, A.~I.,
  {Garc{\'{\i}}a-Hern{\'a}ndez}, D.~A., \& {Yong}, D. 2012, \aap, 540, A44

\bibitem[{{Yang} {et~al.}(2016){Yang}, {Liang}, {Spite}, {Chen}, {Zhao},
  {Zhang}, {Liu}, {Liu}, {Liu}, {Deng}, {Spite}, {Hill}, \& {Zhang}}]{Yang2016}
{Yang}, G.-C., {Liang}, Y.-C., {Spite}, M., {et~al.} 2016, Research in
  Astronomy and Astrophysics, 16, 19

\end{thebibliography}

\begin{appendix}
\section{Figures}
\label{appendix_figs}

We present here the results for the rest of the stars belonging to Group 1. The stars are grouped in the same way as in the main text: stars with the whole heavy elements pattern in agreement with the models belong to subgroup 1a, subgroup 1b includes stars with higher Nb and sometimes Mo and Ru compared to the models, and stars in subgroup 1c show abundances higher than the model predictions in the elements Nb, Mo, Ru, Nd, and Sm.
The $\delta$ values were calculated as described in Sect. \ref{sec:dil_fact_calc}, normalising the models to match the [Ce/Fe] of the individual stars. We only show models with $\delta \leq 0.9$ and with a good match to the observed [Rb/Fe] values.

\subsection{Group 1a: observed abundance pattern in agreement with model predictions}

\begin{figure*}[!ht]
 \caption{Same as Fig. \ref{fig:HD154430} but for HD 58368}
 \label{fig:HD58368}
 \centering
 \includegraphics[width=\hsize]{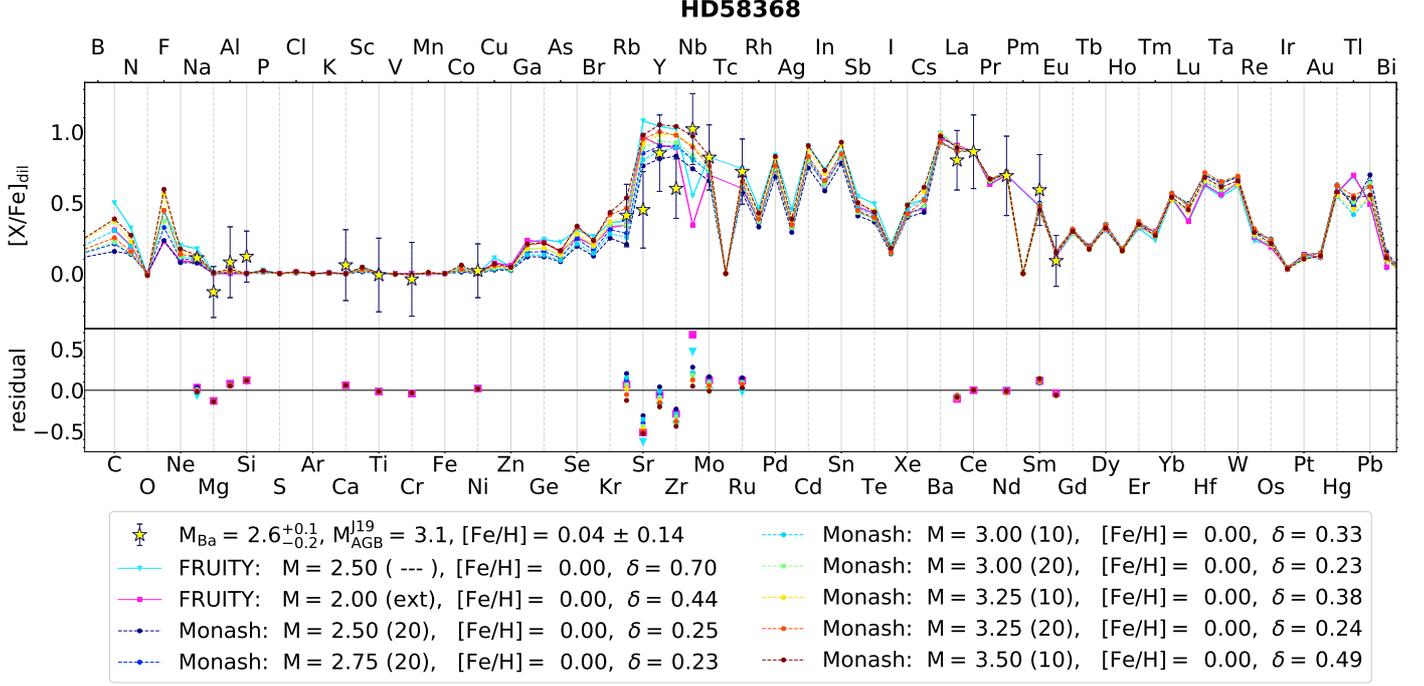}
 \end{figure*}

\subsubsection{HD 58368 (Fig.~\ref{fig:HD58368})}

We show models between 2.0 and 3.5 \msun. 
These models match well the Rb abundance and the second $s$-process peak, although the observed Sr and Zr abundances are lower than all of the model values. The Rb value is indicating that the polluting AGB was around 3 \msun, in agreement with M$^{\rm{J}19}_{\rm{AGB}}$ of 3.1 \msun. Higher mass FRUITY models have higher $\delta$ value than our limit of 0.9.


\begin{figure*}[!ht]
 \caption{Same as Fig. \ref{fig:HD154430} but for HD 91208}
 \label{fig:HD91208}
 \centering
 \includegraphics[width=\hsize]{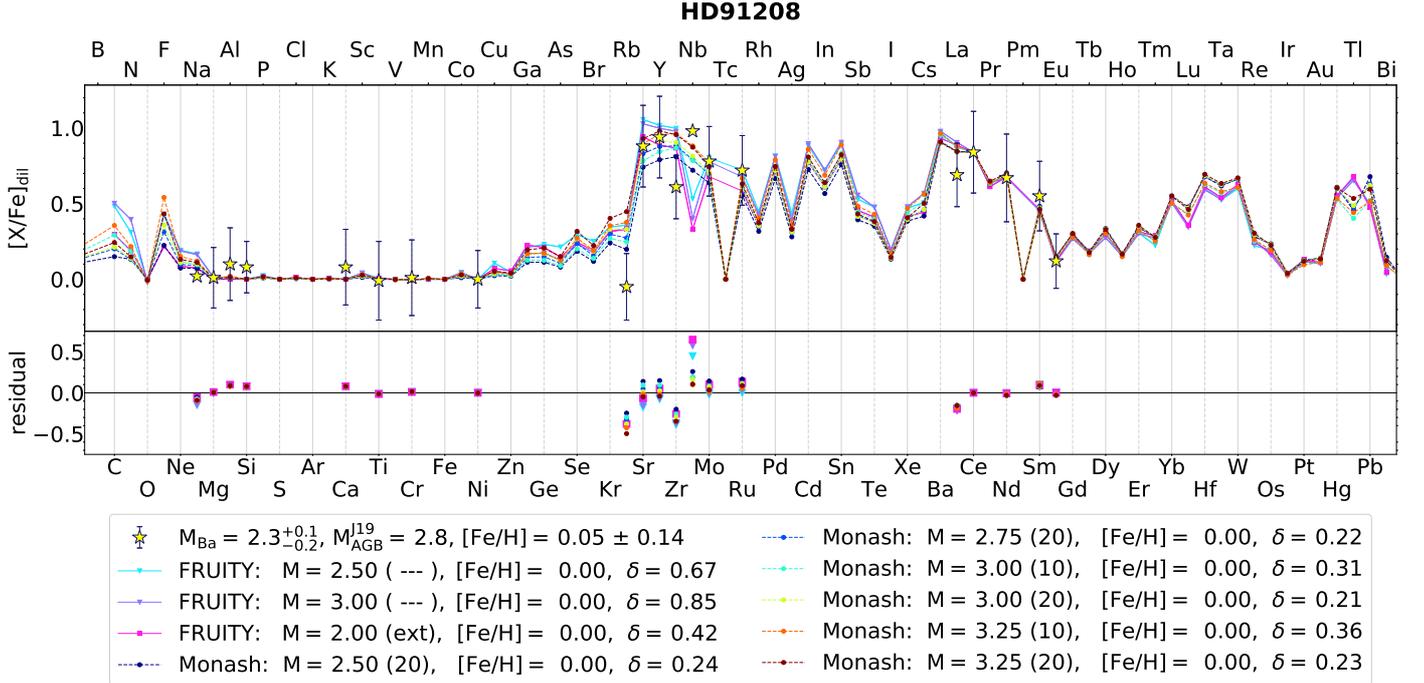}
 \end{figure*}

\subsubsection{HD 91208 (Fig.~\ref{fig:HD91208})}

M$^{\rm{J}19}_{\rm{AGB}}$ is close to 3.0 \msun, and we show models between 2.0 \msun~(FRUITY 'TAIL' and T60) and 3.25 \msun, which we selected as the best compromise to cover both Rb and the first peak elements Sr and Y. Unfortunately, there is no available error bar for Nb, but its proximity to the models allows us to categorise this star in subgroup 1a.
We note that FRUITY models show much larger $\delta$ values than the Monash models for this star. 

\begin{figure*}[!ht]
 \caption{Same as Fig. \ref{fig:HD154430} but for HD 201824}
 \label{fig:HD201824}
 \centering
 \includegraphics[width=\hsize]{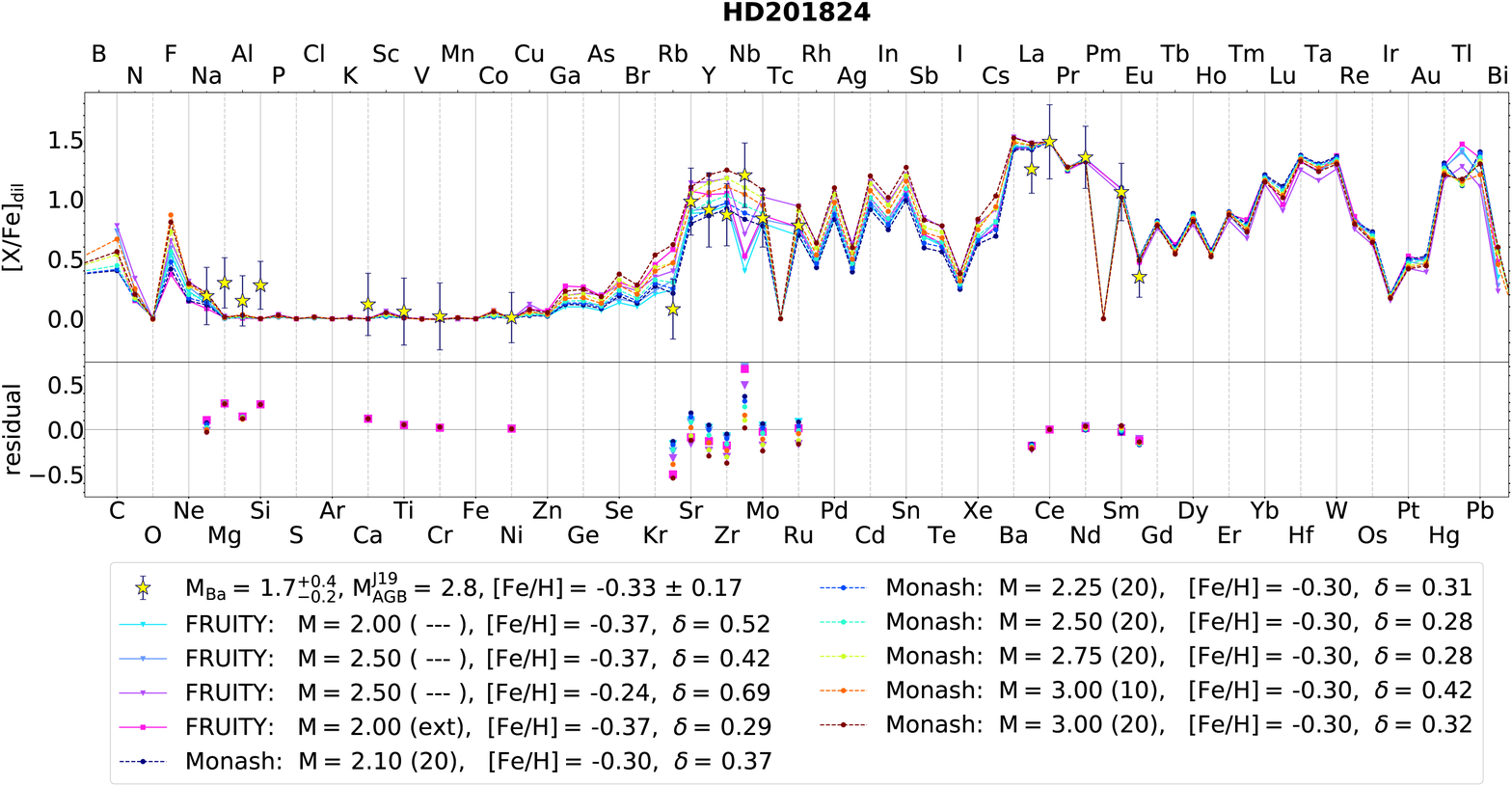}
 \end{figure*}

\subsubsection{HD 201824 (Fig.~\ref{fig:HD201824})}
 
We show FRUITY models between 2.0 and 2.5 \msun~and Monash models between 2.1 and 3.0 \msun. FRUITY models higher than 2.5 \msun~have larger $\delta$ values than our limit of 0.9. The FRUITY 2.5 \msun~model with [Fe/H] $= -$0.37 and Monash models below 3.0 \msun, match the abundance pattern very well, thus we conclude that the polluter AGB was around 2.5 \msun, close to M$^{\rm{J}19}_{\rm{AGB}}$.

\subsection{Group 1b: higher observed Nb (and sometimes Mo and Ru) than model predictions}

\begin{figure*}[!ht]
 \caption{Same as Fig. \ref{fig:HD154430} but for HD 20394}
 \label{fig:HD20394}
 \centering
 \includegraphics[width=\hsize]{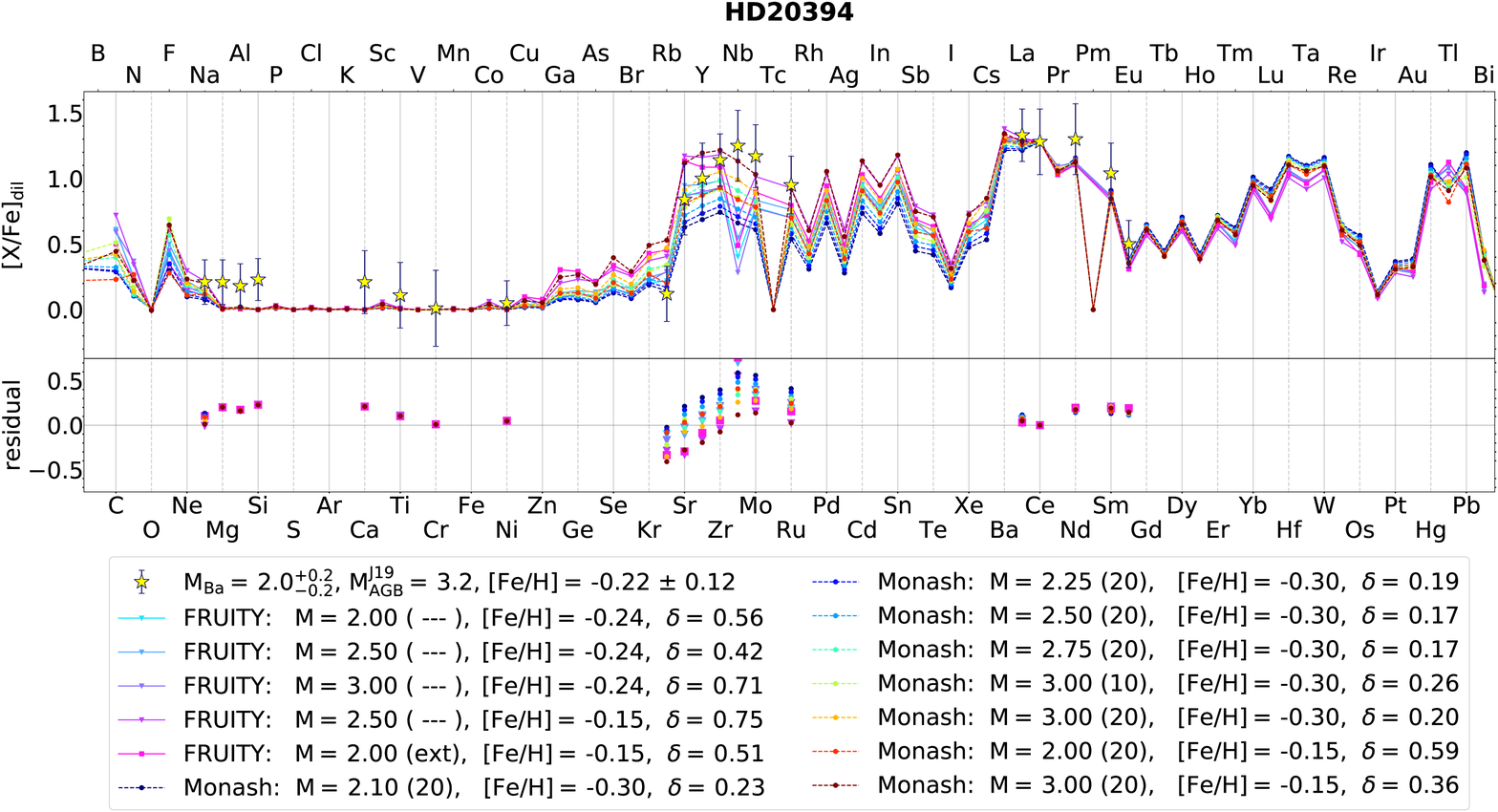}
 \end{figure*}

\subsubsection{HD 20394 (Fig.~\ref{fig:HD20394})}

Both $s$-process peaks are matched well by the models in the mass range of 2--3 \msun. We note that FRUITY models show much larger $\delta$ values than the Monash models for this star. The match of [Rb/Fe] with the models points to an AGB with initial mass of 3 \msun~or lower rather than higher. Although Nb is matched by the 3 \msun~Monash model, that model is far from matching Rb, thus we keep this star in subgroup 1b.

\begin{figure*}[!ht]
 \caption{Same as Fig. \ref{fig:HD154430} but for HD 40430}
 \label{fig:HD40430}
 \centering
 \includegraphics[width=\hsize]{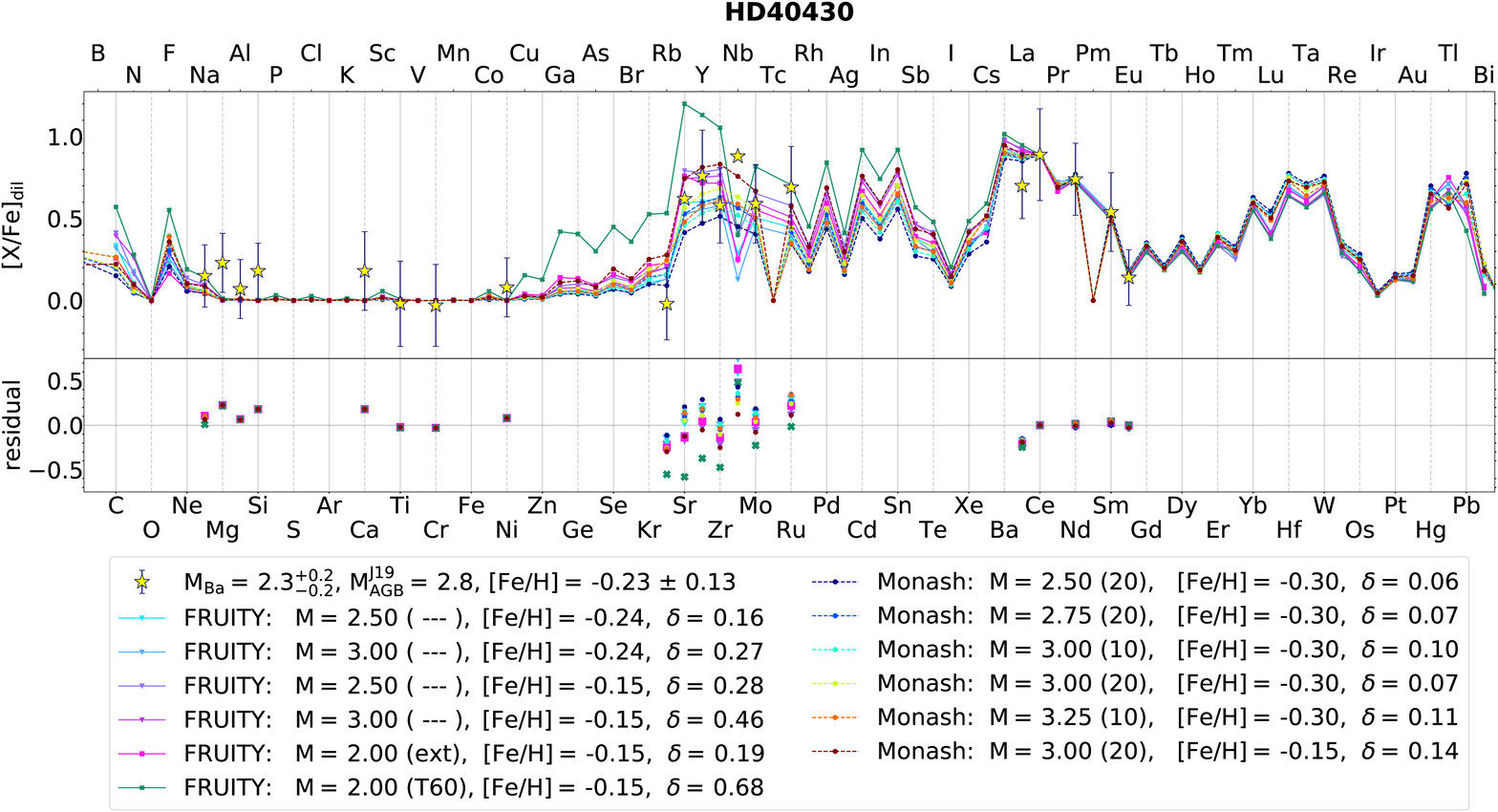}
 \end{figure*}

\subsubsection{HD 40430 (Fig.~\ref{fig:HD40430})}

In agreement with M$^{\rm{J}19}_{\rm{AGB}}$ of 2.8 \msun, models between 2.0 and 3.25 \msun~match well the measured elements of HD 40430. We also show the FRUITY T60 model, which is far from the abundance pattern of the first peak of this Ba star.
The Rb value points to initial AGB masses in the lower mass range, $\approx$2.5 \msun. Nb is higher than the model predictions, however, since an error bar is not available, we keep this star in subgroup 1b.

\begin{figure*}[!ht]
 \caption{Same as Fig. \ref{fig:HD154430} but for HD 53199}
 \label{fig:HD53199}
 \centering
 \includegraphics[width=\hsize]{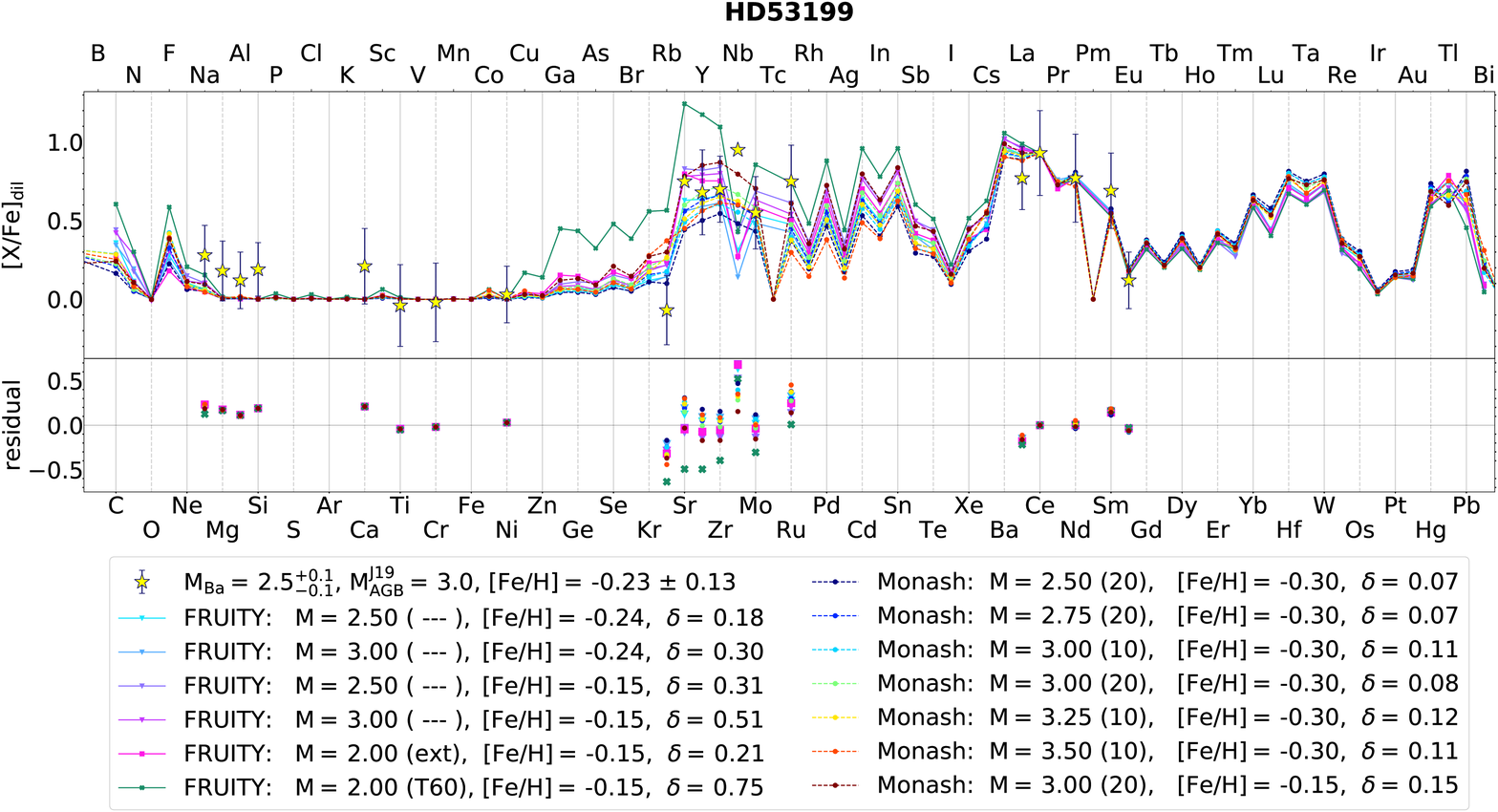}
 \end{figure*}

\subsubsection{HD 53199 (Fig.~\ref{fig:HD53199})}
 
Models between 2.0 and 3.5 \msun~are shown for HD 53199. 
Except from the 2.0 \msun~FRUITY T60 model, all models match well the $s$-process peaks, with the lower than predicted [Rb/Fe] value indicating the lowest initial AGB mass. Models above 3.5 \msun~cannot reach the observed [Rb/Fe] and first $s$-process abundances at the same time, thus we confirm that the polluter AGB is around 3 \msun, in agreement with M$^{\rm{J}19}_{\rm{AGB}}$.

\begin{figure*}[!ht]
 \caption{Same as Fig. \ref{fig:HD154430} but for HD 58121}
 \label{fig:HD58121}
 \centering
 \includegraphics[width=\hsize]{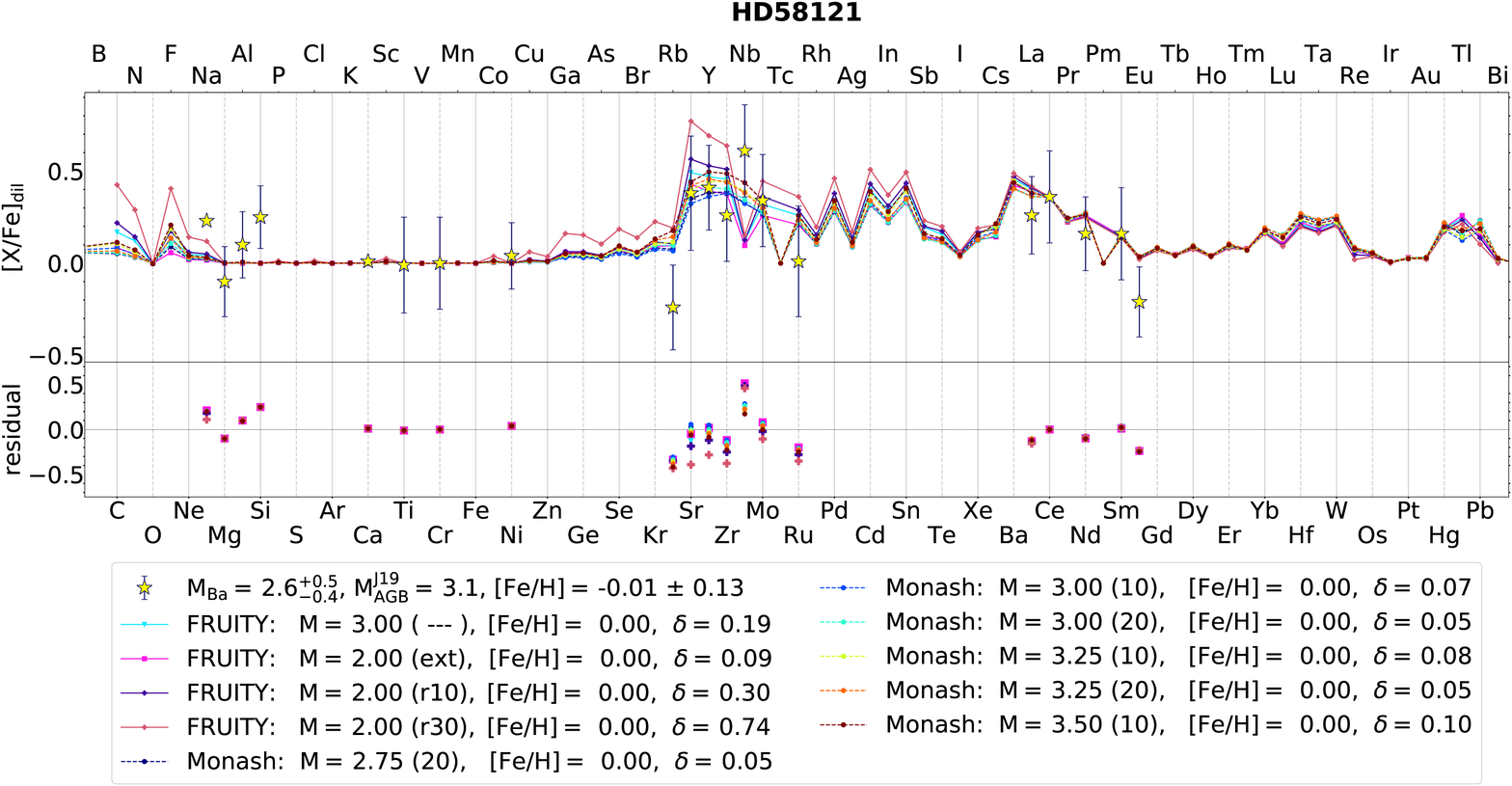}
 \end{figure*}

\subsubsection{HD 58121 (Fig.~\ref{fig:HD58121})}
 
We show different models with masses between 2.0 and 3.5 \msun. Except from the FRUITY r30 model, which shows higher light $s$-elements than other models, all the remaining models are in good agreement with the $s$-process peak elements, although [Rb/Fe] and [Eu/Fe] are lower than 0 and impossible to match by model predictions. \cite{Yang2016} determined Eu abundance for this star using only one line at 6645.11 \AA. They gave [Eu/Fe] = 0.51, which is much higher than the model predictions shown here, and also higher than Eu abundances for normal giant stars in \cite{Forsberg19}.

\begin{figure*}[!ht]
 \caption{Same as Fig. \ref{fig:HD154430} but for HD 59852}
 \label{fig:HD59852}
 \centering
 \includegraphics[width=\hsize]{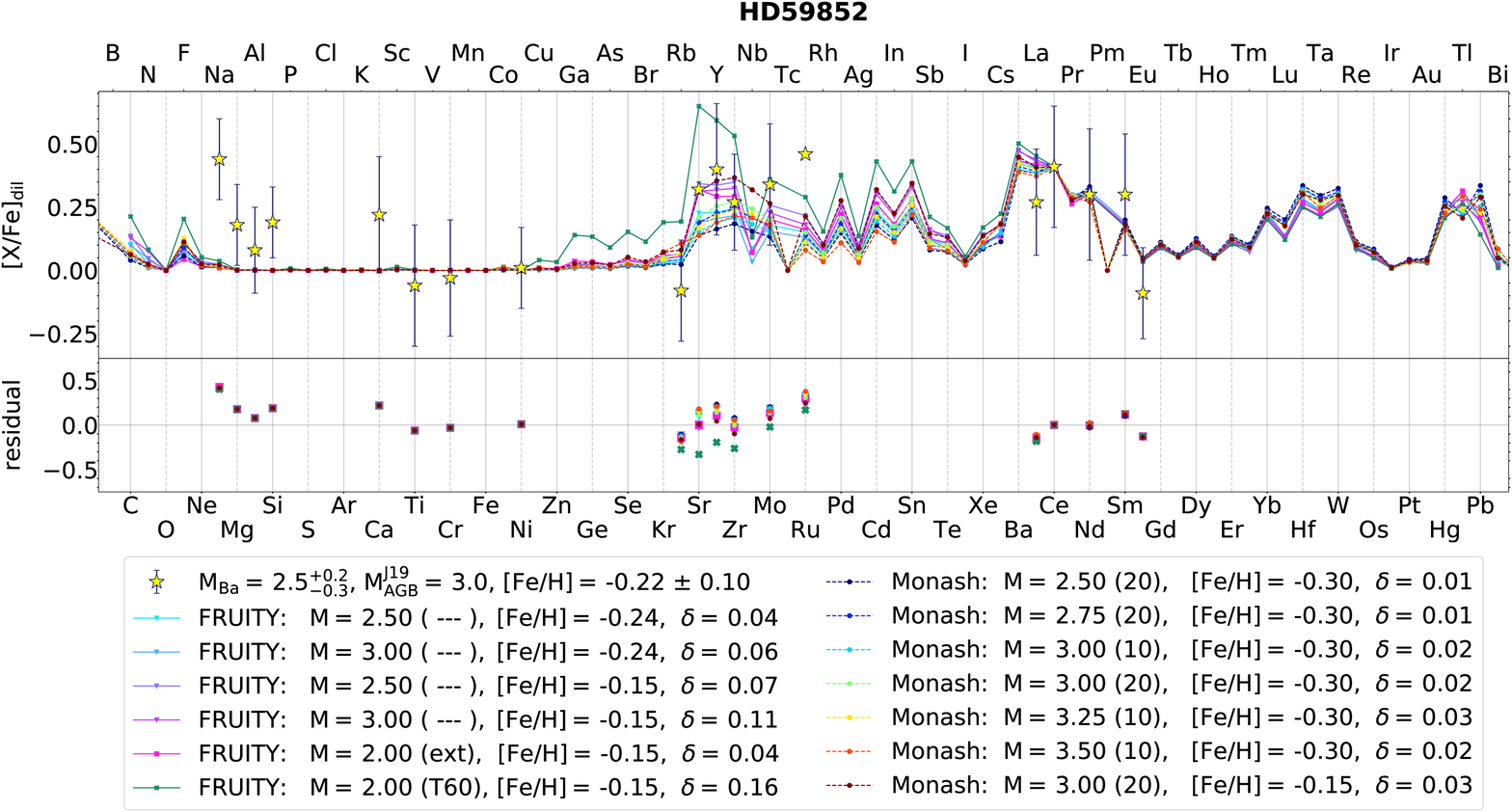}
 \end{figure*}

\subsubsection{HD 59852 (Fig.~\ref{fig:HD59852})}

We show different models between 2.0 and 3.5 \msun~for this star. There are no error bars for three elements (Sr, Zr, Ru), and Nb is not available for this star.
All models, but the FRUITY T60 model, match well the $s$-process abundance pattern, except Ru, thus we keep this star in subgroup 1b. We conclude that an AGB with mass between 2 and 3 \msun~was the polluter for this Ba star, in agreement with M$^{\rm{J}19}_{\rm{AGB}}$.

\begin{figure*}[!ht]
 \caption{Same as Fig. \ref{fig:HD154430} but for HD 95193}
 \label{fig:HD95193}
 \centering
 \includegraphics[width=\hsize]{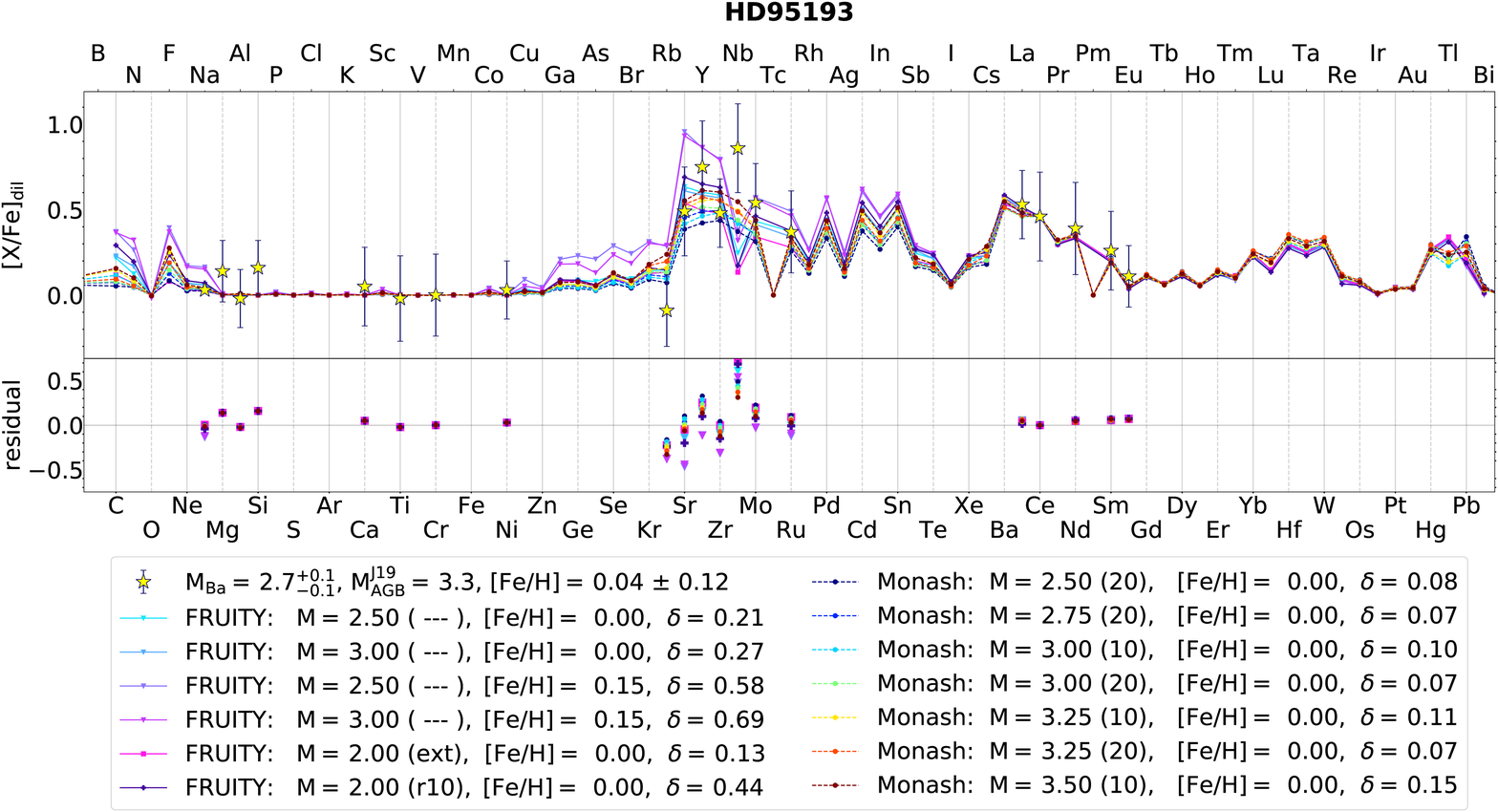}
 \end{figure*}

 \subsubsection{HD 95193 (Fig.~\ref{fig:HD95193})}

We show FRUITY models between 2.0 and 3.0 \msun~and Monash models, with different \iso{13}C pocket sizes, between 2.5 and 3.5 \msun. The FRUITY models with supersolar metallicity ([Fe/H = 0.15], still within the error bar of the Ba star's metallicity) overestimate the determined Sr, Y and Zr abundances compared to other models shown here and these models also have larger $\delta$ values than other model predictions for this star.
All of the other models match the determined abundances well, although the [Rb/Fe] value is lower than most of the model predictions, indicating a lower than 3 \msun~initial AGB mass.


\begin{figure*}[!ht]
 \caption{Same as Fig. \ref{fig:HD154430} but for HD 119185}
 \label{fig:HD119185}
 \centering
 \includegraphics[width=\hsize]{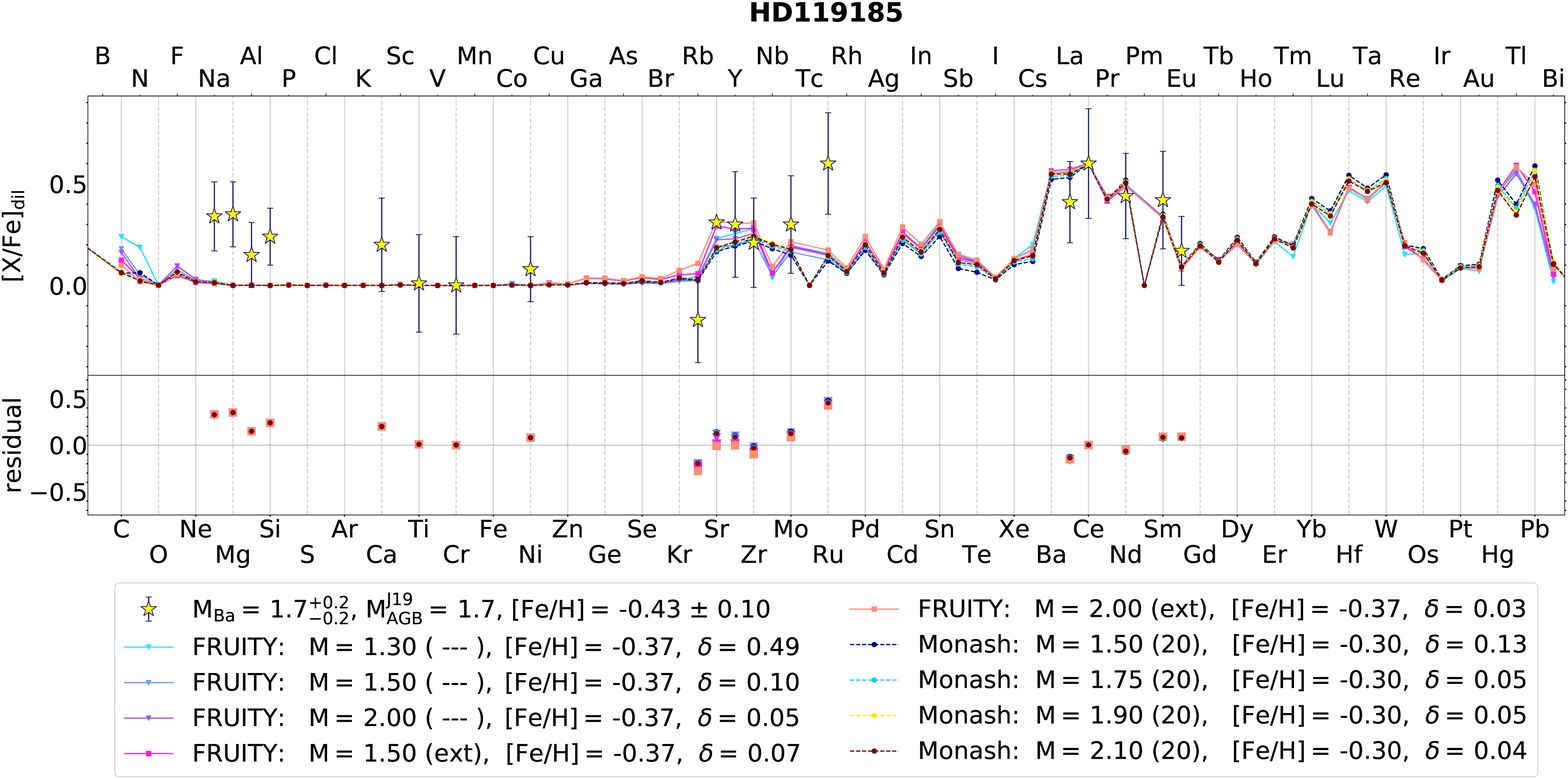}
 \end{figure*}

\subsubsection{HD 119185 (Fig.~\ref{fig:HD119185})}

The abundance pattern of HD 119185 can be matched by low mass (1.5--2.1 \msun) models very well. 
The [Rb/Fe] value is lower than solar, but within the error bar its value is in agreement with a low mass (1.5--2.1 \msun) polluting AGB. As for HD 59852, Nb is absent and [Ru/Fe] is higher than the model predictions.


\begin{figure*}[!ht]
 \caption{Same as Fig. \ref{fig:HD154430} but for HD 143899}
 \label{fig:HD143899}
 \centering
 \includegraphics[width=\hsize]{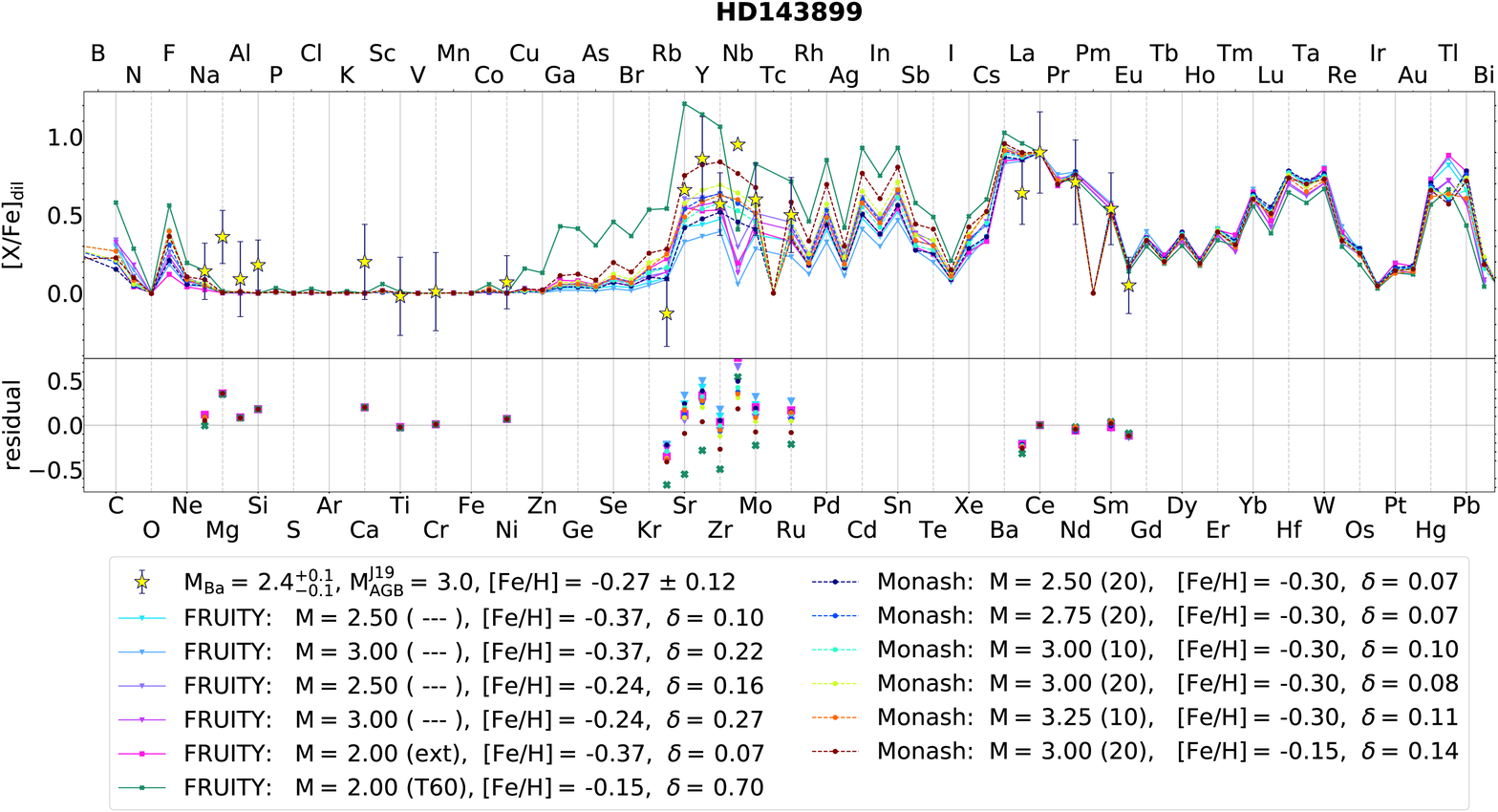}
 \end{figure*}

\subsubsection{HD 143899 (Fig.~\ref{fig:HD143899})}
 
We show models between 2 and 3.25 \msun~with different \iso{13}C pockets. The observed [La/Fe] and [Rb/Fe] values are somewhat lower and, as for the other stars in this subgroup, Nb is higher than the model predictions, but the error bar lacks for this element due to the low number of lines used for the derivation of its abundance. We plot the 2.0 \msun~FRUITY T60 model for completeness, because this model is at the limit of the metallicity range, but it highly overproduces the first peak elements compared to the observations and also the $\delta$ value of this model is much higher than for the others. Models between 2.5 and 3.0 \msun~have the best match with the abundance pattern, which is in agreement with M$^{\rm{J}19}_{\rm{AGB}}$ of 3 \msun.


\begin{figure*}[!ht]
 \caption{Same as Fig. \ref{fig:HD154430} but for HD 200063}
 \label{fig:HD200063}
 \centering
 \includegraphics[width=\hsize]{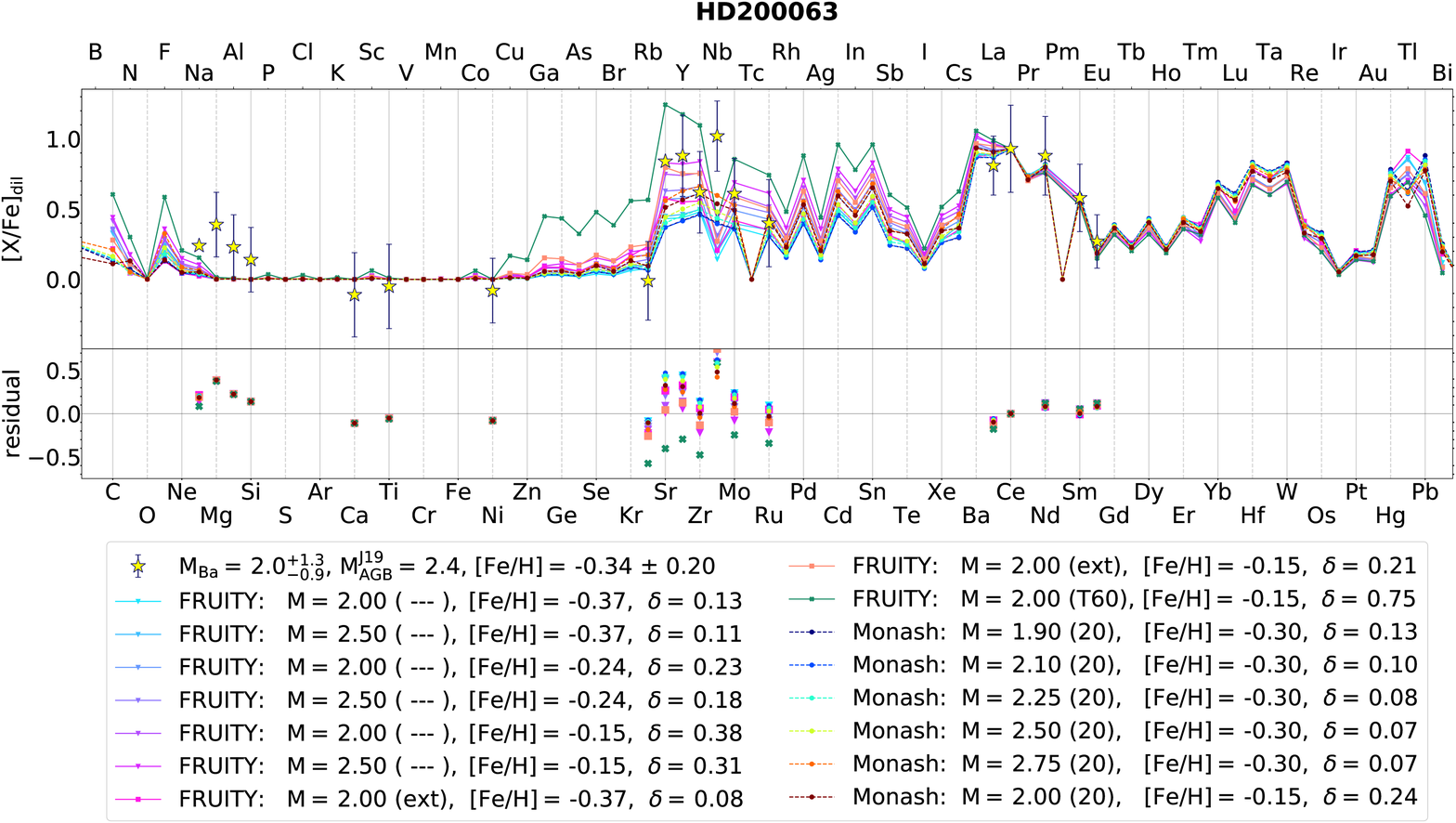}
 \end{figure*}

 \subsubsection{HD 200063 (Fig.~\ref{fig:HD200063})}
 
We show different metallicity models between 1.9 and 2.75 \msun. The uncertain mass of the Ba star and especially the strong assumption of the constant Q when deriving the WD mass could lead to a poorly estimated M$^{\rm{J}19}_{\rm{AGB}}$. All of the models are within the error bar of [Rb/Fe] and, as it was in the case of HD 143899, the FRUITY T60 model largely overproduces the first peak elements. Higher metallicity FRUITY models and the highest mass Monash model with 2.75 \msun~match the abundance pattern within the error bars. This mass range of the models is in agreement with M$^{\rm{J}19}_{\rm{AGB}}$.  Nb is higher than the model predictions, as for any other star in this subgroup. 

\begin{figure*}[!ht]
 \caption{Same as Fig. \ref{fig:HD154430} but for HD 210946}
 \label{fig:HD210946}
 \centering
 \includegraphics[width=\hsize]{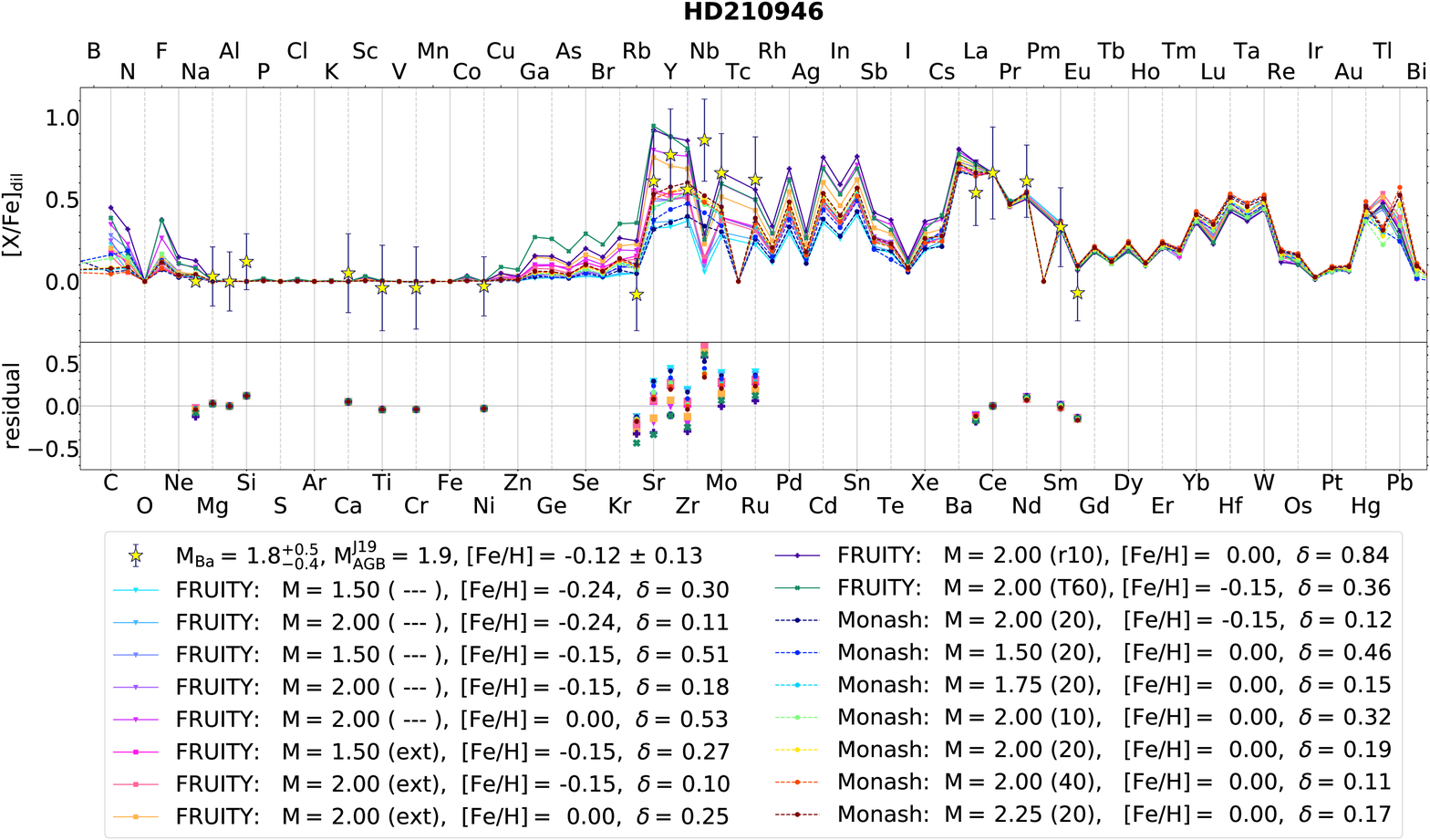}
 \end{figure*}

\subsubsection{HD 210946 (Fig.~\ref{fig:HD210946})}

A large variety of models with different metallicities between 1.5 and 2.25 \msun~are shown. The first $s$-process peak of this star shows higher abundances than the second peak elements. The higher first peak elements along with Rb are in agreement with non-rotating solar metallicity models around 2 \msun. This is in agreement with M$^{\rm{J}19}_{\rm{AGB}}$ of 1.9 \msun. Nb is higher than the model predictions, as this star belongs to subgroup 1b.

\begin{figure*}[!ht]
 \caption{Same as Fig. \ref{fig:HD154430} but for HD 211594}
 \label{fig:HD211594}
 \centering
 \includegraphics[width=\hsize]{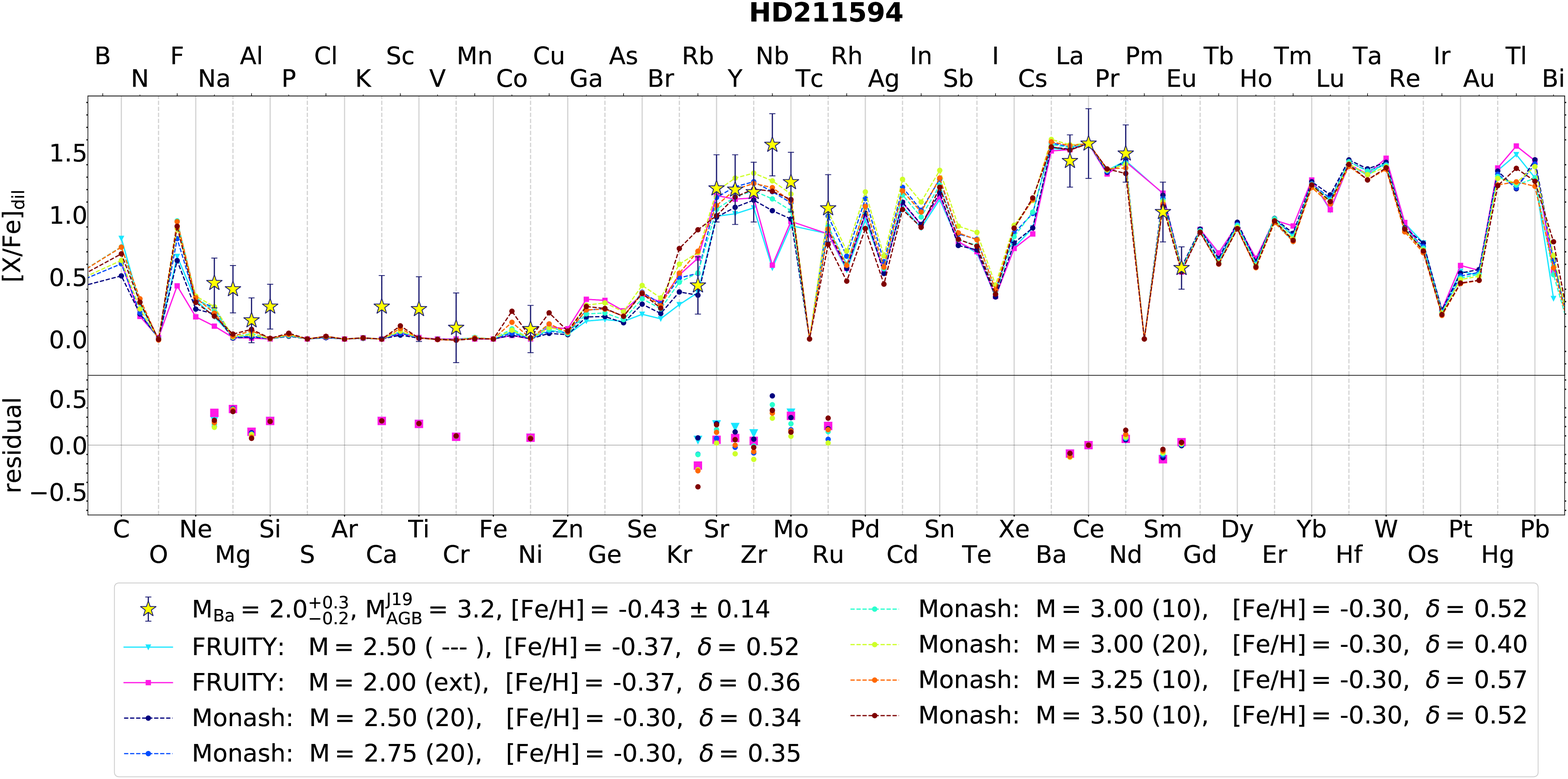}
 \end{figure*}

\subsubsection{HD 211594 (Fig.~\ref{fig:HD211594})}
 
FRUITY models with mass between 2.0 and 2.5 \msun~and Monash models between 2.5 and 3.5 \msun~are shown for this star. Higher mass FRUITY models have larger $\delta$ values than our limit of 0.9. The models match the $s$-process peaks well, except from Nb, which is higher as for the stars in subgroup 1b. The [Rb/Fe] value and the first peak elements point to a $\approx$3 \msun~polluter AGB, in agreement with the estimated initial mass of 3.2 \msun. Note that high $\delta$ values (larger than 0.3) are necessary to match the abundances pattern. 

\subsection{Group 1c: observed Nb, Mo, Ru, Nd, and Sm abundances higher than the model predictions}

\begin{figure*}[!ht]
 \caption{Same as Fig. \ref{fig:HD154430} but for BD $-$14$^{\circ}$2678.\newline}
 \label{fig:BD-142678}
 \centering
 \includegraphics[width=\hsize]{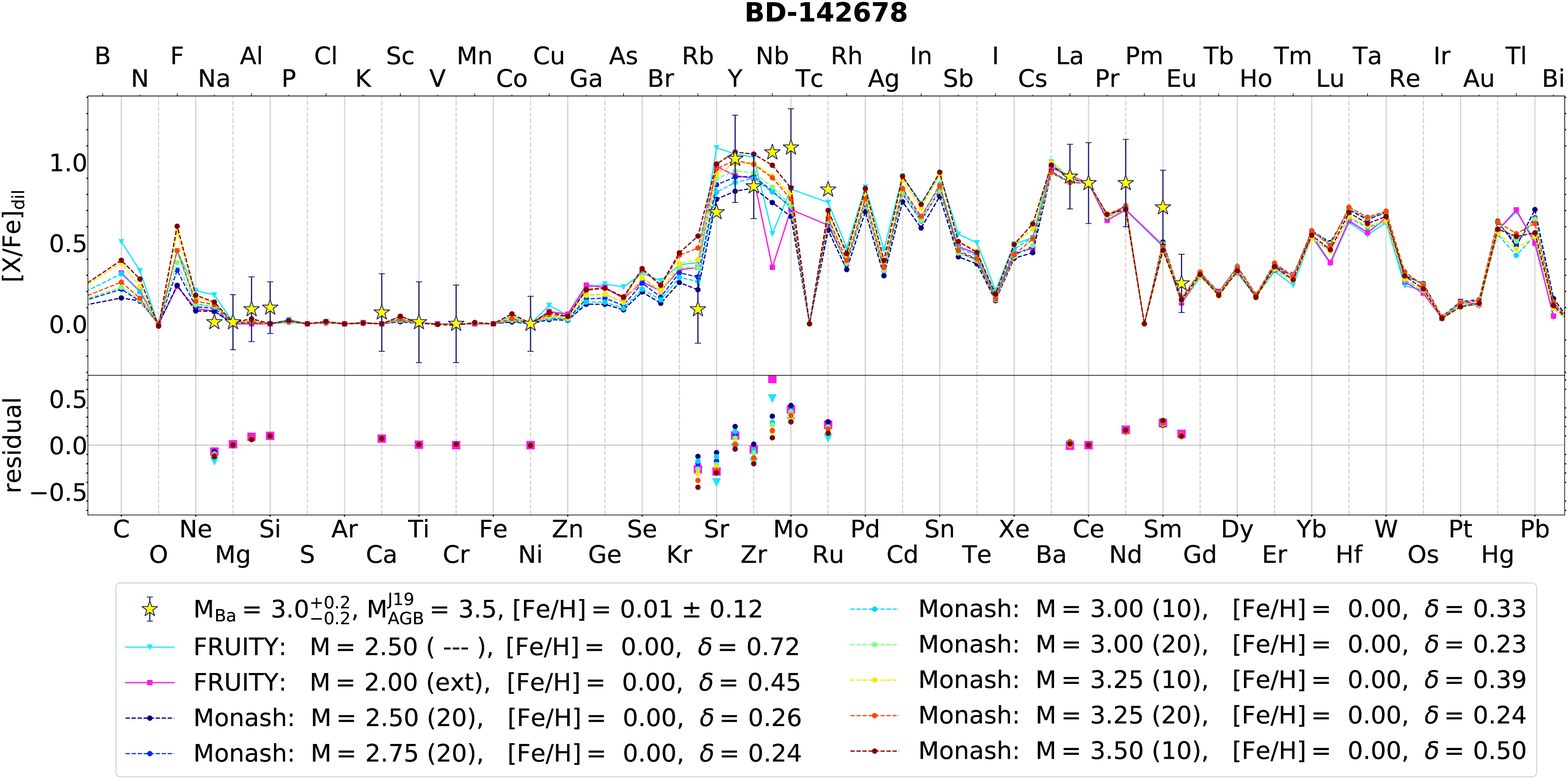}
 \end{figure*}

\subsubsection{BD $-$14$^{\circ}$2678 (Fig.~\ref{fig:BD-142678})}
 
BD $-$14$^{\circ}$2678 has M$^{\rm{J}19}_{\rm{AGB}}$ of 3.5 \msun. The $s$-process peaks, as well as the light elements are in agreement with models in the mass range of 2--3.5 \msun, although based on the determined Rb abundance the polluting AGB might be $\approx$ 3 \msun~or lower, rather than higher. Except from the 2.5 \msun~model, all FRUITY models with standard pocket in the mass range of 2.0--4.0 \msun~have higher $\delta$ values than our limit of $\delta = $ 0.9. We show also 2 \msun~FRUITY 'TAIL' and rotating 'TAIL' models for comparison, although these models overestimate the first peak abundances compared to the observations, and thus are not a good match. As this star belongs to subgroup 1c, Nb, Mo, Ru and Sm are somewhat higher than the model predictions, although Nb and Ru have no error bar.


\begin{figure*}[!ht]
 \caption{Same as Fig. \ref{fig:HD154430} but for CPD $-$64$^{\circ}$4333}
 \label{fig:CPD-644333}
 \centering
 \includegraphics[width=\hsize]{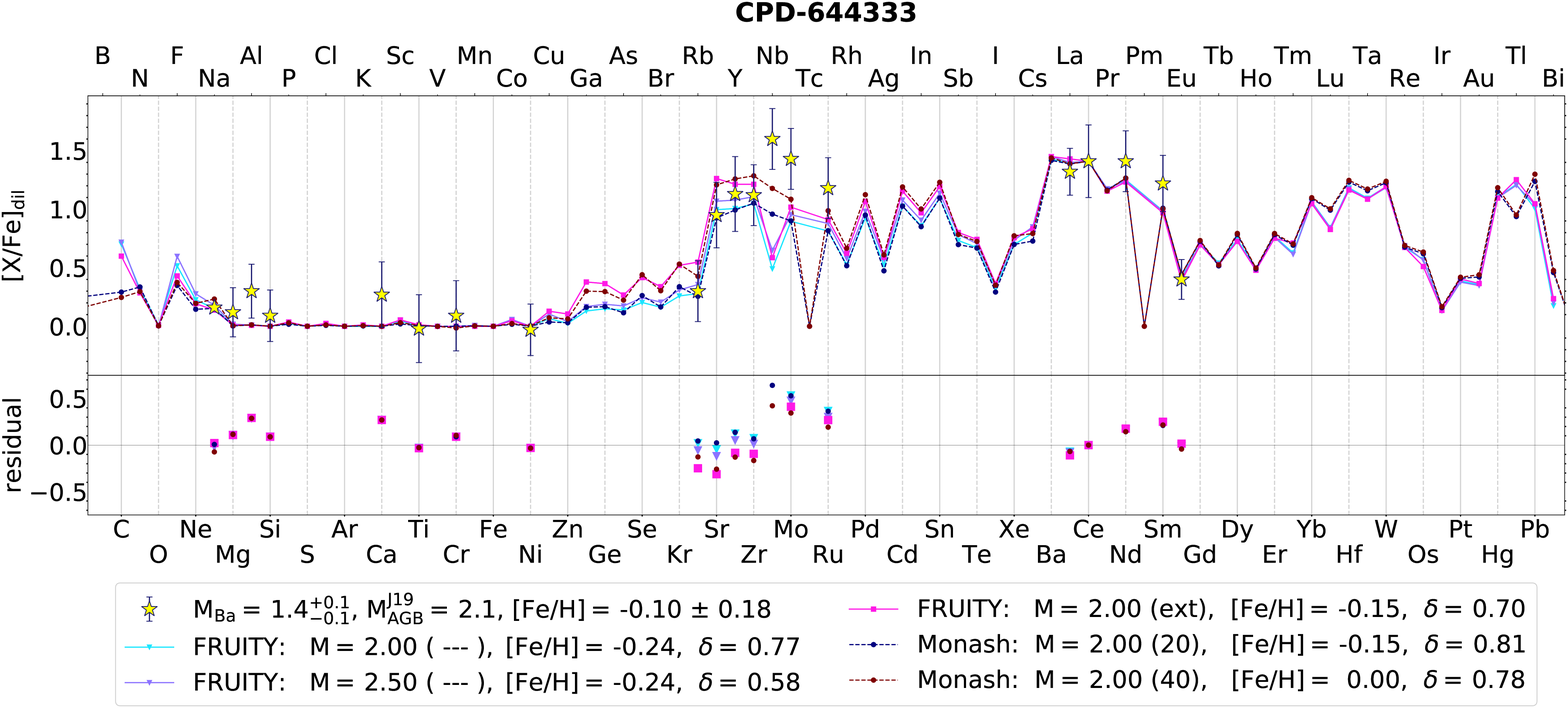}
 \end{figure*}

\subsubsection{CPD $-$64$^{\circ}$4333 (Fig.~\ref{fig:CPD-644333})}

CPD $-$64$^{\circ}$4333 has a very short orbital period close to 1 yr. Since Gaia DR2 parallaxes are calculated from single-star solution, this could result in an unreliable mass of the Ba star and thus M$^{\rm{J}19}_{\rm{AGB}}$ of $\approx$ 2 \msun~can be very uncertain. This is supported by the value of the RUWE for this star of 4.181 \citep{gaia_ruwe}, indicating problems with the single-star astrometric solution due to the short orbital period. Here we show different models with 2 and 2.5 \msun. All of the models are matching the determined abundances well, in agreement with M$^{\rm{J}19}_{\rm{AGB}}$. However, the models show high $\delta$ values, which could be explained by the short period, and thus larger mass transfer efficiency in the system. The elements after the $s$-process peaks show higher values than the model predictions, as for other stars in subgroup 1c.

\begin{figure*}[!ht]
 \caption{Same as Fig. \ref{fig:HD154430} but for HD 24035}
 \label{fig:HD24035}
 \centering
 \includegraphics[width=\hsize]{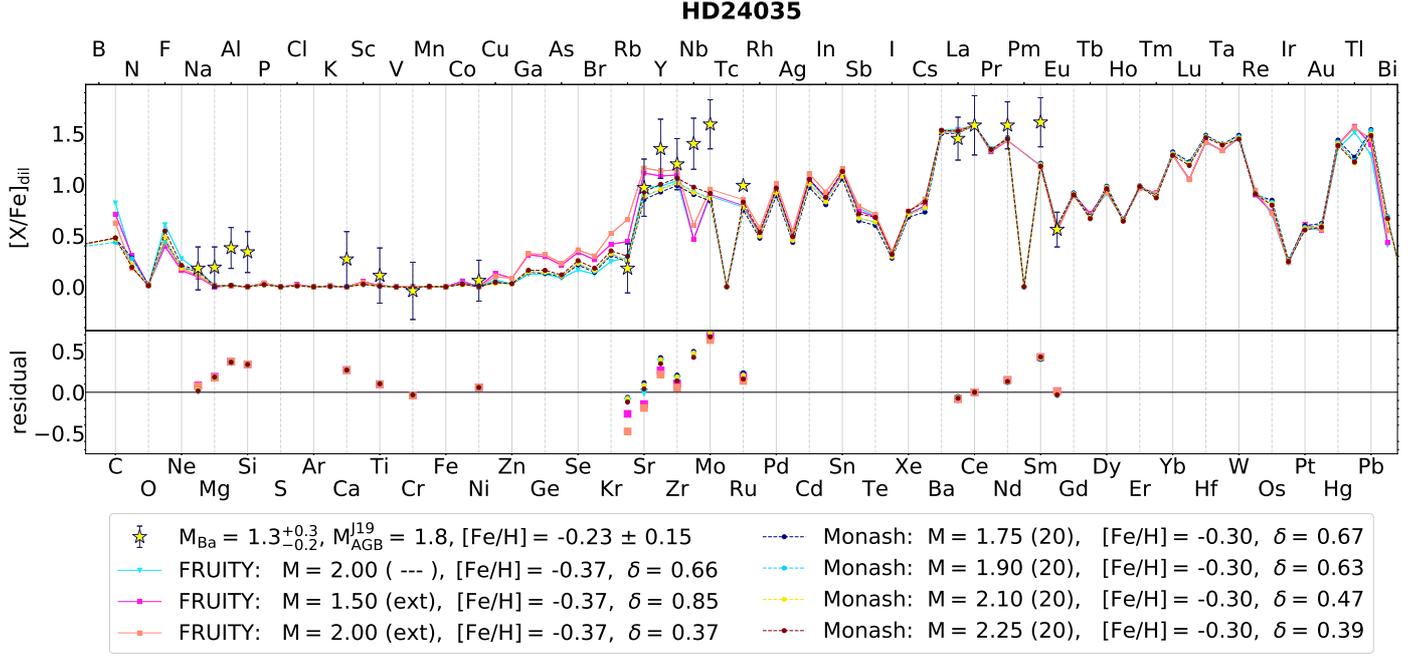}
 \end{figure*}

\subsubsection{HD 24035 (Fig.~\ref{fig:HD24035})}

HD 24035 has the shortest orbital period in our sample with 377.8 days. We show models between 1.5 and a 2.25 \msun.
Both of the $s$-process peaks are matched, but the elements after the peaks show higher abundances than the model predictions. The [Rb/Fe] value suggests that an AGB star with ~2 \msun~was the polluter in the system, in agreement with M$^{\rm{J}19}_{\rm{AGB}}$. The models have high $\delta$ values, but as for CPD $-$64$^{\circ}$4333, the short orbital period of the system could explain this phenomenon.

\cite{Shejeelammal20} also analysed this star and found an initial AGB mass of 2.5 \msun~based on $\chi^2$ minimalisation using FRUITY models. They have found [Fe/H] = $-$0.5 and generally higher $s$-process abundances than the data in this study.


\begin{figure*}[!ht]
 \caption{Same as Fig. \ref{fig:HD154430} but for HD 183915}
 \label{fig:HD183915}
 \centering
 \includegraphics[width=\hsize]{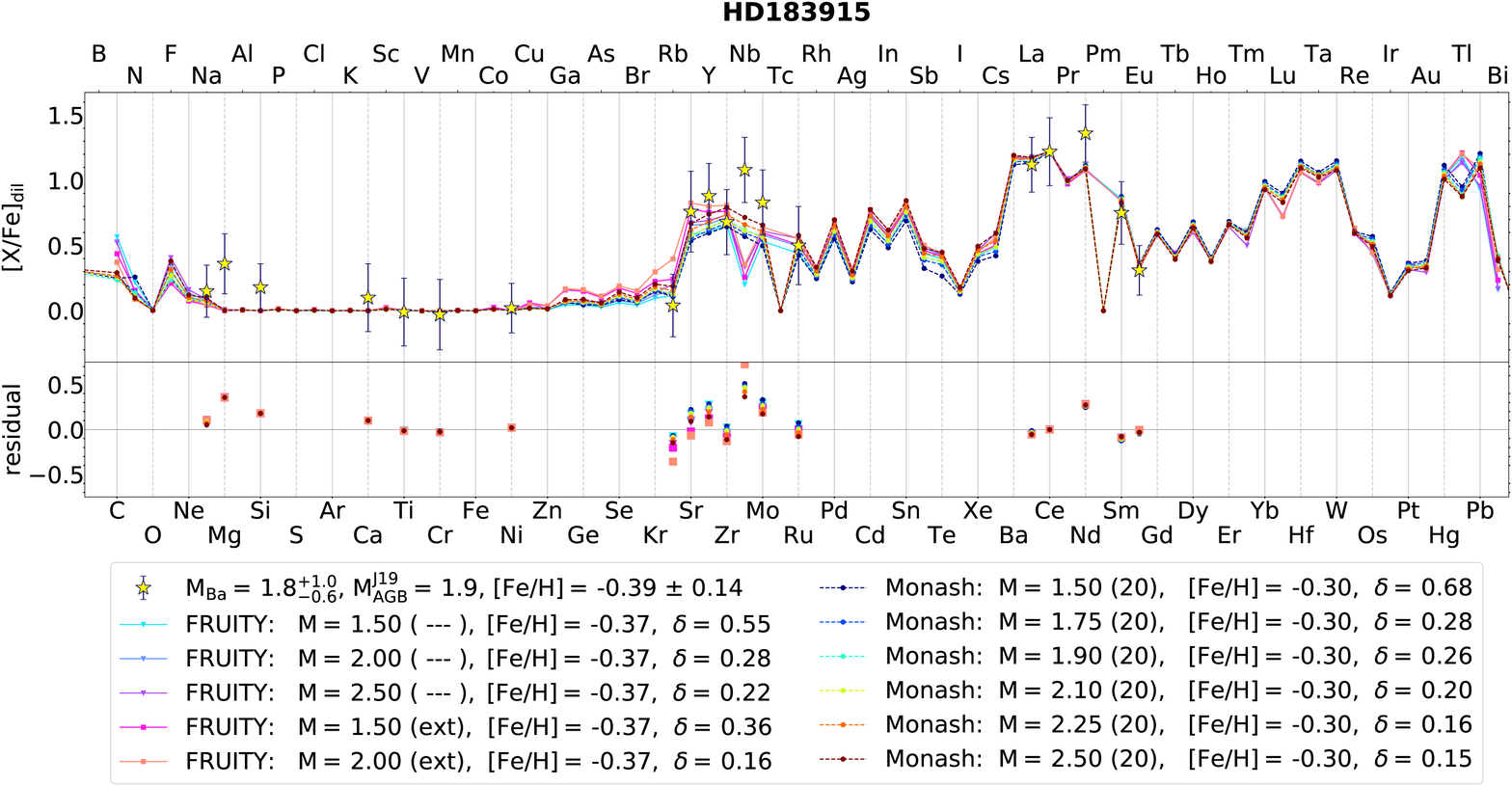}
 \end{figure*}

\subsubsection{HD 183915 (Fig.~\ref{fig:HD183915})}
 
Considering the large uncertainty of the Ba star mass, we plot models in the mass range between 1.5 and 2.5 \msun. All of these models are able to reproduce the abundance pattern at the $s$-process peaks of this Ba star, however, the 2.0 \msun~FRUITY 'TAIL' model is overestimating the [Rb/Fe] value compared to that of the star and Nb, Mo and Nd are higher than the model predictions. 

\clearpage

\section{Tables}
\label{appendix_models}

All models with \iso{13}C pocket in the metallicity range between [Fe/H] = $-$0.67 and 0.15 and between $-$0.7 and 0, and with mass from 1.3 to 3 and from 1.25 to 4.5 \msun~(FRUITY, Table \ref{tab:modelsFRUITY} and Monash, Table \ref{tab:modelsMonash}, respectively) used for comparison for at least one of the Ba stars in the sample. See Sect. \ref{sec:models} for the description of the models.

\begin{table*}[!htbp]
    \centering
    \caption{FRUITY models used in this study.}
    \label{tab:modelsFRUITY}
\begin{tabular}{|c|c|c|c|c|c|c|}
    \hline
    \diagbox[width=2.0cm, font=\Large, innerrightsep=0pt, innerleftsep=0pt]{$\substack{\mathrm{M} \\ (\mathrm{M}_\odot)}$}{$\substack{\mathrm{[Fe/H]} \\ \mathrm{(dex)}}$} & $-$0.67 &  $-$0.37 & $-$0.24 & $-$0.15 & 0.00 & 0.15 \\ \hline
    1.3 & ST & ST & & & & \\ \hline
    1.5 & ST, ext & ST, ext & ST & ST, ext, T60 & ST & \\ \hline
    2 & ST, ext & ST, ext & ST & ST, ext, T60 & ST, ext, r10, r30 & \\ \hline
    2.5 & & ST & ST & ST & ST & ST \\ \hline
    3 & & ST & ST & ST & ST & ST \\ \hline 
    \end{tabular} 
    \tablefoot{ST means the standard \iso{13}C pocket size, while ext, T60, r10 and r30  stands for 'TAIL', rotating 'TAIL' (60 km/s), and rotating models with 10 and 30 km/s initial rotation velocity, respectively.}
\vspace*{0.2 cm}
    \centering
    \caption{Monash models used in this study.}
    \label{tab:modelsMonash}
    \begin{tabular}{|c|c|c|c|c|}
    \hline
    \diagbox[width=2.0cm, font=\Large, innerrightsep=0pt, innerleftsep=0pt]{$\substack{\mathrm{M} \\ (\mathrm{M}_\odot)}$}{$\substack{\mathrm{[Fe/H]} \\ \mathrm{(dex)}}$} & $-$0.70 & $-$0.30 & $-$0.15 & 0.00 \\ \hline
    1.25 & 20, 60 & & & \\ \hline
    1.50 & 20, 60 & 20 & & 20 \\ \hline
    1.75 & 20 & 20 & & 20  \\ \hline
    1.90 &  & 20 & &  \\ \hline
    2.00 & 20, 60 & & 20 & 10, 20, 40 \\ \hline
    2.10 &  & 20 &  & \\ \hline
    2.25 &  & 20 & & 20  \\ \hline
    2.50 &  & 20 & & 20  \\ \hline
    2.75 &  & 20 & & 20  \\ \hline
    3.00 &  & 10, 20 & 20 & 10, 20 \\ \hline
    3.25 &  & 10 & & 10, 20 \\ \hline
    3.50 &  & 10 & & 10  \\ \hline
    3.75 &  & 10 & &  \\ \hline
    4.00 &  & 1, 10 & &  \\ \hline
    4.25 &  &  & & 10  \\ \hline
    4.50 &  &  & & 10 \\ \hline
    \end{tabular}
    \tablefoot{The \iso{13}C pocket size is given in the cells in units of 10$^{-4}$ \msun~for each model.}
\end{table*}

\end{appendix}
\end{document}